\documentclass[12pt]{article}
\usepackage{fullpage,hyperref,amssymb,amsmath}

\numberwithin{equation}{section}

\def\a{\alpha}\def\b{\beta}\def\g{\gamma}\def\d{\delta}\def\e{\epsilon}
\def\l{\lambda}\def\m{\mu}\def\n{\nu}\def\o{\omega}
\def\r{\rho}\def\s{\sigma}\def\t{\tau}\def\th{\theta}
\def\varphi{\varphi}
\def\G{\Gamma}\def\D{\Delta}\def\L{\Lambda}\def\O{\Omega}

\def\CA{\mathcal{A}}\def\CB{\mathcal{B}}
\def\CD{\mathcal{D}}\def\CE{\mathcal{E}}\def\CF{\mathcal{F}}
\def\CG{\mathcal{G}}\def\CH{\mathcal{H}}
\def\CJ{\mathcal{J}}\def\CK{\mathcal{K}}\def\CL{\mathcal{L}}
\def\CM{\mathcal{M}}\def\CN{\mathcal{N}}\def\CO{\mathcal{O}}
\def\CP{\mathcal{P}}

\def\CV{\mathcal{V}}\def\CX{\mathcal{X}}

\def\IZ{\mathbb{Z}}
\def\IR{\mathbb{R}}\def\IC{\mathbb{C}}
\def\IP{\mathbb{P}}

\def\gfrak{\mathfrak{g}}
\def\hfrak{\mathfrak{h}}

\def\zbar{{\bar z}}

\def\pd{\partial}

\def\bb{\mathbf{b}}
\def\bg{\mathbf{g}}\def\bh{\mathbf{h}}

\def\bI{\mathbf{I}}

\def\bgamma{{\boldsymbol{\gamma}}}\def\bomega{{\boldsymbol{\omega}}}
\def\bOmega{{\boldsymbol{\Omega}}}

\DeclareMathOperator{\diag}{diag}
\DeclareMathOperator{\tr}{tr}
\DeclareMathOperator{\Tr}{Tr}

\DeclareMathOperator{\Ad}{Ad}
\DeclareMathOperator{\re}{Re}  
\DeclareMathOperator{\im}{Im}  

\def\tilde{\widetilde}

\def\w{\wedge}

\def\ha{\dfrac12}\def\half{\tfrac12}
\def\qu{\dfrac14}\def\quarter{\tfrac14}
\def\ei{\dfrac18}
\def\iover#1{\tfrac{i}{#1}}

\long\def\symbolfootnote[#1]#2{\begingroup%
\def\thefootnote{\fnsymbol{footnote}}\footnote[#1]{#2}\endgroup}

\def\ltilde{{\tilde\l}}
\def\rtilde{{\tilde\r}}
\def\qtilde{{\tilde q}}
\def\htilde{{\tilde h}}
\def\Gtilde{{\tilde G}}
\def\Riem{\text{Riem}}
\def\phys{\text{phys}}
\def\WZW{\text{WZW}}

\def\gphys{{g_\phys}}
\def\lphys{\l_\phys}
\def\BH{B^\CM}

\begin{document}
\begin{titlepage}
\setcounter{page}{0}
\begin{flushright}
  arXiv:1106.6291\\ MIT-CTP 4276\\ NSF-KITP-11-133\\ UPR 1232-T
\end{flushright}
\vspace*{\stretch{1}}
\begin{center}
  \huge T-folds, doubled geometry,\\ and the $SU(2)$ WZW model
\end{center}
\vspace*{\stretch{0.75}}
\begin{center}\hskip5.5pt  
  \large Michael B. Schulz\symbolfootnote[1]{mbschulz at brynmawr.edu}
\end{center}
\begin{center}
  \textit{Department of Physics, Bryn Mawr College\\
    Bryn Mawr, PA 19010, USA\\
    and\\
    Center for Theoretical Physics, Massachusetts Institute of Technology\\
    Cambridge, MA 02139, USA\\}
    \end{center}
\vspace*{\stretch{1}}
\begin{abstract}
  \normalsize The $SU(2)$ WZW model at large level $N$ can be
  interpreted semiclassically as string theory on $S^3$ with $N$ units
  of Neveu-Schwarz $H$-flux.  While globally geometric, the model
  nevertheless exhibits an interesting \emph{doubled} geometry
  possessing features in common with nongeometric string theory
  compactifications, for example, nonzero $Q$-flux.  Therefore, it can
  serve as a fertile testing ground through which to improve our
  understanding of more exotic compactifications, in a context in
  which we have a firm understanding of the background from standard
  techniques.  Three frameworks have been used to systematize the
  study of nongeometric backgrounds: the T-fold construction,
  Hitchin's generalized geometry, and fully doubled geometry.  All of
  these double the standard description in some way, in order to
  geometrize the combined metric and Neveu Schwarz B-field data.  We
  present the T-fold and fully doubled descriptions of WZW models,
  first for $SU(2)$ and then for general group.  Applying the
  formalism of Hull and Reid-Edwards, we indeed recover the physical
  metric and $H$-flux of the WZW model from the doubled description.
  As additional checks, we reproduce the abelian T-duality group and
  known semiclassical spectrum of D-branes.

  \end{abstract}
  \vspace*{\stretch{5}}
  \flushleft{30 June 2011}
\end{titlepage}


\tableofcontents
\newpage


\section{Introduction}
\label{sec:Intro}

Over the last two decades, a beautiful picture has emerged for the
geometric encoding of low energy quantum field theories in string
theory.  Results from the early 1990s for Calabi-Yau compactifications
related field content to Betti numbers~\cite{Candelas:1985en}, and
interactions to intersection numbers in classical and quantum
cohomology rings~\cite{Strominger:1985it}, with mirror symmetry
relating the
two~\cite{Dixon,Lerche:1989uy,Candelas:1989hd,Greene:1989cf,Greene:1990ud,Aspinwall:1990xe}.
From physical reasoning, we gained early hints at the rich
interrelation between complex and symplectic geometry, which despite
two decades continues to bear fruit as a fertile area of mathematical
investigation integrating many subdisciplines.  (For an overview, see
Refs.~\cite{Hori:2003ic,Bridgeland:2009}.) The duality revolution of
the mid 1990s brought us D-branes as the microscopic carriers of
Ramond-Ramond (RR) charge~\cite{Polchinski:1995mt}.  With it, came
AdS/CFT correspondence~\cite{Maldacena:1997re} and an understanding of
the role in compactifications of microscopically consistent objects
that appear to violate classical energy conditions\footnote{For
  example, orientifold planes and D-branes wrapped on cycles of $\int
  R\w R <0$ display this
  property.}~\cite{Becker:1996gj,Sethi:1996es,Dasgupta:1999ss,Verlinde:1999fy,Chan:2000ms}.
These objects evaded classical no-go theorems forbidding internal
Neveu-Schwarz and Ramond-Ramond fluxes~\cite{Maldacena:2000mw}.  The
fluxes generate a superpotential~\cite{Gukov:1999ya} which was studied
systematically beginning in the early
2000s~\cite{Dasgupta:1999ss,Giddings:2001yu,Kachru:2002he}, and which,
together with instanton effects, generically lifts all moduli in type
IIB string theory~\cite{Kachru:2003aw}.\footnote{The presence of
  moduli immediately renders a compactification unrealistic.  The
  moduli fields generate long range interactions not observed in
  nature, and lead to overclosure of the universe in cosmology.}
Virtually all model building in type II string theory today takes
place in orientifold models with internal flux. (See, e.g.,
Ref.~\cite{Denef:2008wq}).

Yet, this story is not complete.  The combination of flux and mirror
symmetry highlighted the insufficiency of standard \emph{geometric}
compactifications to describe the full set of topological data
describing a string theory compactification.  In the context of
effective field theory, the problem is easy to understand.  For
example, consider KKLT type compactifications of type IIB string
theory~\cite{Kachru:2003aw}.  Ignoring the subsector of the theory
from localized objects (\hbox{D-branes} and singularities), the matter
comes from $H^2$, and the gauge fields from $H^3$.  On the other hand,
the fluxes are cohomology classes in $H^3$.  We cannot achieve the
most general gauge couplings in the low energy effective field theory,
since the counting is only sufficient to couple the gauge fields to a
single matter multiplet (from the universal hypermultiplet).
Moreover, the gauge group is abelian.  The question naturally arises,
how to realize general gaugings (gauge group plus couplings) of the
compactified supergravity theory.  While well understood in the
context of localized objects, this is still poorly understood in the
bulk.

Duality arguments~\cite{Kachru:2002sk,Shelton:2005cf} suggest that the
basic problem is not a fundamental constraint of the microsopic
theory, but an insufficiency of the standard hierarchical choice of
topological data to describe the most general compactifications.  One
cannot first choose the topology of a compact manifold, and then
subsequently choose the topology of various other bundle-like objects
on that manifold.  For the Riemannian structure, one cannot first
choose the curvature of spacetime, and then the curvature of various
fields living in that spacetime.  The two are intrinsically
interrelated, and must be desribed as a whole.  From the point of view
of string theory, this is not so suprising.  After all, everything is
built from one object, strings.  However, this also implies that
everything is \emph{probed} by strings, and the corresponding notion
of spacetime geometry need not agree with our intuition from a
spacetime that can be probed by point particles.

In
Refs.~\cite{Hull:2004in,Hull:2006va,Hull:2007jy,Dall'Agata:2008qz,Hull:2009sg}
it was argued that the appropriate mathematical setting for realizing
the integrated approach to string compactifications motivated in the
previous paragraph is a doubled spacetime roughly thought of as that
seen simultaneously by left and right moving degrees of freedom.  For
example, a string on a circle carries momentum and a winding number,
or equivalently, independent left and right momenta.  The space of
states of a single such string is spanned by
$|p_L,p_R,N_\text{osc}\rangle$, where $N_\text{osc}$ describes the
modes of oscillation.  In a Fourier transformed basis, this becomes
$|x_L,x_R,N_\text{osc}\rangle$, with $x_L$ and $x_R$ independent
parameters, which we would like to interpret as coordinates on a space
of twice the usual dimension.  While this argument invoked a circle,
it is clear that the same momentum and position parameters would
persist as independent integration constants in locally solving the
string wave equation in an open set.  In general, one expects to
obtain a global discription by sewing together these open sets in a
suitable way.

Even in the context of the standard worldsheet sigma model, a similar
doubling is apparent.  For example, in conformal gauge
($\g_{ab}\propto\d_{ab}$), the worldsheet Polyakov action
\begin{equation}\label{eq:Polyakov}
  S = \int d^2\s\CL
  = \frac1{4\pi\a}\int d^2\s \bigl(\sqrt{\g}\g^{ab}
  \pd_a X^\m\pd_b X^\n + i\e^{ab}B_{\m\n}\pd_a x^m\pd_b X^n\bigr)
\end{equation}
gives Hamiltonian density
\begin{equation}\label{eq:HamiltonianDensity}
  \CH = \frac1{4\pi\a'}\Bigl(
  G_{mn}X'^m X'^n + G^{mn}(\a'p_m + B_{mp}X'^p)(\a'p_q + B_{nq}X'^q)
  \Bigr).
\end{equation}
Here, $p_m = (2\pi/\sqrt{\g})\pd\CL/\pd{\dot X^m}$ is the canonical
momentum, $\dot X^m(\s) = dX^m(\s)/d\s^2$, and $X'^m(\s) =
dX^m(\s)/d\s^1$.  If there is even a local description of a
compactification in terms of a standard geometric one, so that we can
locally apply this sigma model, with $X^m(\s)$ embedding the string
worldsheet into an open manifold $M$, then $(X'^m,p_m)$ can be thought
of as valued in the tangent plus cotangent bundle $(T\oplus T^*)M$.
The Hamiltonian density gives a Riemannian structure on $(T\oplus
T^*)M$.  There is a canonical $O(d,d)$ invariant structure as well,
from the quantity $p_mX'^m/(4\pi)$.  Here $d=\dim(M)$.  The integrated
sum and difference of these two quantities give the Virasoro
generators $L_0$ and $\bar L_0$ in the left moving and right moving
worldsheet sectors.  Thus, the local worldsheet geometry of the
compactification furnishes a Riemannian structure and a (constant)
$O(d,d)$ structure on $(T\oplus T^*)M$.

The geometry of $(T+T^*)M$ was explored by Hitchin and Gualtieri in
Refs.~\cite{Hitchin:2004ut,Gualtieri:2003dx}, pioneering a branch of
mathematics called generalized complex geometry.  A foundational
observation is that a complex structure $\CJ^M{}_N$ on $(T+T^*)M$ can
be constructed from either a complex structure $J^m{}_n$ on $M$ (from
$\CJ=\diag(J,-J^T)$), or a symplectic structure $\o_{mn}$ on $M$
(putting $\o$ and $\o^{-1}$ on the off-diagonal).  This allows one to
unify the rich complex and symplectic geometric structures of mirror
symmetry as well as interpolate between the two.  We will refer to the
geometry of $(T+T^*)M$ as generalized geometry, whether or not complex
structures are introduced.

Generalized complex geometry partially rises to the task of supplying
the missing supergravity gaugings of type II
compactifications~\cite{Grana:2005ny,Grana:2006hr}.  Pure spinors on
the doubled tangent bundle correspond to maximal isotropic subbundles,
i.e., $d$-dimensional subbundles that are null with respect to the
$O(d,d)$ metric.  Thus a local pure spinor determines a local
projection to a $d$ dimensional subbundle, which generalizes $TM$.
Pure spinors also define almost complex structures of the generalized
geometry reducing the $O(d,d)$ structure to $U(d/2,d/2)$.  Via a map
to differential forms, the pure spinors corresponding to the
symplectic and complex structures of a Calabi-Yau manifold become
$e^{-\o}$ and the holomorphic $(d,0)$-form $\O$, and the structure
group is reduced to $U(d/2,d/2)_\o\cap U(d/2,d/2)_\O = SU(d/2)\times
SU(d/2)$~\cite{Gualtieri:2003dx}.  Calabi-Yau 3-folds give a
generalized geometry with $SU(3)\times SU(3)$ \emph{holonomy}.  When
the spinors are instead covariantly conserved by a connection with
torsion, there is $SU(3)\times SU(3)$ \emph{structure}.  In the latter
case, the map of pure spinors to differential form need not give $\o$
and $\O$ of definite degree, and the torsion need not map $p$ forms to
$p+1$ forms.  Instead we obtain torsion data furnishing maps of the
form
\begin{equation}\label{eq:maps}
  K_{mnp}\colon\
  \begin{matrix}
    \O^0\to \O^3\\ \O^3\to \O^6
  \end{matrix},
  \qquad f_{mn}{}^p\colon\
  \begin{matrix}
    \O^2\to \O^3\\ \O^3\to \O^4
  \end{matrix},
  \qquad Q^{mn}{}_p\colon\
  \begin{matrix}
    \O^3\to \O^2\\ \O^4\to \O^3
  \end{matrix},
  \qquad R^{mnp}\colon\
  \begin{matrix}
    \O^3\to \O^0\\ \O^6\to \O^3
  \end{matrix}.
\end{equation}
The same data can be used to define a a Roytenberg bracket giving the
generalized geometry the structure of an $\a$-Lie $\b$-algebroid,
where the appropriate prefixes $\a$ and $\b$~\cite{Halmagyi:2009te}
depend on which of $K,f,Q,R$ are nonzero.  Note that
Eq.~\eqref{eq:maps} is exactly the data required to furnish the full
set of $\CN=2$ gauged supergravity
couplings~\cite{Grana:2005ny,Grana:2006hr}.\footnote{What is needed is
  matrix of data to contract the K\"ahler and complex structure
  symplectic period vectors.}  The main deficiency of this approach is
that it generically describes a string theory compactification only
locally when $Q$ is nonzero, and does not appear to apply even locally
when $R$ is nonzero.  As discussed in connection with
Eq.~\eqref{eq:HamiltonianDensity}, generalized geometry applies
locally, when compactification locally has a description in terms of a
standard sigma model.

What, then, is the global structure, and what do we mean by a
\emph{nongeometric} compactification?  In this context, there are two
basic constructions,
monodrofolds~\cite{Hellerman:2002ax,Dabholkar:2002sy,Hellerman:2006tx},
and doubled geometry~\cite{Hull:2007jy,Dall'Agata:2008qz,Hull:2009sg}.
A monodrofold can be thought of a two step compactification.  In the
first step, we are given a compactification having a discrete gauge
symmetry $\G_\text{modular}$.  For example, geometric moduli spaces
take the form of a noncompact Teichm\"uller space quotiented by a
modular group.  For the common NSNS sector of a toroidal string theory
compactification on $T^d$, the metric and $B$-field moduli parametrize
a coset space of the form $O(d,d)/\bigl(O(d)\times O(d)\bigr)$ and are
identified under the action of the T-duality group $O(d,d;\IZ)$.
Then, in further compactifying on a base manifold, in the second step,
it is possible to specify nontrivial monodromies in the duality group,
when traversing noncontractible loops in the base.\footnote{See
  Ref.~\cite{Hellerman:2006tx} for a generalization of the orbifold
  construction, which gives the modular invariant partition functions
  of flat monodrofolds over $S^1$.  Here, flat means that the
  monodromies have fixed points so that the moduli do not have to vary
  over the base $S^1$.}  In respose to the question, ``is the result a
CFT?'' we take the point of view that the monodromies are part the
basic topological data in defining the worldsheet sigma model.
Conformality is a dynamical question for the effective field theory,
whose equations of motion \emph{are} the vanishing beta function
conditions of the worldsheet theory.\footnote{There may or may not be
  static solutions.  The data may give time dependent ``runaway''
  solutions or domain wall solutions where the volume modulus or
  dilaton runs away pathologically, as is true for generic
  supergravity gaugings.  Only certain choices of topological data
  will lead to physically interesting results.}\footnote{When the base
  manifold is large, the monodromies require small spatial gradients
  of fields, and correspondingly small masses for the lifted moduli,
  so that the relevant effective field theory truncations are
  justified.  However, we will not worry about this a priori.  The
  supergravity regime is always in some sense nongeneric in a theory
  in which the basic unit is string scale.  In any realistic
  compactification, one requires a small parameter so that corrections
  are controlled.}

When the T-duality group is used in this construction, the result is
called a T-fold~\cite{Dabholkar:2002sy,Hull:2004in,Hull:2006va}.  In
this case, the mondromies generically mix the metric and $B$-field of
the fiber theory, so that neither is globally well defined.  For
example, consider a $T^2$.  In this case, the T-duality group is
$O(2,2;\IZ)$, which is the same as $SL(2,\IZ)\times SL(2,\IZ)$ up to
$\IZ_2$ identifications.  The latter acts by fractional linear
transformation on the complex structure modulus $\t$ and complexified
K\"ahler modulus $\r = b+iv$, where $b = \int B$ on the $T^2$, and $v$
is the $T^2$ volume.  Now further compactify on a circle.  Monodromies
$\rho\mapsto\rho+N$ or $\t\mapsto\t+N$ correspond to $H$-flux, and a
nontrival $T^2$-bundle, respectively.  However, $-1/\r\mapsto -1/\r +
N$ is also a perfectly good monodromy and mixes the metric and
$B$-field.  In this case, one finds~\cite{Kachru:2002sk}
\begin{equation}
  ds^2 = dz^2 + \frac1{1+(Nz)^2}(dx^2+dy^2),\quad
  B_{xy} = -\frac{Nz}{1+(Nz)^2},
\end{equation}
where $x,y$ are coordinates on the $T^2$ fiber, and $z$ is the
coordinate on the $S^1$ base.  Locally, this gives a perfectly well
defined metric and $B$-field.  Globally, neither is single valued on
$S^1$.\footnote{This particular example can be converted to a
  geometric one by a global T-duality transformation $(\r',\t') =
  (-1/\r,\t)$.  However, we choose monodromy $(\r,\t)\mapsto
  (-1/\r,-1/\t)$ rather than
  $(\r,\t)\mapsto\bigl(-1/(-1/\r+N),\t\bigr)$, then the result is
  nongeometric in all T-duality frames.}  The next simplest T-fold
base after $S^1$ is $\IP^1$.  Via F-theory/heterotic duality, it is
also relatively straightforward to describe T-folds over $\IP^1$ base,
which generalize heterotic compactifications on
K3~\cite{McOrist:2010jw}.

While T-folds do not have a global geometry or topology in the
conventional sense, they do in a doubled sense~\cite{Hull:2004in}.  If
the fiber is doubled to include both the physical torus $T^d$ and
\hbox{T-dual} torus $\tilde T^d$, then the $O(d,d)$ monodromies act
linearly as transition functions on the fiber.  The full specification
of topological data consists of the topology of the doubled fibration,
together with a choice of $B$-field on the base.  The metric and
$B$-field on the fiber are neatly packaged into a single Riemannian
metric of the same form as Eq.~\eqref{eq:HamiltonianDensity}, which is
well defined on the doubled fibration.  Thus, the doubled fibers
carries a very similar structure to that described in the context of
generalized geometry, except that \emph{the space itself is doubled},
not just the tangent bundle.  To locally recover the standard
nondoubled description requires a choice of polarization,\footnote{In
  fact, not only the local recovery, but the formulation of the
  doubled worldsheet theory itself requires a choice of polarization
  in order to specify a gauging (cf.~Sec.~6.1.2 of
  Ref.~\cite{Hull:2004in}).}  that is, a choice of dual $\tilde T^d$
in the $T^{2d}$ fiber, which must be null with respect to the $O(d,d)$
metric.  A global polarization defines a global projection from the
total space of the doubled fibration to the total space of the
physical $T^2$ fibration.  When a global polarization exists, we
recover the standard sigma model description globally.  Polarizations
are thus much like the pure spinors of generalized geometry in that
they define physical subbundles of doubled bundles.

Doubled geometry takes the T-fold construction one step further, by
doubling all $d$
dimensions~\cite{Hull:2007jy,Dall'Agata:2008qz,Grana:2008yw,Hull:2009sg}.
It has been applied to generalizations of toroidal compactifications,
and furnishes the gaugings of the common NSNS sector of their
supergravity theories.  Just as gauged analogs of a Calabi-Yau
compactification have been realized via generalized geometry with
$SU(3)\times SU(3)$ structure, so too in the trivial holonomy case, in
the doubled geometry context, we seek a geometry with \emph{identity}
structure, i.e., a parallelizable manifold.  For technical reasons we
further restrict to ``consistent absolute
parallelism''~\cite{Grana:2008yw} where the metric is constant in the
frame basis.  The possibilities are $S^7$ or a Lie group, and we
discard the former since it is not $2d$ dimensional.
Compactifications based on a fully doubled group manifold, as opposed
to the torus-fiber doubling of Ref.~\cite{Hull:2004in} or circle-base
doubling of Ref.~\cite{Dabholkar:2005ve}, were first considered by
Dall'Agata and Prezas in Ref.~\cite{Dall'Agata:2008qz}.  They were
subsequently studied by Gra\~na et al.\ in Sec.~5.3 of
Ref.~\cite{Grana:2008yw}, and in depth by Hull and Reid-Edwards in
Refs.~\cite{Hull:2007jy,Hull:2009sg}.

Starting from the doubled space
\begin{equation}
  \CX_{2d} = \G\backslash\CG_{2d},
\end{equation}
where $\G$ is a discrete subgroup, Hull and Reid-Edwards build a
formalism for describing compactifications that yield gauged
supergravities with gauge group
$\CG_{2d}$~\cite{Hull:2007jy,Hull:2009sg}.  Given a Lie algebra frame
$T_M$, the left-invariant forms $\CP^M$ furnish a coframe.  The
additional required data beyond $\CG_{2d}$ and $\G$ is a constant,
signature $(d,d)$ symmetric metric $L_{MN}$ which must be invariant
under the action of $\CG_{2d}$.

The formulation of a worldsheet theory~\cite{Hull:2009sg} on this
doubled space is a true generalization of the Polyakov
action~\eqref{eq:Polyakov}.\footnote{See also
  Ref.~\cite{Dall'Agata:2008qz} for a complementary approach.}  To
locally recover the standard action, when this is possible, requires a
choice of polarization on the tangent bundle of $\CX_{2d}$.  Given a
choice of polarization, it is convenient to choose a frame $T_M =
(Z_m,X^m)$ putting the $O(d,d)$ metric $L_{MN}$ into canonical form,
with identity matrices off diagonal.  One then defines a Riemannian
metric, of the form~\eqref{eq:HamiltonianDensity}.  The metrics in
other polarizations are locally related by $O(d,d)$ transformation.
$\CG_{2d}$-invariance of $L_{MN}$ implies that the Lie algebra takes
the form
\begin{equation*}
  \begin{split}
    [Z_m,Z_n] & =  K_{mnp}X^p + f_{mn}{}^pZ_p,\\
    [Z_m,X^n] & =  f_{pm}{}^nX^p + Q^{np}{}_mZ_p,\\
    [X^m,X^n] & =  Q^{mn}{}_pX^p + R^{mnp}Z_p.
  \end{split}
\end{equation*}
In the case that $R^{mnp} = 0$, the $X^m$ close to form a subgroup
$\tilde G_d\subset\CG_{2d}$, which can be quotiented out, to leave the
physical $d$-dimensional geometry $X_d$.  This is a local statement,
and whether or not $R^{mnp}=0$ depends upon the local choice of
polarization.  In this case, the worldsheet model of Hull and
Reid-Edwards indeed agrees with the standard Polyakov action.  For
such polarizations, it is natural to seek a relation between the
doubled geometry of Hull and Reid-Edwards, and the corresponding
generalized geometry.  This has been done in detail for the special
cases with $f,K$ or $f,Q$ nonzero in
Ref.~\cite{ReidEdwards:2010vp}. In these cases, the Lie algebra of the
doubled gauge group $\CG_{2d}$ indeed agrees with the Roytenberg
bracket on $(T+T^*)X_d$.
\bigskip

In this paper we present the T-fold and doubled geometry desciptions
of Wess-Zumino-Witten (WZW) models at large level $n$, emphasizing the
special case of $SU(2)$.  The $SU(2)$ case describes strings
propagating on a 3-sphere of radius $\sqrt{n\a'}$ with $n$ units of
Neveu-Schwarz $H$-flux.  Our motivations for studying WZW models in
this context are as follows.  (For earlier work on WZW models in
doubled and generalized geometry, see
Refs.~\cite{Dabholkar:2005ve,Dall'Agata:2008qz,Halmagyi:2009te}.)

\begin{enumerate}
\item We have outlined three approaches to nongeometric string theory
  compactifications above: T-folds, generalized geometry, and doubled
  geometry.  It would be interesting to further clarify the relation
  between them.  Additional examples are needed to elucidate the
  generalized geometry and doubled geometry approaches, particularly
  compact examples, and ideally one with a clear CFT description.  A
  noncontroversial example yielding interesting generalized and
  doubled geometries is well suited to this goal, even if geometric.
  And in this case, the generalized geometry is applicable globally.
\item The $SU(2)$ WZW model is somewhat counterintuitive.  The naive
  expectation is that it would have $f\ne0$, since $f$ is
  conventionally associated with a space whose \hbox{1-forms} close
  with torsion.  It would also have $K\ne0$, conventionally associated
  with $H$-flux.  The remaining $Q$ and $R$ would naively vanish,
  since these are conventionally thought of as obstructions to global
  and local geometry, respectively.  However, these structure
  constants do not describe an $SU(2)\times SU(2)$ gauge algebra.
  Instead, setting the $SU(2)$ generators equal to the sum and
  difference of $Z_m,X^m$ gives nonvanishing $K$ and
  $Q$.\footnote{That the $SU(2)$ WZW model has only $K$ and $Q$
    nonzero, or alternatively $f$ and $R$ nonzero, can be found in
    Refs.~\cite{Hull:2004in,Dall'Agata:2008qz}.  See also
    Ref.~\cite{Avramis:2009xi} for an interesting follow-up on
    \cite{Dall'Agata:2008qz}, which matches a worldsheet analysis to
    the gaugings of $\CN=4$ supergravity, and describes the gauging
    for all cases with one or two of $K,f,Q,R$ nonzero.}  Where did
  the naive intuition fail?
\item There are at least two discrete groups involved.  What fixes
  them?  The doubled geometry is proported to encompass not only the
  discrete abelian T-duality group that is a symmetry of string
  theory, but also the nonabelian and Poisson-Lie T-dualities, which
  generically fail beyond tree level.  What characterizes the
  restriction of the polarization choices to those of abelian
  T-duality?  And what fixes the discrete group~$\G$ in the definition
  of the doubled space?  For a WZW model at level $n$, it seems
  natural that the integer $n$ should show up in the defining
  topological data (for example, through a $\IZ_n$ factor in $\G$) and
  not just in the choice of polarization.
\item There has been only a modest amount of work on D-branes in the
  context of T-folds and doubled
  geometry~\cite{Hull:2004in,Albertsson:2008gq,Lawrence:2006ma}.
  However, they do have a relatively straightforward description in
  doubled geometry.  They wrap maximal isotropic submanifolds.  Do the
  D-brane predictions of doubled geometry agree with the well known
  results for WZW models?
\item Ultimately we are interested in an analogous doubled geometry
  for gauged analogs of Calabi-Yau compactifications.  Mirror pairs of
  Calabi-Yau manifolds are $T^3$ fibrations.  We know how to double
  the $T^3$ fiber, and even how to twist this
  fibration~\cite{Tomasiello:2005bp}.  How do we double the base?  The
  base is a rational homology sphere, which we can think of as
  analogous to an $S^3$.  For the quintic, it is $S^3/\IZ_5$.  The
  $SU(2)\cong S^3$ WZW model provides a case study of the doubled
  geometry of one such $S^3$, albeit a very special one with $H$-flux.
\end{enumerate}

\noindent An outline of the paper is as follows:\medskip

In Sec.~\ref{sec:PhysBg}, we fix notation, and describe the physical
metric and $H$-flux of a WZW model, first for $SU(2)$, and then for
general gauge group.  The reader is referred to App.~\ref{app:LieAlg}
for conventions and basic facts about Lie Algebras.  We allude to the
worldsheet description only minimally thoughout the paper.
App.~\ref{app:Ws} contains a review of the basic worldsheet and CFT
aspects of WZW models.

Sec.~\ref{sec:Tfold} is devoted to T-folds.  We review the definition
of a T-fold, including its $O(d,d)$ and Riemannian metrics, and the
procedure for recovering a physical background given a choice of
polarization.  The T-fold description of the WZW model for $SU(2)$ and
then for general gauge group are presented in Secs.~\ref{sec:TfoldSU}
and~\ref{sec:TfoldG}.  For $SU(2)$, the T-fold is a doubled Hopf
fibration quotiented by $\IZ_n$.  The physical $S^1$ fibration gives a
sphere, and dual $\tilde S^1$ fibration gives a Lens space, with
T-duality interchanging the two.  For a general group, the physical
fiber is the Cartan torus $T^r$, and half the dual coordinate is
valued in $(T^r)^*/(\IZ_n)^r$, where $(T^r)^*$ denotes the Cartan
torus of the dual gauge group.\footnote{For simply connected groups,
  $G^*$ is $G$ quotiented by its center.  Therefore, $SU(2)^* =
  SU(2)/\IZ_2$ and the two factors of 2 cancel in this case.}
T-duality again simply introduces a factor of $\IZ_n$ quotienting the
physical space for each $U(1)$ dualized.  The total space of the
T-fold is interpretated as the group manifold
$\big(U(1)^r\big)_L\times G^\WZW_R$, which suggests that for WZW
models, the T-fold is embedded in the fully doubled space as a
subgroup.\footnote{This contrasts to the example of a $T^3$ with
  $H$-flux, for which the T-fold appears to arise by partial projection
  from the fully doubled space to the physical base.}

Sec.~\ref{sec:Doub} is the heart of the paper.  We refer the reader to
the introduction of Sec.~\ref{sec:Doub} for a more complete overview
of the results of this section.  The first half covers generalities.
Subsecs.~\ref{sec:DoubGenSpace} through~\ref{sec:DoubGenEFT} provide a
review of the formalism of Hull and Reid-Edwards.
Sec.~\ref{sec:DoubGenRecovery} describes the recovery of the physical
from doubled geometry.  Here, we emphasize that \mbox{$Q$-flux} alone
is not an obstruction to global geometry, but rather the interplay of
$Q$ and $\G$; the condition for global geometry is $\G$-invariant~$Q$.
The double geometry of WZW models is discussed in
Sec.~\ref{sec:DoubG}.  The doubled space takes the form
\begin{equation*}
  \CX_{2d} = \G\backslash\bigl(G_1\times G_2),
\end{equation*}
where $G_1$ and $G_2$ are two copies of the physical WZW group
$G^\WZW$.  Global polarizations are choices of maximal isotropic
subgroup $\Gtilde$ conjugate to $G_{\diag}$.  Quotienting the doubled
space $\CX_{2d}$ by $\Gtilde$ gives the physical space $G^\WZW$.  In
the diagonal polarization, the projection to the physical target space
is $\pi\colon (g_1,g_2)\mapsto g_\phys = g_1^{-1}g_2$.  The symmetry
under right multiplication in $\CG_{2d}$ gives the $g_\phys\mapsto
\O_1 g_\phys \O_2^{-1}$ symmetry of the physical model.  We show that
the correct physical metric and $H$-flux are indeed recovered,
including the condition $r^2 = n\a'$.  In Secs.~\ref{sec:DoubGPol}
and~\ref{sec:DoubGTduality}, we consider general polarizations and
describe abelian T-duality.  The known results for semiclassical
D-branes in WZW models are reproduced in Sec.~\ref{sec:DoubGDbranes}.
Finally, in Sec.~\ref{sec:DoubGDiscrete}, we describe restrictions on
the discrete group $\G$, but do not fully resolve the question of what
$\G$ is at level $n$ for each choice of modular invariant.  In
Sec.~\ref{sec:Conclusions}, we conclude with a summary of results and
discussion of open questions.


\section{The 3D physical background}
\label{sec:PhysBg}


\subsection{Target space description of the $SU(2)$ WZW model}
\label{sec:PhysBgSU}

Consider a 3-sphere $S^3$ of radius $\sqrt{n\a'}$ (in string frame)
and $n$ units of $H$-flux,
\begin{equation}\label{eq:PhysBg}
  ds_\phys^2 = r^2 ds^2_{S^3},
  \quad r^2 = n\a',
  \quad\text{and}\quad
  H = 2N\a'\omega_{S^3}.
\end{equation}       
Here, $ds^2_{S^3}$ and $\o_{S^3}$ are the metric and volume form on a
unit $S^3$,
\begin{align}\label{eq:UnitThreeSphere}
    ds_{S^3}^2 &= \qu\Bigl(
      (d\phi^1)^2 + (d\phi^2)^2 + (d\phi^3)^2 + 2\cos\phi^1 d\phi^2 d\phi^3
    \Bigr),\\
    \o_{S^3} &= \ei\sin\phi^1\, d\phi^1\w d\phi^2\w d\phi^3,\quad
    \int_{S^3}\omega_{S^3} = 2\pi^2.
\end{align}
The polar angle $\phi^1$ takes values in the interval $I_1=[0,\pi]$
and the sum and difference of the azimuthal angles, $\phi^2\pm\phi^3$,
are periodic modulo $4\pi$.\footnote{\label{foot:FundDomain}Thus, a
  fundamental domain of the azimuthal angles is $0\le\phi^2<2\pi$ and
  $0\le\phi^3<4\pi$.}  The normalization of $H$ follows from the
quantization condition
\begin{equation}\label{eq:Hquant}
  \frac1{2\pi\a'}\int_{S^3}H = 2\pi n,\qquad n\in\IZ,
\end{equation}
which ensures that the phase
\begin{equation}\label{eq:SingVal}
  \exp\bigl(\frac{i}{2\pi\a'}\int_\Sigma B\bigr)
\end{equation}
is single valued in the string path integral.  A convenient gauge
choice for $B$ is
\begin{equation}\label{eq:Bgauge}
  B = -\qu n\a'\cos\phi^1\, d\phi^2\w d\phi^3.  
\end{equation}

As discussed in App.~\ref{app:Ws}, the background \eqref{eq:PhysBg}
arises as the semiclassical description of the $SU(2)$
Wess-Zumino-Witten (WZW) model, at large level $n$
\cite{Gepner:1986wi}.  The group manifold $SU(2)$ has the topology of
a 3-sphere, and can be parametrized as
\begin{equation}\label{eq:SUg}
  g(\phi^1,\phi^2,\phi^3)
  = e^{-i\phi^2\s_3/2} e^{-i\phi^1\s_1/2} e^{-i\phi^3\s_3/2},
\end{equation}
where the $\s_m$ are the Pauli spin matrices.  Then, $g^{-1}dg =
-\iover{2}\s_m\l^m$, in terms of the left-invariant 1-forms,
\begin{equation}\label{eq:Lforms}
  \begin{split}
    \l^1 &=  \cos\phi^3 d\phi^1 + \sin\phi^3\sin\phi^1 d\phi^2,\\
    \l^2 &= -\sin\phi^3 d\phi^1 + \cos\phi^3\sin\phi^1 d\phi^2,\\
    \l^3 &= d\phi^3 + \cos\phi^1 d\phi^2.
  \end{split}
\end{equation}
The $\l^p$ satisfy $d\l^p + \half \e_{mnp} \l^m\w \l^n = 0$ and their
product is
\begin{equation}\label{eq:Vol}
  \l^1\w \l^2\w \l^3 
  = \sin\phi^1 d\phi^1\w d\phi^2\w d\phi^3
  = 8\o_{S^3},
\end{equation}
where $\o_{S^3}$ is the volume form on a unit $S^3$.  Aside from the
factor of $n\a'$, the metric \eqref{eq:PhysBg} is the unit $SU(2)$
metric metric $ds^2 = \quarter\bigl((\l^1)^2+(\l^2)^2+(\l^3)^2\bigr)$.


\subsection{Target space description of the WZW model for general
  group}
\label{sec:PhysBgG}

For a general group $G^\WZW$, with Lie algebra
\begin{equation}\label{eq:LieGWZW}
  [t_m,t_n] = c_{mn}{}^p t_p,
\end{equation}
and left-invariant Maurer-Cartan form 
\begin{equation}
  g^{-1} dg = \l = \l^pt_p,
  \quad\text{where}\quad
  d\l^p + \frac12 c_{mn}{}^p\l^m\w\l^n = 0,
\end{equation}
the target space description of the WZW model at level $n$ is as
follows.  The metric of Eq.~\eqref{eq:PhysBg} generalizes to
\begin{equation}\label{eq:PhysMetricG}
  ds^2_\phys = r^2 ds^2_G,\quad r^2 = n\a',
\end{equation}
in terms of ``unit'' $G^\WZW$ metric
\begin{equation}
  ds^2_G = -\frac14 \tr'(\l\l) = \frac14 \psi^2 d_{mn} \l^m\l^n,
\end{equation}
and the $H$-flux becomes
\begin{equation}\label{eq:WZWHflux}
  H = -\frac{n}{12}\Tr'(\l\w\l\w\l)
    = \frac{\hat n}{12} c_{mnp}\l^m\w\l^n\w\l^p,
  \quad\text{where}\quad
  c_{mnp} = c_{mn}{}^q d_{qp}. 
\end{equation}
Here, $d_{mn}$ is the normalized Killing form, $\psi^2$ is the length
squared of a long root, and $\hat n = \psi^2 n/2$.  Note that $d_{mn}$
is related to the nonnormalized Killing form
\begin{equation}
  \tilde d_{mn} = - c_{mp}{}^q c_{nq}{}^p
\end{equation}
via $\tilde d_{mn} = h^\vee\psi^2 d_{mn}$, where $h^\vee$ is the dual
Coxeter number of $G^\WZW$.  We refer the reader to
App.~\ref{app:LieAlg} for additional Lie algebra conventions and to
App.~\ref{app:Ws} for the worldsheet description of a WZW model.


\section{The 4D T-fold description: doubled Hopf fibration}
\label{sec:Tfold}

A T-fold is a generalization of a $T^n$ fibration, in which the
transition functions are allowed to lie in the T-duality group
$O(n,n)$ rather than its geometric subgroup.  In this section, we
review the definition of a T-fold and then present the T-fold
description of the large level $SU(2)$ WZW model as an $S^1\times
\tilde S^1$ fibration over $S^2$.  The physical $S^1$ fibration is the
Hopf fibration of the physical space $SU(2)\cong S^3$ and the dual
$\tilde S^1$ fibration defines the Lens space $SU(2)/\IZ_n\cong
L_{(n,1)}$ T-dual to this background.  (It is also possible to give a
T-fold description as a $T^2\times\tilde T^2$ fibration over the
interval~$I_1$.  However, the latter is somewhat less natural since
the fibers degenerate.  See App.~\ref{app:5dTfold}.)  Finally, in
Sec.~\ref{sec:TfoldG} we generalize from $SU(2)$ to arbitrary group.


\subsection{T-fold generalities}
\label{sec:TfoldGen}


\subsubsection{T-fold backgrounds vs.\ geometric backgrounds with
  $B$-field}
\label{sec:TfoldGenTvsGeoB}

Recall that a Riemannian manifold is a differentiable manifold $M$
endowed with a metric $G_{mn}$. A \hbox{$B$-field} is conventionally
thought of as a $U(1)$ gerbe connection, that is, a 2-form potential
for the $H$-flux, analogous to the 1-form potential $A$ of
electromagnetism.\footnote{In electromagnetism, the local 1-form $A$
  (defined in each coordinate patch of the manifold~$M$) is a
  connection on a $U(1)$ principle bundle.  The curvature of this
  bundle is the global 2-form $F=dA$, i.e., the field strength.
  Finally, the topology of the bundle is characterized by the
  cohomology class $[F]\in H^2(M,\IZ)$.  Similarly, a gerbe of
  connection $B$ is characterized by curvature $H$ and topology
  $[H]\in (2\pi)^2\a'H^3(M,\IZ)$.}

Of particular interest in string theory is the case that $M$ is a
$T^n$ fibration over some base manifold $\CB$.  The simplest
supersymmetry preserving string theory backgrounds ($T^n$, K3, CY$_n$
and products thereof) are generically of this type (due to the special
Lagrangian $T^n$ fibration of generic CY$_n$).  A manifold $M$ is a
$T^n$ fibration if it admits a projection to a base $\CB$ and is
diffeomorphic to $T^n\times U$ over sufficiently small patches
$U\subset \CB$ away from singular fibers.  Globally, the patches are
sewn together via transition functions in $GL(n)$ acting on the
fibers.\footnote{To be precise, the transition functions of the
  tangent bundle are in $GL(n)$.  For the coordinates, the homogeneous
  part of the transition functions lies in the same $GL(n)$, the
  inhomogenous part (translations in $T^n$) is a semidirect $U(1)^n$,
  and the full structure group is $GL(n)\ltimes U(1)^n$.}

As defined by Hull \cite{Hull:2004in,Hull:2006va}, a T-fold is a
generalization of a $T^n$ fibration with $B$-field, which permits not
only geometric transition functions in $GL(n)$ but also T-duality
transition functions in $O(n,n)$.  Since these transition functions
can mix the metric and $B$-field, neither $G$ nor $B$ necessarily has
a global interpretation as a metric or gerbe connection on any
$\dim\CB+n$ dimensional space; nor does there necessarily exist a
global $\dim\CB+n$ dimensional topology or associated gerbe topology.
From a $\dim\CB+n$ dimensional point of view, a T-fold is
nongeometric.

However, a T-fold does always have a global \emph{doubled} geometry
and topology.  For a \hbox{T-fold} (in contrast to the construction of
Sec.~\ref{sec:Doub}), this doubling refers to the fiber only.  One
simply considers the product of the physical torus $T^n$ fiber and
\hbox{T-dual} torus fiber $\tilde T^n$ over each patch $U\subset\CB$.
Since $O(n,n)\subset GL(2n)$, the \hbox{T-duality} transition
functions become ordinary transition function on the doubled $T^{n}$
fiber, $T^{2n}$.  The topology and curvature of this $T^{2n}$
fibration characterize all topological and curvature information that
one would seek in the pair $(M,B)$ of the geometric case, except for
the $B$-field $B_\CB$ on the base, which must be specified
separately.\footnote{To be precise, $B_\CB$ is the \emph{pullback} of
  a gerbe connection on the base to a gerbe connection on the total
  space of the $T^{2n}$ fibration.}


\subsubsection{Metrics and polarizations}
\label{sec:TfoldGenMetPol}

On a T-fold, we define two metrics: a constant $O(n,n)$ invariant
fiber metric, and a Riemannian metric on the total space,
\begin{subequations}
  \begin{align}
    ds^2_{O(n,n)} &= \CL_{IJ} \eta^I \eta^J,\label{eq:OnnMetric}\\
    ds^2_\text{T-fold} &= ds^2_\CB + \CH_{IJ}\eta^I\eta^J.\label{eq:Rmetric}
  \end{align}
\end{subequations}
Here, the $\eta^I = dx^I + \CA^I$ are the global fiber 1-forms, where
$\CA^I$ is the $T^{2n}$ connection.  The Riemannian fiber metric
$\CH_{IJ}(y)$ is a symmetric matrix such that
\begin{equation}
  \CH^T\CL^{-1}\CH = \CL.
\end{equation}
Given a \emph{polarization} or choice of null decomposition $T^{2n} =
T^n\times \tilde T^n$ into physical and dual subspaces over each patch
$U\subset\CB$, it is convenient to choose a basis of fiber 1-forms
$\eta^I = (\eta^i,\tilde\eta_i)$ that respects the decomposition.  We
similarly choose a coordinate decomposition so that the $x^i$ ($\tilde
x_i$) are coordinates on the physical subspace $T^n$ (dual subspace
$\tilde T^n$).  Then,
\begin{equation}\label{eq:Loffdiagonal}
  \CL_{IJ} = 
  \begin{pmatrix}
    0 & L_i{}^j\\
    (L^T)^i{}_j & 0
  \end{pmatrix},
  \quad
  \CH_{IJ} =
  \begin{pmatrix}
    G + B^TG^{-1}B & B^TG^{-1}L\\
    L^TG^{-1}B & L^TG^{-1}L
  \end{pmatrix}_{IJ}.
\end{equation}
In a canonical basis with coordinate periodicities $x^i\cong x^i +
2\pi\n$ and $\tilde x_i\cong \tilde x_i + 2\pi\tilde\n$, the former
becomes
\begin{equation}\label{eq:Lcanonical}
  \CL_{IJ} = \frac{1}{\n\tilde\n}
  \begin{pmatrix}
    0 & \d_i{}^j\\
    \d^i{}_j & 0
  \end{pmatrix},
  \quad\text{i.e.,}\quad
  L_i{}^j = \frac1{\n\tilde\n}\d_i{}^j.
\end{equation}
With these definitions, the two metrics become
\begin{subequations}\label{eq:SimplifiedTfoldMetrics}
  \begin{align}
    ds^2_{O(n,n)} &= 2L_i{}^j\eta^i \tilde\eta_j,
    \label{eq:OnnMetricII}\\
    ds^2_\text{T-fold} &= ds^2_\CB(y) + G_{ij}\eta^i \eta^j 
       + G^{ij}
       \bigl(L_i{}^k\tilde\eta_k + B_{ik}\eta^k\bigr)
       \bigl(L_j{}^l\tilde\eta_l + B_{jl}\eta^l\bigr).
    \label{eq:RmetricII}
  \end{align}
\end{subequations}
Here, and in all subsequent sections, we set $\a'=1$ for simplicity.
The field $B_{ij}$ parametrizes the off diagonal components of the
$T^n\times\tilde T^n$ Riemannian metric.  Given a choice of doubled
fiber coordinates $x^I$, we can write $\eta^I = dx^I + \CA^I$, where
the connection $\CA^I(y)$ dependes only on the base coordinates $y^a$.
Then, given a polarization, we decompose $\CA^I$ as $\CA^I =
\bigl(A^i,(L^{-1})_i{}^j B_j\bigr)$, so that
\begin{align}\label{eq:eetilde}
  \eta^i = dx^i + A^i\quad\text{and}\quad
  \tilde\eta_i = d\tilde x_i + (L^{-1})_i{}^jB_j.
\end{align}


\subsubsection{Recovery of the physical background}
\label{sec:TfoldGenRecovery}

Given a choice of gerbe connection $B_\CB$ on the base, and a choice
of polarization over a patch $U\subset\CB$, the \emph{physical} metric
and $B$-field in this patch are~\cite{Fidanza:2003zi}
\begin{subequations}\label{eq:TfoldRecovery}
  \begin{align}
    ds^2 &= ds^2_\CB(y) + G_{ij}\eta^i\eta^j,\label{eq:PhysMetric}\\
    B &= B_\CB + (dx^i+\half A^i)\w B_i 
    + \half B_{ij} (dx^i + A^i)\w (dx^j + A^j).\label{eq:FormOfB}
  \end{align}
\end{subequations}
Here, $B_\CB$ is a local 2-form on the base and the $B_i$ are local
1-forms on the base.  Note that the polarization need not be defined
globally, so that the recovery of the physical background and standard
sigma model, is only patchwise.  If a global polarization exists, then
the compactification is geometric and described globally by a standard
sigma model.  In this case, the total space of the T-fold is dual
torus fibration over the physical space, and it possible to recover
the physical background by global projection.  Otherwise, the
compactification is only locally geometric, and is globally
\emph{nongeometric}.

From the worldsheet point of view, a T-fold background is not really
so different from a globally geometric compactification with
$B$-field: In each patch, we have a standard sigma model description
and the usual $\b$-function equations.  Globally, for either a
manifold or a T-fold, transition functions are necessary in order to
relate the sigma model Lagrangian in overlaps between coordinate
patches.


\subsubsection{T-duality action on fiber}
\label{sec:TfoldGenTduality}

T-duality acts by $O(n,n;\IZ)$ transformation as
\begin{equation}\label{eq:Taction}
  \eta^i \mapsto \CO^i{}_j \eta^j,\quad
  \CH \mapsto \CO^{-1T}{}_i{}^k\CH_{kl}\CO^{-1}{}^l{}_j,\quad
  \CO\in O(n,n;\IZ).
\end{equation}
The $B$-field component $B_\CB$ in Eq.~\eqref{eq:FormOfB} is a
T-duality invariant.  Here, the $O(n,n)$ condition is $O^T\CL O =
\CL$, and the restriction to $\IZ$ indicates that $O^i{}_j$ preserves
the lattice defining the doubled torus.  This is the ``active'' point
of view.  For the passive transformations, the 1-forms $\eta^a$ and
Riemannian metric are held fixed, and we consider different choices of
polarization on the same doubled fibration.


\subsection{T-fold description of the $SU(2)$ WZW model}
\label{sec:TfoldSU}

We now show that the $SU(2)$ WZW model admits a T-fold description as
the $\IZ_n$ quotient of a doubled Hopf fibration over $S^2$.  In this
section, we consider only the semiclassical
background~\eqref{eq:PhysBg}.  In Sec.~\ref{sec:Doub}, we also briefly
comment on the CFT interpretatation.

A 3-sphere can be thought of as the Hopf fibration of $S^1$ over $S^2$
with $-1$ unit of Euler class.  Following this interpretation, we
write Eq.~\eqref{eq:UnitThreeSphere} as
\begin{equation}\label{eq:Hopf}
  ds^2_{S^3} = \qu\Bigl(ds^2_{S^2} + (d\phi^3 + A^3)^2\Bigr)
  \quad\hbox{with}\quad
  A^3 = \cos\phi^1\,d\phi^2,
\end{equation}
where the metric and volume form on a unit 2-sphere are
\begin{equation}\label{eq:TwoSphere}
  ds^2_{S^2} = (d\phi^1)^2 + \sin^2\phi^1\,(d\phi^2)^2
  \quad\text{and}\quad
  \o_{S^2} = \sin\phi^1\,d\phi^1\w d\phi^2. 
\end{equation}
The topology of this fibration is characterized by its Euler class
$dA^3 = -\o_{S^2}$, viewed as an element of $H^2(S^2,\IZ)$.  Here, the
coordinate $\phi^3$ is periodic modulo $4\pi$ on each fiber, and
$\phi^2$ is the usual $S^2$ azimuthal angle periodic modulo $2\pi$ on
the base.  (See Footnote~\ref{foot:FundDomain}.)

Now consider the background \eqref{eq:PhysBg}.  In the decomposition
of Eq.~\eqref{eq:FormOfB}, the $B$-field \eqref{eq:Bgauge} has only a
1-form component,
\begin{equation}\label{eq:BdecompI}
  B_3 = \qu n\cos\phi^1 d\phi^2
  \quad\hbox{with}\quad
  dB_3 = -\qu n\o_{S^2}.
\end{equation}
The NS sector topological data (the spatial $S^3$ topology, and
cohomology class $[H]\in(2\pi)^2 H^3(M,\IZ)$), and Riemannian data
(the choice of a physical spatial metric and gerbe connection), are
completely encoded in the topology and Riemannian metric of the T-fold
background, once a polarization is specified.

The $O(1,1)$ metric \eqref{eq:OnnMetric} and Riemannian metric
\eqref{eq:Rmetric} are
\begin{subequations}
  \begin{align}
    ds^2_{O(1,1)} &= \frac12\eta^3\tilde\eta_3,\label{eq:OnnMetricI}\\
    ds^2_\text{T-fold} &= \qu\Bigl(
    n \bigl( (\eta^1)^2 +  (\eta^2)^2 +  (\eta^3)^2 \bigr)  
    + \frac1{n}(\tilde\eta_3)^2\Bigr),\label{eq:RmetricI}
  \end{align}
\end{subequations}
where $\eta^i = \l^i$ of Sec.~\ref{sec:PhysBgSU}, and 
\begin{equation}\label{eq:threetilde}
  \tilde\eta_3 = d\tilde\phi_3 +n\cos\phi^1 d\phi^2,
  \quad\text{with}\quad
  d\tilde\eta_3 = -n\o_{S^2}.
\end{equation}
In contrast, the 1-form on the Hopf fiber of the physical $S^3$ is
\begin{equation}
  \eta_3 = d\phi^3 + \cos\phi^1 d\phi^2,
  \quad\text{with}\quad
  d\eta^3 = -\o_{S^2}.
\end{equation}
Here, we are working in the convention $\phi^3\cong\phi^3+4\pi$ and
$\tilde\phi_3\cong\tilde\phi_3+4\pi$, so that $\n=\tilde\n=2$ in the
notation of Sec.~\ref{sec:TfoldGenMetPol}.

The physical metric is obtained by dropping the $\tilde\eta_3$ term
from Eq.~\eqref{eq:RmetricI}.  T-duality inversion of the physical
$S^1$ fiber interchanges $\eta^3$ and $\tilde\eta_3$, and thus
exchanges $-1$ unit of Euler class and $n$ units of $H$-flux with $-n$
units of Euler class and $1$ unit of $H$-flux.  The T-dual background
is a Lens space $L_{(n,1)} = S^3/\IZ_n$ with the minimal quantum of
$H$-flux.  Indeed, it is known that at level $n$, the $SU(2)$ WZW
model and $SU(2)/\IZ_n$ WZW orbifold are exactly equivalent as
CFTs\footnote{For $n = n_1n_2$, the equivalence of the
  $SU(2)/\IZ_{n_1}$ and $SU(2)/\IZ_{n_2}$ CFTs has also been
  demonstrated \cite{Maldacena:2001ky}.  This is the T-duality
  equivalence of $S^3/\IZ_{n_1}$ with $n_2$ units of $H$-flux and
  $S^3/\IZ_{n_2}$ with $n_1$ units of $H$-flux, which describe the
  near horizon angular geometry of a system of $n_1$ KK monopoles and
  $n_2$ NS5-branes, and $n_2$ KK monopoles and $n_1$ NS5-branes,
  respectively.}~\cite{Gaberdiel:1995mx,Maldacena:2001ky,Distler:2006}.

The $\IZ_n$ quotient can be seen more clearly by writing
\begin{equation}
  \tilde\l'^3 = d\tilde\phi'^3 + \cos\phi^1 d\phi^2,
\end{equation}
with $\tilde\phi'^3= \tilde\phi_3/n$ periodic modulo $4\pi/n$ instead
of mod $4\pi$.  The metrics then takes the form
\begin{align}
    ds^2_{O(1,1)} &= \frac{n}2\l^3\tilde\l'^3,\\
  ds^2_\text{T-fold} &= \frac{n}4\Bigl(
  (\l^1)^2 +  (\l^2)^2 +  (\l^3)^2 + (\tilde\l'^3)^2 \Bigr),
\end{align}
with
\begin{equation}
  \l^3 = d\phi^3 + \cos\phi^1 d\phi^2,
  \quad\text{and}\quad
  \tilde\l'^3 = d\tilde\phi'^3 +\cos\phi^1 d\phi^2.
\end{equation}
On the $\IZ_n$ covering space
($\tilde\phi'^3\cong\tilde\phi_3'+4\pi$), the physical and dual $S^1$
fibrations each define a total space $S^3$.  The $\IZ_n$ acts freely
by translation on the dual Hopf fiber:
$\tilde\phi'^3\mapsto\tilde\phi'^3 + 4\pi/n$.\footnote{Note that the
  linear combination $\l^3-\tilde\l_3$ is closed.  A linear
  combination of the Hopf fibers is therefore trivially fibered, and
  topologically, the covering space of the T-fold factors as
  $S^3\times S^1$.  This will come up again in Sec.~\ref{sec:TfoldG}.}


\subsection{T-fold description of the WZW model for general group}
\label{sec:TfoldG}

The $SU(2)$ discussion of the previous section readily generalizes to
an arbitrary compact semisimple Lie group $G^\WZW$.  The Cartan
subalgebra generates a $U(1)^r$ isometry, which endows $G^\WZW$ with
the structure of a $T^r$ Cartan torus fibration over the coset
$G^\WZW/U(1)^r$.  The $B$-field of the WZW model defines a formal
$\tilde T^r$ fibration over the same base.  In
Sec.~\ref{sec:TfoldGPhysFib}, we describe the physical fibration of
the WZW background and present the metric and $B$-field in the form
required by Eqs.~\eqref{eq:TfoldRecovery}.  Then, in
Sec.~\ref{sec:TfoldGDoubFib}, we show how the same data is encoded the
doubled fibration of the T-fold.


\subsubsection{The physical $T^r$ fibration}
\label{sec:TfoldGPhysFib}

In terms of the Chevalley basis\footnote{In the notation used here and
  in App.~\ref{app:LieAlgChevalley}, the $e_\a$ ($f_{-\a}$) include
  both the fundamental Chevalley generators $e_i$ ($f_i$) of the
  defining commutation relations and the descendents obtained from
  multiple commutators of the $e_i$ ($f_i$).} of $\gfrak^\WZW$
(defined in App.~\ref{app:LieAlgChevalley}), write
\begin{equation}\label{eq:ZtfromChevalley}
  t_i = -i h_i,\quad t_{1\a} = -i(e_\a+f_{-\a}),
  \quad\text{and}\quad t_{2\a} = -(e_\a-f_{-\a}),
\end{equation}
and let $a$ denote an index that runs over $1i$ and $2i$.  Then, the
Lie algebra takes the form
\begin{equation}\label{eq:TrFibLieAlg}
  [t_i,t_j] = 0,\quad
  [t_a,t_i] = c_{ai}{}^jt_j,\quad
  [t_a,t_b] = c_{ab}{}^it_i,
\end{equation}
and the normalized Killing form takes the block diagonal form
$\diag(d_{ij},d_{ab})$.  The $G^\WZW$ invariance of the Killing form
implies that the lowered index structure constant is completely
antisymmetric.  Therefore, the two structure constants of
Eq.~\eqref{eq:TrFibLieAlg} are related by
\begin{equation}
  c_{ia}{}^cd_{cb} = c_{ab}{}^jd_{ji} = c_{iab}.
\end{equation}
Since the $t_i$ generate an abelian subalgebra, the group manifold
$G^\WZW$ is fibered by Cartan tori $T^r = U(1)^r$.  A generic element
can be parametrized by coordinates $(x^i,y^a)$ as
\begin{equation}
  g(x,y) = \exp(y^at_a)\prod_{m=1}^r\exp(x^it_i).
\end{equation}
The Killing isomorphism (c.f.~App.~\ref{app:LieAlgCartan}) maps the
Cartan subalgebra $\hfrak$ to the dual space $\hfrak^*$ by trading
Chevalley generators $h_j$ for coroots $\a_{(j)}^\vee$.  Thus
$ix^jt_j$ maps to $x$, where
\begin{equation}
  x = x^j\a_{(j)}^\vee = x_jw^{(j)} 
  \quad\text{and}\quad x_j = d_{jk}x^k.
\end{equation}
Here the $w^{(i)}$ are the basis of weights, dual to the $\a_{(j)}$.
Taking into account the periodic identifications, we have
\begin{equation}
  x\in \hfrak^*/(2\pi\L^\vee)\cong T^r,
\end{equation}
i.e., $x\cong x+2\pi\a^\vee$, where $\a^\vee$ is any coroot, or in
components, $x^j\cong x^j + 2\pi$ and $x_j\cong x_j + d_{jk}N^k$,
where the $N^k$ are integers.

The metric~\eqref{eq:PhysMetricG} can be written in the fibration form
\begin{equation}
  ds^2_\phys = ds^2_\CB + \frac{\hat n}2 d_{ij}(dx^i+A^i)(dx^j+A^j),
\end{equation}
where the $x$-dependence drops out of the metric on the base
$\CB=G^\WZW/U(1)^r$,
\begin{equation}
  ds^2_\CB(y) = \frac{\hat n}2 d_{ab}\l^a\l^b,
\end{equation}
and the fiber 1-forms satisfy
\begin{equation}
  d\l^i = dA^i = -\half c_{ab}^i\l^a\w\l^b.
\end{equation}  
The $H$-flux~\eqref{eq:WZWHflux} becomes
\begin{equation}
  H = \frac{\hat n}{4}c_{iab}\l^i\w\l^a\w\l^b,
\end{equation}
and can be obtained from a $B$-field of the form~\eqref{eq:FormOfB},
with
\begin{equation}
  B_\CB = \frac{\hat n}4 d_{ij}A^i\w A^i,\quad
  B_i = \frac{\hat n}2 d_{ij}A^j
  \quad\text{and}\quad
  B_{ij} = 0.
\end{equation}


\subsubsection{The doubled fibration}
\label{sec:TfoldGDoubFib}

In the T-fold description, the torus fibration of $G^\WZW$ is promoted
to a doubled fibration by including both the Cartan torus $T^r$ and
dual Cartan torus $\tilde T^r$ fibers.  The description follows
straightforwardly from Secs.~\ref{sec:TfoldGen}
and~\ref{sec:TfoldGPhysFib} once we specify the $O(r,r)$ fiber
metric~\eqref{eq:Loffdiagonal}.

Let $\tilde x_i$ and $\tilde t^i$ denote the dual coordinates and
generators, respectively.  Then $i\tilde x_mt^m$ canonically maps to
$\tilde x\in\hfrak^*$, where
\begin{equation}
  \tilde x = \tilde x_m w^{(m)} = \tilde x^mw_{(m)}
  \quad\text{and}\quad\tilde x^m = d^{mn}\tilde x_n.
\end{equation}
Taking into account the periodic identifications, we have
\begin{equation}
  \tilde x\in \hfrak^*/(2\pi(\L^\vee)^*)\cong \tilde T^r,
\end{equation}
i.e., $\tilde x\cong\tilde x+2\pi w^\vee$, where $w^\vee$ is any
weight, or in components, $\tilde x_j\cong\tilde x_j + 2\pi$ and
$\tilde x^j\cong\tilde x^j + d^{ik}N_k$, where the $N_k$ are integers.

Since the coordinates $(x^i,\tilde x_i)$ have the canonical $2\pi$
periodicities, $L^i{}_j$ of Eq.~\eqref{eq:Loffdiagonal} takes the
canonical form $\d^i{}_j$ in this basis.

The doubled metrics~\eqref{eq:SimplifiedTfoldMetrics} become
\begin{subequations}
  \begin{align}
    ds^2_{O(n,n)} &= 2L_i{}^i\eta^i \tilde\eta_j,\\
    ds^2_\text{T-fold} &= ds^2_\CB(y) + \frac{\hat n}2 d_{ij}\eta^i \eta^j 
       + \frac2{\hat n} d^{ij}\tilde\eta_i\tilde\eta_j,
     \end{align}
\end{subequations}
where
\begin{equation}
  \eta^i = \l^i = dx^i + A^i
  \quad\text{and}\quad
  \tilde\eta_i = d\tilde x_i + \frac{\hat n}2 d_{ij} A^i.
\end{equation}

Proceeding as in the $SU(2)$ case, it is natural to define new
coordinates on the dual fiber, $\tilde x'^{i} = (2/\hat n)d^{ij}\tilde
x_j$.  Then,
\begin{subequations}
  \begin{align}
    ds^2_{O(n,n)} &= 2L_{ij}\l^i \tilde\l'^j,
    \quad L_{ij} = \frac{\hat n}2 d_{ij},\\
    ds^2_\text{T-fold} &= ds^2_\CB(y) + \frac{\hat n}2 d_{ij}
    \bigl(\l^i \l^j + \tilde\l'^i\tilde\l'^j\bigr),
     \end{align}
\end{subequations}
where
\begin{equation}\label{eq:landltilde}
  \l^i = dx^i + A^i,\quad
  \tilde\l'^i = d\tilde x'^i + A^i.
\end{equation}
This puts the T-fold Riemannian metric in the form of doubled Cartan
torus fibration with the same connection for either factor, as was
obtained in Sec.~\ref{sec:TfoldSU} for the $SU(2)$ case.  However, we
still need to account for the modified periodicites of the $\tilde
x'^i$.  The coordinate $\tilde x' = (2/\hat n) \tilde x$ satisfies
\begin{equation}\label{eq:TfoldGDualTorus}
  \frac{\tilde x'}2\in \frac{\hfrak^*}{(2\pi/\hat n)(\L^\vee)^*}
  = \biggl(\frac{\hfrak^*}{(2\pi/\hat n)\L}\biggr)/C,
\end{equation}
where we have used the fact that the ratio of the weight lattice to
the root lattice is $C$, the center of the group.  For $G^\WZW$ a
simply laced group, we have $(\psi^2/2)\L = \L^\vee$.  In this case,
the two coordinate periodicities are given by
\begin{equation}\label{eq:TfoldGBothTori}
  x \in \biggl(\frac{\hfrak^*}{2\pi\L^\vee}\biggr)
  \quad\text{and}\quad
 \frac{\tilde x'}2 \in \biggl(\frac{\hfrak^*}{(2\pi/n)\L^\vee}\biggr)/C.
\end{equation}

For $SU(2)$, with $C = \IZ_2$, this becomes
\begin{equation}
  x \in \biggl(\frac{\hfrak^*}{2\pi\L^\vee}\biggr)
  \quad\text{and}\quad
  \tilde x' \in \biggl(\frac{\hfrak^*}{(2\pi/n)\L^\vee}\biggr)
  = \biggl(\frac{\hfrak^*}{2\pi\L^\vee}\biggr)/\IZ_n,
\end{equation}
so we indeed obtain two copies of the same Cartan torus fibration (in
this case, the Hopf fibration), up to a quotient by $\IZ_n$ on the
second factor, in agreement with Sec.~\ref{sec:TfoldSU}.

Let us interpret what we have done.  The group $G^\WZW$ represents
both the physical space and the left and right action on the group
manifold $g_\phys \mapsto \O_L g_\phys \O_R^{-1}$.  Let us view the
Lie algebra of the previous section as that of the right action.
There is no harm in considering a slightly larger group, which also
includes the left action of the Cartan subalgebra.  Let us add
superscripts $L$ and $R$ to Cartan generators to distinguish between
left and right actions.  The Lie algebra becomes
\begin{equation}
  [t^R_i,t^R_j] = 0,\quad
  [t_a,t^R_i] = c_{ai}{}^j t^R_j,\quad
  [t_a,t_b] = c_{ab}{}^i t^R_i,\quad
  [t^L_i,t^R_j] = 0,\quad 
  [t^L_i,t_a]=0,
\end{equation}
where the new generators just contribute an abelian
$\bigl(U(1)^r\bigr)_L$.  It is convenient to define physical and dual
Cartan generators $t_i$ and $\tilde t'_i$ via
\begin{equation}
  t^L_i = t_i-\tilde t'_i,\quad t^R_i = t_i + \tilde t'_i.
\end{equation}
In this basis, the Lie algebra becomes
\begin{equation}
  [t_i,t_j] = [\tilde t'_i,\tilde t'_j] = 0,\quad
  [t_a,t_i] = [t_a,\tilde t'_i] = c_{ai}{}^j(t_j+\tilde t'_i),\quad
  [t_a,t_b] = c_{ab}{}^i(t_i+\tilde t'_i).
\end{equation}
The doubled fibration described in this section \emph{is} the group
manifold of this Lie algebra.

The appearance of $\tilde x'/2$ rather than $\tilde x'$ in
Eq.~\eqref{eq:TfoldGBothTori} has the following interpretation.  The
Killing form on the enlarged algebra, in the $(t^L_i,t^R_i,t_a)$ basis
is $\diag(d_{ij},d_{ij},d_{ab})$.  In the $(t_i,\tilde t'_i,t_a)$
basis it is $\diag(\half d_{ij},\half d_{ij},d_{ab})$.  Thus, the
natural dual generator with upper index is $2d^{ij}\tilde t'_j =
2\tilde t'^i$ with conjugate coordinate $\tilde x'_i/2$.

Define left and right fiber coordinates by $x^i = x_L^i+x_R^i$ and
$\tilde x'^i = -x_L^i + x_R^i$.  Then,
\begin{equation}
  \l^it_i + \tilde\l'^it_i = \l_L^i t^L_i + \l_R^i t^R_i,
\end{equation}
where
\begin{equation}
  \l_L^i = \half(\l^i - \tilde\l'^i) = dx_L^i
  \quad\text{and}\quad
  \l_R^i = \half(\l^i + \tilde\l'^i) = dx_R^i + A^i.
\end{equation}
Thus, it is no coincidence that the difference between the 1-forms in
Eq.~\eqref{eq:landltilde} is trivially fibered.  The total space of
the T-fold factorizes as the produce of a left $U(1)^r$ and a right
$G^\WZW$.

T-duality interchanges factors of $n$ between the two denomenators of
Eq.~\eqref{eq:TfoldGBothTori}.  A straightforward generalization of
this result, starting with an orbifold of the original space by
$\bigoplus_{i=1}^r\IZ_{n_i}$, is
\begin{equation}
  x \in \biggl(\frac{\hfrak^*}{2\pi\L^\vee}\biggr)/
  \bigoplus_{i=1}^r\IZ_{n_i}
  \quad\text{and}\quad
 \frac{\tilde x'}2 \in \biggl(\frac{\hfrak^*}{2\pi\L^\vee}/C\biggr)/
 \bigoplus_{i=1}^r\IZ_{\tilde n_i},
\end{equation}
where $n_i\tilde n_i = n$ (no sum) for $i=1,\dots,r$.  T-duality on
the $i$th $U(1)$ interchanges $n_i$ and $\tilde n_i$.

Looking back at the $SU(2)$ example, and comparing Eqs.~\eqref{eq:SUg}
with the results of Sec.~\ref{sec:TfoldSU}, we see that the effect of
T-duality at level $n$ is a right quotient, replacing $SU(2)$ by
$SU(2)/\IZ_n$.  Thus, in the present context, we expect the discrete
groups to quotient the $\bigl(U(1)^r)_R$ fiber of the
$\bigl(G^\WZW\bigr)_R$ factor of the T-fold topology, leaving the
$\bigl(U(1)^r\bigr)_L$ factor unchanged.


\section{The 6D fully doubled description:  $\G\backslash(S^3\times S^3\bigr)$}
\label{sec:Doub}

In this section we describe the doubled geometry of Hull and
Reid-Edwards, focusing on the doubled geometry of WZW models, and then
the special case of the $SU(2)$ WZW model at level $n$.  The first
half of the section covers generalities.
Subsecs.~\ref{sec:DoubGenSpace} through~\ref{sec:DoubGenEFT} provide a
careful review of the formalism of Hull and Reid-Edwards.
Sec.~\ref{sec:DoubGenRecovery} deals with the recovery of the physical
from doubled geometry.  The general idea was already sketched in the
introduction.  When $R\ne0$, there is a closed subgroup $\Gtilde$ with
structure constants $Q$, by which we can quotient to obtain the
physical geometry.  This always works locally, but a potential
obstruction is the discrete group $\G$.  The relevant condition for
global geometry is $\G$-invariant~$Q$: conjugation by elements of $\G$
should preserve $\Gtilde$.  This guarantees that there is a global
polarization. Thus, we see that $Q$-flux alone is not an obstruction
to global geometry, but rather the interplay of $Q$ and $\G$.  Finally
we describe the procedure of Hull and Reid-Edwards for defining local
horizontal and vertical 1-forms, and locally extracting the physical
metric and $B$-field from the doubled geometry.

This brings us to the doubled geometry of WZW models in
Sec.~\ref{sec:DoubG}.  The doubled space takes the form
\begin{equation*}
  \CX_{2d} = \G\backslash\bigl(G_1\times G_2).
\end{equation*}
Here $G_1$ and $G_2$ are two copies of the physical WZW group
$G^\WZW$.  Global polarizations are choices of maximal isotropic
subgroup $\Gtilde$ conjugate to $G_{\diag}$.  Let us focus on the
choice $G_{\diag}$.  Then, the projection to the physical target space
is $\pi\colon (g_1,g_2)\mapsto g_\phys = g_1^{-1}g_2$.  In the doubled
sigma model, $g_1^{-1}(z,\zbar)$ and $g_2(z,\zbar)$ are analogs of the
chiral fields $g_L(z)$ and $g_R(\zbar)$ in the physical model.  A
gauging of the left action of $G_{\diag}$ ensures that the correct
chiral coordinate dependence is restored.  The symmetry under right
multiplication in $\CG_{2d}$ gives the $g_\phys\mapsto \O_1 g_\phys
\O_2^{-1}$ symmetry of the physical model.  By expressing the local
procedure of Hull and Reid-Edwards in terms of global 1-forms, we show
that the correct physical metric and $H$-flux are indeed recovered,
including the condition $r^2 = n\a'$.  When the total space is viewed
as a fibration over the physical base, the horizontal and vertical
1-forms are
\begin{equation*}
  \lambda_\phys = \l_2 - g_\phys^{-1}\l_1 g_\phys
  \quad\text{and}\quad
  \o = \l_2 + g_\phys^{-1}\l_1 g_\phys,
\end{equation*}
where $\l$ denotes a left-invariant form.  The simpler linear
combinations $\l_1\pm\l_2$ define the totally antisymmetric structure
constants $H,f,Q,R$, with $H,Q$ nonzero.  However, they are twisted
relative to the natural forms on the fiber and base.  This resolves
the naive confusion about why $f$ vanishes.

In Secs.~\ref{sec:DoubGPol} and~\ref{sec:DoubGTduality}, we consider
polarizations $\tilde G_\bb= \bb G_{\diag}\bb^{-1}$ and interpret
ordinary abelian T-duality on the Cartan torus in terms of a
restricted subgroup of $\bb\in\CG_{2d}$.  The same maximal isotropic
subspaces furnish possible D-brane worldvolumes in the doubled
description.  Using this observation, in Sec.~\ref{sec:DoubGDbranes}
we reproduces the known results for semiclassical D-branes in WZW
models.  Finally, in Sec.~\ref{sec:DoubGDiscrete}, we describe
restrictions on the discrete group $\G$, but do not fully resolve the
question of what $\G$ is at level $n$ for each choice of modular
invariant.


\subsection{Doubled geometry generalities}
\label{sec:DoubGen}

In Ref.~\cite{Hull:2009sg} (based on earlier
work~\cite{Hull:2004in,Hull:2006va,Hull:2007jy}, see also Sec.~5.3 of
Ref.~\cite{Grana:2008yw}), Hull and Reid-Edwards present a framework
to describe compactifications that are analogs of torus reductions,
twisted by general NSNS sector discrete data.  The basic idea is a
natural extension of the previous section: We would like to extend the
doubled fibration of Sec.~\ref{sec:Tfold} to a fully doubled space.
The topological choice is then that of the doubled manifold
$\CX_{2d}$.  In addition, we require a locally flat $O(d,d)$ invariant
metric and a compatible Riemannian metric.  Then, given a choice of
polarization, we can recover the conventional sigma model description
in each patch, provided there is no ``R-flux'' (defined below) locally
obstructing the projection from the total space to a physical base.

In this paper, we will stick to the purely bosonic WZW model, however,
the natural expectation is that in a supersymmetric context the
\emph{$G$-structure} (structure group of the frame bundle) of the
doubled space $\CX_{2d}$ determines the amount of supersymmetry
preserved by the low energy action\footnote{Since the twisting (NSNS
  discrete data) gauges the supergravity theory, the vacua will
  spontaneously break some or all of the supersymmetry, and preserve
  less supersymmetry than the action.} compared to that of a toroidal
compactification of the same dimension, i.e., on physical space $T^d$
or doubled space $T^{2d}$:\footnote{For analogs of K3 or CY, this
  group structure is the expectation for the Hitchin \emph{generalized
    geometry}
  \cite{Hitchin:2004ut,Gualtieri:2003dx,Jeschek:2004wy,Grana:2005ny,Grana:2006hr}.
  (See also Ref.~\cite{Gauntlett:2003cy} for an authoritative
  discussion of $G$-structures in string theory compactifications,
  preceding their application to generalized geometry.) One might
  question whether analogous statements should hold for a suitable
  supersymmetric generalization of the \emph{doubled geometry} of Hull
  and Reid-Edwards.  A piece of evidence to the affirmative, is the
  agreement verified in Ref.~\cite{ReidEdwards:2010vp} between Lie
  brackets on the doubled geometry and twisted Courant brackets on the
  generalized geometry, for the special case of backgrounds with $K$
  and $f$ flux only.}
\begin{align*}
  \text{unbroken supersymmetry} &\quad\to\quad\text{identity structure
    (i.e., parallelizable $\CX_{2d}$),}\\
  \text{1/2 supersymmetry} &\quad\to\quad\text{$SU(2)\times
    SU(2)$ structure,}\\
  \text{1/4 supersymmetry} &\quad\to\quad\text{$SU(3)\times
    SU(3)$ structure,}
\end{align*}
\noindent and so on.  The corresponding compactifications generalize
purely geometric compactifications on physical spaces $X_d = T^d$,
$K3\times T^{d-4}$ and $\text{CY}_3\times T^{d-6}$ of trivial, $SU(2)$
and $SU(3)$ holonomy, respectively, with no flux, by introducing NSNS
data that twists the compactification and gauges the low energy
supergravity theory.

The work of Hull and Reid-Edwards focuses on the bosonic sector in the
case that the action preserves the same amount of supersymmetry as
flat space.  The NSNS sector topological data includes the doubled
space $\CX_{2d}$, a constant $O(d,d)$ metric, and a choice of
polarization, each of which we now describe.  The NSNS sector
continuous moduli appear in a Riemannian metric on $\CX_{2d}$.


\subsubsection{Doubled space $\CX_{2d}$}
\label{sec:DoubGenSpace}

In the framework of Hull and Reid-Edwards, the manifold $\CX_{2d}$ is
a \emph{twisted doubled torus}, defined as the coset of a group
manifold $\CG_{2d}$ by some discrete subgroup $\G\subset\CG_{2d}$,
\begin{equation}
  \CX_{2d} = \G\backslash\CG_{2d},
\end{equation}
As we will see, the spacetime gauge symmetry arises from the right
$\CG_{2d}$ action on $\CX_{2d}$.  Since this isometry is preserved by
the left quotient by $\G$, the discrete group $\G$ is part of the data
that needs to be specified in order to globally define the model.
Such a quotient is necessary when $\CG_{2d}$ is noncompact, in order
to obtain a compact physical space $X_d$ (or its suitable nongeometric
generalization), i.e., finite 4D Planck mass.  However, $\G$ is part
of the topological data that needs to be specified even when
$\CG_{2d}$ is compact.  Our convention is that $\CG_{2d}$ is simply
connected.  An arbitrary Lie group $\CG'_{2d}$ can be written as the
quotient of its universal cover $\CG_{2d}$ by a discrete normal
subgroup $\G\in\CG_{2d}$.\footnote{The universal covering group
  $\CG_{2d}$ is simply connected.  For $\G$ a normal subgroup of
  $\CG_{2d}$ the coset $\G\backslash\CG_{2d} = \CG_{2d}/\G$ is a
  subgroup.  For $\G$ discrete, $\pi(\G\backslash\CG_{2d})\cong\G$.}
Thus, $\G$ can be nontrivial, even when the doubled target space
$\CX_{2d}$ is a group manifold.
%

Given a basis for the Lie algebra $\gfrak_{2d}$, the correponding left
invariant vector fields form a \emph{frame} $\{T_M\}$ trivializing the
tangent bundle $T\CG_{2d}$, and the dual left invariant 1-forms
$\CP^M$ defined by $g^{-1}dg = \CP=T_M\CP^M$ form a \emph{coframe}
trivializing the cotangent bundle $T^*\CG_{2d}$.  The frame and
coframe satisfy
\begin{equation}
  [T_M,T_N] = t_{MN}{}^P T_P, \quad
  d\CP^P + \tfrac12 t_{MN}{}^P \CP^M\w\CP^N,
\end{equation}
for the same structure constants $t_{MN}{}^P$.


\subsubsection{$O(d,d)$ invariant metric}
\label{sec:DoubGenOdd}

The next piece of data we require is a locally flat $O(d,d)$ invariant
metric on $\CX_{2d}$,
\begin{equation}\label{eq:DoubOddMetric}
  ds^2_{O(d,d)} = \CL_{MN}\CP^M\CP^N,
\end{equation}
or equivalently, an inner product
\begin{equation}
  \langle T_M,T_N\rangle = \CL_{MN}
\end{equation}
on the Lie algebra $\gfrak_{2d}\colon [T_M,T_N] = t_{MN}{}^P T_P$.  By
locally flat, we mean that $\CL_{MN}$ is a constant matrix of
signature $(d,d)$.  A restriction on the choice of $\CG_{2d}$ is that
its action must preserve the $O(d,d)$ metric.  This means that
$\langle [T_P,T_M],T_N\rangle + \langle T_M,[T_P,T_N]\rangle = 0$,
from which the lowered index structure constants $t_{MNP} =
t_{MN}{}^Q\CL_{QP}$ are totally antisymmetric.\footnote{From the point
  of view of the low energy effective field theory, $\CG_{2d}$ must
  have a well defined action on the scalars.  Therefore, it must be
  the semidirect product of a subgroup of $O(d,d)$ (the isometry group
  of the scalar manifold described in Sec.~\ref{sec:DoubGenRiem}) and
  a group under which the scalars are not charged. }


\subsubsection{Polarization and $K,f,Q,R$-flux}
\label{sec:DoubGenPol}

A choice of polarization is a choice of projection defining the
physical subbundle of the tangent bundle (or equivalently, cotangent
bundle) over each patch $U\subset\CX_{2d}$.  The projection must be
null with respect to the $O(d,d)$ metric.  The choice of polarization
is as much part of the defining data of the string compactification as
$\CX_{2d}$, in that different polarizations can, but need not
necessarily, define equivalent vacua.\footnote{As discussed in
  Refs.~\cite{ReidEdwards:2009nu,ReidEdwards:2010vp}, the vacua are
  generically expected to be related by a Poisson-Lie or nonabelian
  T-duality, which is a symmetry at tree level in string theory, but
  not at higher loop.  Inequivalent polarizations are potentially even
  more interesting than equivalent ones, in that they should
  correspond to inequivalent quantum completions of the same classical
  theory.} Given a polarization, it is natural to choose a
corresponding basis $T_M = (Z_m,X^m)$ and dual basis $\CP^N =
(\CP^m,\tilde\CP_m)$, so that $\CL_{MN}$ takes the form
\begin{equation}\label{eq:LMNcanonical}
  \CL_{MN} = 
  \begin{pmatrix}
    0 & L_m{}^n\\
    (L^T)^m{}_n & 0
  \end{pmatrix}.
\end{equation}
In this basis, the Lie alebra $\gfrak_{2d}$ takes the form
\begin{equation}\label{eq:Lalg}
  \begin{split}
    [Z_m,Z_n] & =  K_{mnp}X^p + f_{mn}{}^pZ_p,\\
    [Z_m,X^n] & =  f_{pm}{}^nX^p + Q^{np}{}_mZ_p,\\
    [X^m,X^n] & =  Q^{mn}{}_pX^p + R^{mnp}Z_p,
  \end{split}
\end{equation}
where the same $f$ and $Q$ appear twice due to the antisymmetry of
$t_{MNP}$.  The structure constants are referred to as $K,f,Q,R$-flux,
a name motivated by their appearance in the 3-form
\begin{equation}\label{eq:CalK}
  \CK = -\frac13 \CL_{MN}\CP^M\w d\CP^N = \frac16 t_{MNP}\CP^M\w\CP^N\w\CP^P.
\end{equation}  

The isometry group $\CG_{2d}$ of $\CX_{2d} = \G\backslash\CG_{2d}$
acts by right action and completely geometrizes the gauge group.  (In
contrast, in the non-doubled physical background, the gauge group is
half due to Kaluza-Klein gauge fields $A^m{}_\mu$ and half due to
winding gauge fields $B_{m\mu}$, with generators $Z_m$ and $X^m$,
respectively.)  In the parallelizable case, this observation can be
taken as a definition of the doubled geometry given the gauge group:
the doubled geometry $\CX_{2d}$ is a continuous representation of
group $\CG_{2d}$.  The gauge transformations are just the translations
on $\CX_{2d}$.


\subsubsection{Riemannian metric}
\label{sec:DoubGenRiem}

Finally, we define a Riemannian metric on $\CX_{2d}$,
\begin{equation}\label{eq:DoubRiemMetric}
  ds^2 = \CM_{MN} \CP^M \CP^N = 
  g_{mn}\CP^m\CP^n + g^{mn}
  (L_m{}^p\CP_p + b_{mp}\CP^p)(L_n{}^q\CP_q + b_{nq}\CP^q).
\end{equation}
As for standard toroidal compactifications, the NSNS sector
deformation space includes a coset space $\G_\text{modular}\backslash
O(d,d)/O(d)\times O(d)$, parametrized by a symmetric $2d\times2d$
matrix,
\begin{equation}
  \CM_{MN} = 
  \begin{pmatrix}
    g_{mn} + b^T_{mp}g^{pq}b_{qn}  &  b^T_{mp}g^{pq}L_q{}^n\\
    (L^T)^m{}_pg^{pq}b_{qn} & (L^T)^m{}_pg^{pq}L_q{}^n
  \end{pmatrix}, 
\end{equation}
which we can also write in terms of a vielbein as
\begin{equation}
  \CM = \CE^T\CE
  \quad\text{where}\quad
  \CE^A{}_N =
  \begin{pmatrix}
    e^a{}_n & 0\\
    (e^{-1T})_a{}^p b_{pn} &  (e^{-1T})_a{}^p L_p{}^n
  \end{pmatrix},
\end{equation}
where the lowercase $g_{mn}$ and $b_{mn}$ denote moduli with no
$\CX_{2d}$ coordinate dependence, and where $e^a{}_n$ is a vielbein
for $g_{mn}$.  Here, $g_{mn}$ and $b_{mn}$ are \emph{not} the physical
metric and $B$-field, but do parametrize them, as described at the end
of Sec.~\ref{sec:DoubGenRecovery}.  The Riemannian metric is
\emph{compatible} with the $O(d,d)$ metric in the sense that
$\CE^A{}_N\in O(d,d)$, and therefore a single coframe $\CE^A
=\CE^A{}_M\CP^M$ suffices to put both metrics in unit form,
\begin{equation}
  ds^2_{O(d,d)} = \CL_{AB}\CE^A\CE^B
  \quad\text{and}\quad
  ds^2_\Riem = \d_{AB}\CE^A\CE^B,
  \quad\text{where}\quad
  \CL_{AB} = 
  \begin{pmatrix}
    0 & \d_a{}^b\\
    \d^a{}_b & 0
  \end{pmatrix}.
\end{equation}


\subsubsection{Scalar potential and effective field theory}
\label{sec:DoubGenEFT}

For toroidal compactifications ($t_{MN}^P = 0$), $g_{mn}$ and $b_{mn}$
are exact moduli.  More generally, some of these moduli are lifted by
a scalar potential\footnote{Depending on context (bosonic vs. type II
  vs.~heterotic), there is a possible additional $LLLtt$
  term~\cite{Hohm:2011ex,Schulz:2011}.  However, this term vanishes
  for the case $\CG_{2d} = SU(2)\times SU(2)$ of interest in this
  article.}
\begin{equation}\label{eq:scalarpot}
  V(\CM) = \Bigl(\frac{1}{12}\CM^{MQ}\CM^{NR}\CM^{PS}
  -\frac{1}{4}\CM^{MQ}L^{NR}L^{PS}\Bigr)t_{MNP}t_{QRS}.
\end{equation}
This generalizes the more familiar potential due to $H$-flux.  We
recognize the coefficient of $\CM^{MQ}$ in the second term as $-1/4$
times the nonnormalized\footnote{The conventionally normalized Killing
  form is $\CD_{MQ} = \tilde \CD_{MQ}/(\psi^2 h^\vee)$, where $\psi^2$
  is the length squared of any long root, and $h^\vee$ is the dual
  Coxeter number of $\CG$.  (See App.~\ref{app:LieAlgDef}.)} Killing
form of $\CG_{2d}$
\begin{equation}\label{eq:CartanKillingTilde}
  \tilde\CD_{MQ} = - t_{MN}{}^Pt_{QP}{}^N.
\end{equation}
We will treat the moduli matrix $\CM_{MN}$ as a space of (constant)
deformation \emph{parameters} rather than (noncompact-coordinate
dependent) low energy \emph{fields}, though they are of course
promoted to fields in the low energy action.  In the compactified
supergravity theory, the scalar potential might lift some of these
moduli, or they might acquire dependence on the noncompact spacetime
coordinates (as in cosmological or domain wall solutions).  For
compactification on the $SU(2)_n$ WZW model studied in this article,
the potential lifts some of the moduli, and the linear dilaton
compensates for the nonzero vacuum energy.  In the supersymmetric
context, this is the near horizon $\IR^{6,1}\times S^3$ throat
geometry of a stack of $n$ NS5-branes.  (See
App.~\ref{app:Fivebrane}.)


\subsubsection{Recovery of the physical background}
\label{sec:DoubGenRecovery}


\subsubsection*{Philosophy}

To recover the physical background and standard sigma model in each
patch, we would like to eliminate all reference to the dual
coordinates $\tilde x$ and the dual directions in the tangent and
cotangent bundles.  When this is possible, the background is not
really so different globally from that described by a standard sigma
model (defined patchwise on a \emph{manifold} with $B$-field).  In
this case, the background is said to be locally geometric.  For
globally nongeometric compactifications, there are two qualitatively
different cases: the \emph{tame} case, which is locally geometric, and
the \emph{wild} case, in which there is an obstruction to recovering
the standard sigma model description even locally~\cite{Hull:2009sg}.
The tame/wild distinction is a polarization-dependent statement, as is
the decomposition of the structure constants $t_{MN}{}^P$ of
$\gfrak_{2d}$ into $K$, $f$, $Q$, and $R$.  As noted in
Refs.~\cite{Shelton:2005cf,Dabholkar:2005ve,Lawrence:2006ma,Hull:2007jy,Grana:2008yw},
and elucidated in this context in Ref.~\cite{Hull:2009sg},
nonvanishing $R^{mnp}$ obstructs local geometry.

Roughly speaking, a background is locally geometric if on each patch
$U\in\CX_{2d}$ the polarization allows us to canonically define fields
$G_{mn}(x,\tilde x)$ and $B_{mn}(x,\tilde x)$ that are independent of
the dual coordinates $\tilde x$.  (Otherwise the background is a
coherent state that includes winding modes and not just momentum
modes).  Intuitively, this is the statement that the doubled geometry
is fibered over the physical geometry, so there exists a projection
from $\CX_{2d}$ to a physical base $X_d$.  An embedding
$X_d\hookrightarrow\CX_{2d}$ wouldn't do, since we would then expect
$G_{mn}(x,\tilde x)$ and $B_{mn}(x,\tilde x)$ to depend on the
transverse location $\tilde x_m$ of the embedding, whereas a geometric
background should be independent of dual coordinates.  This intuition
turns out to be exactly right.


\subsubsection*{Implementation}

When $R=0$, the choice of polarization defines a projection from
doubled to physical geometry
\begin{equation}
  \pi\colon\ \CX_{2d} \to X_{d},
\end{equation}
at least patchwise.  When $Q$ is $\G$-invariant, the projection is
globally well defined and the compactification is geometric, otherwise
it is only locally geometric.


\subsubsection*{Global geometry: $R=0$ and $\G$-invariant $Q$}

When $R=0$, the generators $X^m$ form a closed subalgebra
\begin{equation}
  [X^m,X^n] = Q^{mn}{}_pX^p,
\end{equation}
and corresponding subgroup $\tilde G_d \subset\CG_{2d}$.  This
suggests that we can quotient $\CX_{2d}$ by $\tilde G_d$ to obtain a
physical compactification manifold $X_d$,
\begin{equation}
  X_d = \tilde G_d\backslash\CX_{2d}.
\end{equation}
If this succeeds, the doubled space $\CX_{2d}$ is globally a principle
$\tilde G_{d}$-bundle over the physical space
$X_d$,
\begin{equation}
  \tilde G_d \hookrightarrow \CX_{2d} \to X_d.
\end{equation}
This is implemented in the doubled sigma model of Hull and
Reid-Edwards~\cite{Hull:2009sg} by gauging the left action of the
group $\tilde G_d$.
 
A possible global obstruction is the left coset by $\G$ in the
definition $\CX_{2d} = \G\backslash\CG_{2d}$.  There are two natural
conditions one might seek to impose to ensure the persistence of a
global projection to physical base after quotienting by $\G$: (i) the
$\G$-action on $\Gtilde\backslash\G_{2d}$ is well defined, or (ii) the
$\Gtilde$-action on $\G\backslash\CG_{2d}$ is well
defined.\footnote{\label{foot:TwoConditions}As noted in
  Ref.~\cite{Hull:2009sg}, condition (i) requires that for every
$\g\in\G$ and $\tilde h\in\Gtilde$, $\g\tilde h\g^{-1} = \tilde h'$,
for some $\tilde h'\in\tilde G$.  For (ii), the roles of $\Gtilde$ and
$\G$ are reversed: for every $\tilde h\in \Gtilde$ and $\g\in\G$,
$\tilde h\g\tilde h^{-1} = \g'$, for some $\g'\in\G$.  In the latter
case, since $\G$ is discrete and $\tilde G$ is continuously connected
to the identity, deforming $\tilde g$ to the identity shows that
$\g'=\g$; thus, $\tilde h\g\tilde h = \g$, which shows that $\tilde h$
and $\g$ commute, i.e., $\G$ lies in the commutant of $\Gtilde_{d}$ in
$\CG_{2d}$.}  Which is the appropriate condition?

Provided there is a global polarization with $R=0$, the projection
exists globally.  The group $\Gtilde_d$ defines such a polarization on
$\CG_{2d}$, since its left action gives a null decomposition of every
tangent space into the Lie algebra of $\Gtilde_d$ and its complement.
It also defines a global polarization on the quotient $\CX_{2d} =
\G\backslash\CG_{2d}$, provided the null tangent subspaces generated
by left $\Gtilde$ action agree at identified points $a$ and $\g a$ in
$\CG_{2d}$, where $\g\in\G$.  For agreement, conjugation by an
arbitrary $\g\in\G$ must preserve the subgroup $\Gtilde_d\subset\CG$.
This is condition~(i) (see Footnote~\ref{foot:TwoConditions}).

Condition (ii) is too strong.  For a global polarization, we require
only that the $\G$-quotient preserve the \emph{group} $\tilde G_d$,
not the individual \emph{elements.}


\subsubsection*{Local geometry: $R=0$ and $\G$-noninvariant $Q$}

The $\G$-invariance condition just described guarantees the existence
of a \emph{global} projection.  When it fails, the $\tilde G$ actions
on $\G$-identified points do not agree.  The Lie algebra of $\tilde
G_d$ determines a different $d$-dimensional null subspace
$T^\phys_p\CX_{2d} \subset T_p\CX_{2d}$, depending on which preimage
of the point $p\in\CX_{2d}$ we choose in $\CG_{2d}$.  So, in each
contractible open set $U\subset\CX_{2d}$, $\tilde G$ does not define a
unique polarization, but up to $\dim\G$ of them.  We are free to make
an arbitrary choice to break this ambiguity.  Then, we have a valid
polarization in each open set, and well defined projection to physical
subspace in each open set.  Thus we are able to recover a
$d$-dimensional geometry locally but not globally.  As in the previous
case, the choice of polarization defines a gauging of the doubled
sigma model of Hull and Reid-Edwards~\cite{Hull:2009sg}, now patchwise
in each open set of $\CX_{2d}$.  The gauging permits patchwise
recovery of the standard sigma model.


\subsubsection*{Physical metric and $H$-flux}

When $R=0$, we can locally recover the physical background as follows.
The remainder of this subsection is primarily a review of Sec.~2.6 of
Ref.~\cite{Hull:2009sg}.  The reader is encouraged to consult
Ref.~\cite{Hull:2009sg} for further details.  Given a choice of
polarization, and corresponding left invariant frame $(Z_m,X^m)$, an
arbitrary element $g\in\CG_{2d}$ can be written
\begin{equation}\label{eq:gdecomp}
  g(x,\tilde x) = \tilde h(\tilde x) h(x),
  \quad\text{where}\quad
  \tilde h(\tilde x) = \exp(\tilde x_m X^m)
  \quad\text{and}\quad
  h(x) = \exp(x^m Z_m).
\end{equation}
The left invariant 1-forms $\CP^M$ are the components of the
Maurer-Cartan form $\CP = \CP^MT_M = g^{-1}dg$.  It is convenient to
define a 1-form $\Phi = h\CP h^{-1} = \tilde h^{-1} d(\tilde h
h)h^{-1}$, which is a left invariant with respect to $\tilde h$ and
right invariant with respect to $h$.  Then,\footnote{Here, our
  convention for $\CV^M{}_N(x)$ differs by a transpose from
  Ref.~\cite{Hull:2009sg} and agrees with
  Ref.~\cite{ReidEdwards:2010vp}.}
\begin{equation}
  \CP^M = (\CP^m,\CP_m) = \CV^M{}_N(x)\Phi^N,
\end{equation}
where $\CV^M{}_N = (\Ad_{h^{-1}(x)})^M{}_N$ is the adjoint action of
$h^{-1}(x)$, and
\begin{equation}
  \Phi = dh\,h^{-1} + \tilde h^{-1} d\tilde h.
\end{equation}
In the $(Z_m,X^m)$ basis, we can expand the Lie valued 1-forms on the
right hand side as
\begin{align}
  r &= dh\,h^{-1} = r^m Z_m + r_m X^m,\\
  \tilde\ell &= \tilde h^{-1} dh = \tilde\ell^m Z_m + \tilde\ell_m X^m.
\end{align}
Thus, $\Phi^M = (p^m,q_m)$, where
\begin{align}
  p^m &= r^m(x) + \tilde\ell^m(\tilde x),\\
  \tilde q_m &= \tilde\ell_m(\tilde x) + r_m(x),
\end{align}
From the Lie algebra~\eqref{eq:Lalg}, we see that
\begin{align}
  \text{$\tilde\ell^m = 0$ } 
  & \text{ if $R^{mnp}=0$ and the $X^m$ close to generate a group $\tilde G_d$}\\
  \text{$r_m = 0$ } 
  & \text{ if $K_{mnp}=0$ and the $Z_m$ close to generate a group $G_d$}.
\end{align}
The former is the case of interest here.  In this case, the
\emph{physical 1-forms} $p^m(x) = r^m(x)$ depend strictly on the
physical coordinates $x$, while the second term in $q_m =
\tilde\ell_m(\tilde x) + r_m(x)$ can be thought of as encoding the
fibration of the doubled geometry over the physical base (in a
coordinate patch).

The Riemannian metric on $\CX_{2d}$ is
\begin{equation}
  ds^2_\text{Riem} = \CM_{MN}\CP^M\CP^N = \CH_{MN}(x)\Phi^M\Phi^N,
\end{equation}
where
\begin{equation}
  \CH_{MN}(x) = \CM_{PQ}\,\CV(x)^P{}_M\CV(x)^Q{}_N,
\end{equation}
with no $\CX_{2d}$ coordinate dependence in $\CM_{PQ}$.

Like $\CM_{MN}$, the Riemannian metric $\CH_{MN}(x)$ also is symmetric
$O(d,d)$ matrix satisfying
\begin{equation}
  \CH^T\CL^{-1}\CH = \CL.
\end{equation}
Therefore, it defines a $d$ dimensional symmetric and antisymmetric
tensor fields
\begin{equation}
  G_{mn} = G_{mn}(x)\quad\text{and}\quad \BH_{mn}=\BH_{mn}(x),
\end{equation}
such that
\begin{equation}\label{eq:HmetricGB}
  \CH(x) =
  \begin{pmatrix}
    G + (\BH)^TG^{-1}\BH & (\BH)^TG^{-1}L\\
    L^TG^{-1}\BH & L^TG^{-1}L
  \end{pmatrix}.
\end{equation}
That is, we can locally extract from $\CH_{MN}(x)$ a metric and
$B$-field living in the physical subpace of $(T^*\CX_{2d})^2$ with
respect to our choice of polarization:
\begin{equation}\label{eq:PhysMetricHull}
  ds^2_\text{phys} =  G_{mn}\,p^m\otimes p^n,
  \quad \BH = \ha\BH_{mn}\,p^m\w p^n.
\end{equation}
When $R\ne 0$, we have $p^m = p^m(x,\tilde x)$, so these fields still
functionally depend $\tilde x$, even if their tensorial components lie
strictly in the physical directions.  However, in the case that $R=
0$, the functional dependence on $\tilde x$ drops out, and these
fields can be thought of as pullbacks of quantities defined on the
physical base $X_d = \tilde G_{d}\backslash \CX_{2d}$, or at least the
patchwise analog of this in the local geometry discussion earlier in
this section.

When $R=0$, the field $G_{mn}p^mp^n = G_{mn}p^m{}_rp^n{}_s dx^r dx^s$
is the physical metric in the local geometric description of the
background.  The field $\BH$, while convenient in this framework, is
not quite the standard $B$-field, but only the moduli dependent part.
It determines (the pullback of) the $H$-flux via
\begin{equation}\label{eq:HfluxHull}
  H = d\BH - \frac12 d\bigl(L_m{}^np^m\w\tilde q_n\bigr) - \ha\CK,
\end{equation}
where $d$ is the exterior derivative on the doubled space $\CX_{2d}$,
and $\CK$ was defined in Eq.~\eqref{eq:CalK}.  This expression for $H$
was derived entirely from the worldsheet description of Hull and
Reid-Edwards,
%
in the course of verifying the equivalence of the doubled and standard
sigma model descriptions when $R=0$ (cf.~Sec.~6.2 of
Ref.~\cite{Hull:2009sg}).  The worldsheet description of Hull and
Reid-Edwards takes the form of a gauged sigma model on the doubled
space, with the polarization determining the particular gauging in
each patch.\footnote{It would be interesting to rederive this equation from a
spacetime point of view, and to provide intuition on why this 3-form
has physical components only, i.e., can locally be viewed as a
pullback of a 3-form on physical space when $R=0$.}


\subsection{Doubled description of the WZW model for general group}
\label{sec:DoubG}

We now apply the formalism of the previous subsection to the doubled
description of the level~$n$ WZW model for arbitrary group $G^\WZW$.
As we will see, the doubled space is
\begin{equation*}
  \CX_{2d} = \G\backslash\CG_{2d},\quad\text{where}\quad
  \CG_{2d} = G_1\times G_2,
\end{equation*}
with $G_1$ and $G_2$ two copies of $G^\WZW$, associated with
the left and right moving worldsheet sectors.  The physical space
$X_{d} = G^\WZW_\phys\cong G^\WZW$ is obtained by identifying elements
of $\CG_{2d}$ under the left action of the diagonal subgroup,
\begin{equation*}
  G^\WZW_\phys \cong \tilde G\backslash\CX_{2d}
  \quad\text{where}\quad
  \tilde G = G_{\diag}\subset G_1\times G_2.
\end{equation*}
Other choices of $\tilde G$ conjugate to $G_{\diag}$ give other global
polarizations.  Up to subtleties involving $\G$, the doubled
worldsheet theory of Hull and Reid-Edwards is a gauged WZW model with
group $G_1\times G_2$ and gauge group $G_{\diag}$.

At level $n$, the doubled metrics in the chiral basis are
\begin{align}\label{eq:DoubMetricsWZW}
  ds^2_{O(d,d)} &= -\frac{n}2\tr'\bigl(\l_2\l_2 - \l_1\l_1\bigr)
  = \frac{\hat n}2 d_{mn}\bigl(\l_2^m\l_2^n - \l_1^m\l_1^n\bigr),\\ 
  ds^2_\text{Riem} &= -\frac{n}{2}\tr'\bigl(\l_2\l_2 + \l_1\l_1\bigr)
  = \frac{\hat n}{2}d_{mn}\bigl(\l_2^m\l_2^n + \l_1^m\l_1^n\bigr), 
\end{align}
where $\l_i = g_i^{-1}dg_i$ is the left-invariant Maurer-Cartan form
on $G^\WZW_i$, and $\hat n = n\psi^2/2$.  (See App.~\ref{app:Ws} for
Lie Algebra conventions.)  The second line assumes the WZW point in
moduli space $g_{mn} = \half\hat n d_{mn}$ and $b_{mn} = 0$.  The
formalism of~Sec.~\ref{sec:Doub} gives the metric at a general point
in moduli space, including would-be moduli that are lifted by the
potential.  The global horizontal and vertical forms adapted to the
$\Gtilde$ fibration are given in Eq.~\eqref{eq:12toHorVert}, and in
this basis,
\begin{equation}
    ds^2_\text{Riem} 
    = \frac{\hat n}{2} d_{mn}\Bigl(\frac{r^2}{n}\l_\phys^m\l_\phys^n +
    \frac{n}{r^2} \o^m\o^n\Bigr),
\end{equation}
when the overall radial modulus is allowed to vary away from $n$.  For
generic moduli $g_{mn}$ not proportional to $d_{mn}$, the Riemannian
metric in this basis is position dependent.

We show that the local procedure described in
Sec.~\ref{sec:DoubGenRecovery} indeed gives the correct physical
metric~\eqref{eq:PhysMetricG} and $H$-flux~\eqref{eq:WZWHflux} when
expressed in terms of global 1-forms.  Allowing only the overall
volume modulus to vary, we review the worldsheet arguments for $r^2 =
n$ and show that the effective potential~\eqref{eq:scalarpot} indeed
stabilizes the radial modulus to this value.  In addition to the
standard polarization with $\tilde G = G_{\diag}$, other polarizations
are obtained by conjugation, $\tilde G_\bb = \bb^{-1}G_{\diag}\bb$,
and we relate the abelian \hbox{T-duality} group to a restricted
subgroup of $\bb$.  These are closely related to the spectrum of
D-branes of the theory, as we explain.  Finally, in
Sec.~\ref{sec:DoubGDiscrete}, we discuss the choice of discrete group
$\G$, tentatively identifying $\G/C_{\diag}$ with $\IZ_n$ at level
$n$, where $C$ is the center of $G^\WZW$.


\subsubsection{Symmetries}
\label{sec:DoubGSym}

First consider $G^\WZW=SU(2)$.  The symmetry of the physical target
space $SU(2)_\phys\cong S^3$ is $O(4)$.  The subgroup $SO(4)\subset
O(4)$ is realized as left and right multiplication on points
$\gphys\in SU(2)_\phys$:
\begin{equation}\label{eq:gaction}
  \gphys \mapsto \O_1\gphys\,\O_2^{-1},
  \quad\text{where}\quad
  \O_1\in SU(2)_1
  \quad\text{and}\quad
  \O_2\in SU(2)_2.
\end{equation}
Since the center $-1_i\in SU(2)_i$ gives the same action $k\mapsto -k$
for left or right multiplication, we have $SO(4) \cong
\IZ_2^\text{diag}\backslash\bigl(SU(2)_1\times SU(2)_2\bigr)$, where
$\IZ_2^\text{diag}$ is generated by $(-1,-1)$, and we have chosen for
later convenience to write the quotient on the left.\footnote{A
  different $\IZ_2$, orientation reversal $k\mapsto k^{-1}$ of
  $SU(2)$, promotes $SO(4)$ to $O(4)$.  This $\IZ_2$ interchanges
  $SU(2)_1$ and $SU(2)_2$.  For the oriented string, this is not a
  symmetry of the theory.  }  This is the gauge group in the spacetime
description of the $SU(2)$ WZW model at level $n$.  The $SU(2)_i$
subgroups for $i=1$ and $2$ are associated with currents in the left
and right moving worldsheet sectors.  We use subscripts $1$ and $2$
rather than $L$ and $R$ to avoid confusion with the left and right
$\CG_6$-action on the doubled space $\CX_6$, which does not correspond
to left and right worldsheet sector.

Analogous statements hold when $SU(2)$ is generalized to an arbitrary
Lie group $G^\WZW$.  In this case $\IZ_2^\text{diag}$ becomes
$C_\text{diag}$, where $C$ is the center of $G^\WZW$.  The physical
gauge group from left and right multiplication of points $g_\phys\in
G^\WZW_\phys$ is $C^{\diag}\backslash\bigl(G_1\times G_2\bigr)$.


\subsubsection{The doubled space}
\label{sec:DoubGSpace}

The fully doubled space $\CX_{2d}$ must form a continuous representation
of the gauge group $C^{\diag}\backslash\bigl(G_1\times G_2\bigr)$,
where $G_1$ and $G_2$ are two copies of $G^\WZW$.  Therefore, in the
notation of the previous section, we expect that
\begin{equation}
  \CX_{2d} = \G\backslash\CG_{2d}
  \quad\text{where}\quad
  \CG_{2d} = G_1\times G_2,
\end{equation} 
where $\G\subset\CG_{2d}$ is a discrete subgroup.  For simplicity of
exposition, we will assume trivial $\G$, and consider the modification
(and motivation) for more general $\G$ in
Sec.~\ref{sec:DoubGDiscrete}.

Let us write
\begin{equation}
  \bg = (g_1,g_2)\quad\text{for}\quad
  \bg\in \CG_{2d}= G_1\times G_2, 
\end{equation}
and choose the standard product group composition law on $\CG_{2d}$,
\begin{equation}\label{eq:composition}
  \bg\circ \bg' = (g_1,g_2)\circ (g'_1,g'_2) = (g_1 g'_1,g_2 g'_2).
\end{equation}
Right $\CG_{2d}$-action on $\CX_{2d}=\CG_{2d}$
\begin{equation}
  \bg\mapsto \bg\bOmega^{-1} = (g_1\O_1^{-1},g_2\O_2^{-1})
\end{equation}
should project to the gauge action
\begin{equation}
  \gphys\mapsto \O_1\gphys\O_2^{-1},\quad \gphys\in G^\WZW_\text{phys}, 
\end{equation}
on the physical space.  This implies that the projection from doubled
to physical space is
\begin{equation}\label{eq:ProjDoubledPhys}
  \pi\colon\quad
   \CG_{2d}\to G^\WZW_\phys,
  \quad
  \bg = (g_1,g_2) \mapsto \gphys = g_1^{-1}g_2.
\end{equation}  
To recover the physical geometry $X_d = G^\WZW_\phys$ from the doubled
geometry $\CG_{2d}$, we can quotient $\CG_{2d}$ by the left action of
the diagonal subgroup $\tilde G = G_\text{diag}\subset G_1\times G_2$,
which leaves $g_\phys = g_1^{-1}g_2$ invariant.  Here, $G_\text{diag}$
consists of elements $(\tilde h,\tilde h)\in\CG_{2d}$.  Thus,
\begin{equation}
  G^\WZW_\phys = G_{\diag}\backslash
  \bigl(G_1\times G_2\bigr).
\end{equation}

As noted in App.~\ref{app:Ws}, the general solution to the
classical equations of motion of the WZW model is
\begin{equation}
  g_\phys(z,\bar z) = g_L(z)g_R(\zbar)\qquad\text{(on-shell).}
\end{equation}
Comparing to Eq.~\eqref{eq:ProjDoubledPhys}, we obtain the
on-shell identifications
\begin{equation}\label{eq:g1g2vsgLgR}
  g_1^{-1}(z,\zbar) = g_L(\zbar),\quad
  g_2(z,\zbar) = g_R(z),\qquad
  \text{(on-shell).}
\end{equation}
The doubled sigma model of Ref.~\cite{Hull:2009sg}, promotes $g_L(z)$
and $g_R(\zbar)$ to fields $g_1^{-1}(z,\zbar)$ and $g_2(z,\zbar)$,
each of which has the full $z,\zbar$ dependence, thus doubling the
target space from $G^\WZW_\phys$ to $\CG_{2d}=G_1\times
G_2$.  The choice of polarization in each patch of $\CG_{2d}$
defines a gauging of the doubled sigma model.  In the case at hand,
the gauge identifications precisely implement the left quotient by
$\tilde G = G_{\diag}$ globally, so the formalism of Hull and
Reid-Edwards indeed reproduces the standard physical sigma model.


\subsubsection{Recovery of the physical WZW background}
\label{sec:DoubGRecovery}


\subsubsection*{Embeddings and local trivialization}

Consider an arbitrary simply connected group $G^\WZW$, with doubled
group $\CG_{2d} = G_1\times G_2$ with the group composition
law~\eqref{eq:composition}.  In addition to the standard left and
right embeddings, it is useful to define submanifolds $\Gtilde,G
\subset \CG_{2d}$ composed of diagonal and anti-diagonal diagonal
elements, $(a,a)$ and $(a^{-1},a)$, respectively.  Both map
bijectively to $G^\WZW$, via
\begin{align}
  \tilde\iota\colon\quad& G^\WZW\to \tilde G,\quad a\mapsto (a,a),\\
  \iota\colon\quad & G^\WZW\to G,\quad a\mapsto (a^{-1},a).
\end{align}
Under the $\CG$ group composition law~\eqref{eq:composition}, the
former closes and the latter does not:
\begin{align*}
  \text{On\ }\Gtilde\colon\quad
  & (a,a)\circ (b,b) = (ab,ab),\\
  \text{On\ }G\colon\quad
  & (a^{-1},a)\circ (b^{-1},b) 
  = (a^{-1}b^{-1},ab)\not\equiv \bigl((ab)^{-1},ab).
\end{align*}
Thus, $\Gtilde=G_{\diag}\subset\CG$ is a subgroup of $\CG$ isomorphic
to $G^\WZW$, whereas $G$ is \emph{not} as subgroup, though
topologically both are embeddings $G^\WZW\hookrightarrow\CG_{2d}$.
  
Following Sec.~\ref{sec:DoubGenRecovery}, we write an arbitrary
element of $\CG_{2d}$ as\footnote{The product group of the present
  section presents a notational inconvenience.  To avoid a excess of
  bold type earlier in the paper, $h$ ,$\tilde h$ and $g$ of
  Sec.~\ref{sec:DoubGenRecovery} correspond to $\bh$, $\tilde\bh$ and
  $\bg$ here.}
\begin{equation}\label{eq:ghtildeh}
  \bg(x,\tilde x) = \tilde\bh(\tilde x)\circ \bh(x)
                = (\tilde h h^{-1},\tilde h h).
\end{equation}
Here,
\begin{equation}
  \tilde\bh = \tilde\iota(\tilde h) = (\tilde h,\tilde h) \in \Gtilde
  \quad\text{and}\quad
  \bh = \iota(h) = (h^{-1},h) \in G,
\end{equation}
where $\tilde h,h\in G^\WZW$, parametrized by coordinates $\tilde x$
and $x$, respectively.

The projection
\begin{equation}
  \pi\colon\quad\CG_{2d}\to G^\WZW_\phys,
  \quad\text{mapping}\quad
  \bg = (g_1,g_2)\mapsto \gphys = g_1^{-1}g_2 = h^2,
\end{equation}
gives $\CG$ the structure of a $G^\WZW$ fibration over $G^\WZW$.  This
is the map from doubled group $\CG_{2d}=G_1\times G_2$ to the physical
group $G^\WZW_\phys$ of the standard WZW model. The map
\begin{equation}
  \phi = \iota\circ\tilde\iota\colon\quad G^\WZW\times G^\WZW\to \CG,
  \quad h,\tilde h \mapsto (\tilde h h^{-1},\tilde h h),
\end{equation}
gives a local trivialization of $\CG$ as a $G^\WZW$ fibration, where
we view $\tilde h(\tilde x)$ as the fiber coordinate and $h(x)$ as the
base coordinate.\footnote{\label{foot:LocalSections}Equivalently,
  $\s\colon g_\phys \mapsto (h,h) = (g_\phys^{1/2},g_\phys^{1/2})$
  gives a local section, and then a generic element of $\CG_{2d}$ can
  be written $\bg = \tilde\iota(\tilde h)\circ\s(g_\phys)$.  The
  section is not global, since the function $g_\phys^{1/2}$ can be
  defined locally but not globally on $G_\phys$.  In contrast, the
  left and right embeddings $\s_L\colon g_\phys \mapsto
  (g_\phys^{-1},1)$ and $\s_R\colon g_\phys\mapsto (1,g_\phys)$ do
  give global sections, in terms of which the generic element of
  $\CG_{2d}$ can be written $\bg=\tilde\iota(g_2)\circ\s_L(g_\phys)$
  and $\bg=\tilde\iota(g_1)\circ\s_R(g_\phys)$ respectively.} It is
surjective, but not one-to-one.  However, the many-to-oneness is
entirely due to using $h$ rather than $g_\phys=h^2$ to parametrize the
base.


\subsubsection*{Polarization and $O(d,d)$ metric}

To recover the physical WZW background, we must first choose a
polarization on $\CX_{2d}$ and specify the $O(d,d)$
metric~\eqref{eq:DoubOddMetric}.  The polarization has already been
implicitly defined by Eq.~\eqref{eq:ghtildeh}.  We now make this more
explicit.  In the present context, it is convenient to write
\begin{equation}
  T_M = (Z_m,X_m),\quad x^M = (x^m,\tilde x^m)
\end{equation}
without flipping the index placement for the dual generators and
coordinates.  A chiral basis\footnote{Recall that $g_L = g_1^{-1}$ and
  $g_R = g_2$ from Eq.~\eqref{eq:g1g2vsgLgR}.  Thus the generators in
  the left-moving and right-moving worldsheet sectors are $-T_1$ and
  $T_2$, respectively.} for $\gfrak_{2d}$ is $T^1_m = (t_m,0)$ and
$T^2_m = (0,t_m)$, where the $t_m$ generate the Lie
algebra~\eqref{eq:LieGWZW} of $G^\WZW$.  In terms of these generators,
the polarization is specified by writing
\begin{align}\label{eq:TvsZX}
  -T^1_m &= Z_m - X_m\quad\text{left-moving worldsheet sector,}\\
   T^2_m &= Z_m + X_m\quad\text{right-moving worldsheet sector.}
\end{align}
Then,
\begin{align}
  \tilde\bh(x) &= \exp(\tilde x^mX_m),\quad 
  \tilde h(\tilde x) = \exp(\half\tilde x^mt_m),\\
  \bh(x) &= \exp(x^mZ_m),\quad h(x) = \exp(\half x^mt_m),
  \quad\text{and}\quad g_\phys = h^2 = \exp(x^mt_m).
\end{align}
With these definitions, the Lie algebra of $\CG_{2d}$ in the
$(T^1_m,T^2_m)$ basis is
\begin{equation}\label{eq:LalgChiral}
  [T^1_m,T^1_m] = c_{mn}{}^p T^1_p,\quad
  [T^2_m,T^2_m] = c_{mn}{}^p T^2_p,\quad
  [T^1_m,T^2_n] = 0.
\end{equation}
In the $(Z_m,X_m)$ basis, the Lie algebra is of the general
form~\eqref{eq:Lalg}, with the simpler index placement
\begin{equation}\label{eq:LalgSimpler}
  \begin{split}
    [Z_m,Z_n] & =  K_{mn}{}^p X_p + f_{mn}{}^p Z_p,\\
    [Z_m,X_n] & =  f_{mn}{}^p X_p + Q_{mn}{}^p Z_p,\\
    [X_m,X_n] & =  Q_{mn}{}^p X_p + R_{mn}{}^p Z_p.
  \end{split}
\end{equation}
From Eqs.~\eqref{eq:TvsZX} and~\eqref{eq:LalgChiral}, we find
\begin{equation}\label{eq:DoubWZWKfQR}
  K_{mn}{}^p = Q_{mn}{}^p = \tfrac12 c_{mn}{}^p,\quad
  f_{mn}{}^p = R_{mn}{}^p = 0.
\end{equation}

The root lattice of $\CG$ is the vector sum of the root lattices of
$G_1$ and $G_2$.  Thus, the nonnormalized Killing
form on $\gfrak_{2d}$ in this basis is
\begin{equation}\label{eq:CartanKillingTildeSum}
  \tilde\CD_{MN} = \diag(\tilde d_{mn},\tilde d_{mn}\bigr)
  \qquad\text{($(T^1_m,T^2_m)$ basis).}
\end{equation}
The dual Coxeter numbers of all three algebras is the same,
$h^\vee(\gfrak_{2d}) = h^\vee(\gfrak_1) = h^\vee(\gfrak_2)$, and we
choose the natural convention in which the length squared of long
roots $\psi^2$ is the same as well.  Then, the normalized
Killing form $\CD_{mn} = h^\vee\psi^2\tilde\CD_{mn}$ similarly
satisfies
\begin{equation}\label{eq:CartanKillingSum}
  \CD_{MN} = \diag(d_{mn},d_{mn}\bigr)
  \qquad\text{($(T^1_m,T^2_m)$ basis).}
\end{equation}
In the $(Z_m,X_m)$ basis, these last two equations become
\begin{align}
  \tilde\CD_{MN} &= \half\diag(\tilde d_{mn},\tilde d_{mn}\bigr)
  \qquad\text{($(Z_m,X_m)$ basis),}\label{eq:CartanKillingTildeSumZX}\\
  \CD_{MN} &= \half\diag(d_{mn},d_{mn}\bigr)
  \qquad\text{($(Z_m,X_m)$ basis).}\label{eq:CartanKillingSumZX}
\end{align}

The $O(d,d)$ metric in the $(Z_m,X_m)$ basis is given by
Eq.~\eqref{eq:DoubOddMetric}, with
\begin{equation}\label{eq:LforWZW}
  \CL_{MN} = 
  \begin{pmatrix}
    0 & L_{mn}\\
    L^T_{mn} & 0
  \end{pmatrix},
  \quad\text{where}\quad
  L_{mn} = \tfrac14 n\psi^2 d_{mn} = \tfrac12 \hat n d_{mn}.
\end{equation}
This choice is unique, since $\tfrac12 \hat n d_{mn}$ is the only
quantity with the correct index structure that appears in the
definition~\eqref{eq:WZWaction} of the physical WZW model.  The
flux~\eqref{eq:WZWHflux} $H_{mnp} = \tfrac12 \hat n d_{pq}c_{mn}{}^q$
and (on-shell) metric~\eqref{eq:PhysMetricG} $G_{mn} = \tfrac12\hat n
d_{mn}$ are determined solely by the tensors $\tfrac12 \hat n d_{mn}$
and $c_{mn}{}^p$ on the group manifold $G^\WZW$.


\subsubsection*{Doubled metrics and local fibration structure}

We now perform the local analysis of Sec.~\ref{sec:DoubGenRecovery},
to obtain the local horizontal and vertical 1-forms $p^m$ and
$\qtilde_m$ and the doubled Riemannian and $O(d,d)$ metrics in these
coordinates.  When the radial modulus is stabilized to $r=n$, we show
that these give the quoted results~\eqref{eq:DoubMetricsWZW}.

The $\CG$ left-invariant 1-form is
\begin{equation}\label{eq:CPDoubWZW}
  \begin{split}
    \CP = \bg^{-1} d\bg
   &= (\htilde h^{-1},\htilde h)^{-1}\circ \bigl(d(\htilde h^{-1}),d(\htilde h)\bigr)\\
   &= \bigl(h(\ltilde -\l)h^{-1},h^{-1}(\ltilde + \r)h\bigr)\\
   &= \iota h^{-1}\circ\bigl(\ltilde-\l,\ltilde+\r\bigr)\circ \iota h,
  \end{split}
\end{equation}    
where $\iota h = (h^{-1},h)$, and $\l,\r,\ltilde,\rtilde$ are the
invariant 1-forms constructed from $h(x),\tilde h(x)\in G^\WZW$,
\begin{equation}
  \l(x) = h^{-1}dh,\quad \r(x) = dh\,h^{-1},\quad
  \ltilde(\tilde x) = \htilde^{-1}d\htilde,\quad 
  \rtilde(\tilde x) = d\htilde\,\htilde^{-1}.
\end{equation}
Thus, 
\begin{equation}\label{eq:PhiDoubWZW}
  \Phi = \bigl(\ltilde-\l,\ltilde+\r\bigr)
  = (\ltilde -\l)^m T^1_m + (\ltilde+\r)^m T^2_m
  \quad\text{and}\quad
  \CV = \Ad_{\iota h^{-1}},
\end{equation}
in the notation of Sec.~\ref{sec:DoubGenRecovery}.  In the $(Z_m,X_m)$
basis,
\begin{equation}
  \begin{split}
    \Phi = p^m Z_m + \qtilde^m X_m,
  \end{split}
\end{equation} 
where
\begin{equation}
  \begin{split}
    p^m &= -(\ltilde -\l)^m + (\ltilde +\r)^m = (\l + \r)^m,\\
    \tilde q^m
    &= (\ltilde - \l)^m +(\ltilde +\r)^m = (2\ltilde + \r - \l)^m.
  \end{split}
\end{equation}

The doubled Riemannian metric was given in terms of the metric and
$B$-field moduli $g_{mn}$ and $b_{mn}$ in
Eq.~\eqref{eq:DoubRiemMetric}.  For simplicity, let us set $b_{mn} =
0$ and allow only the overall radial modulus to vary
\begin{equation}\label{eq:ModuliChoice}
  b_{mn} = 0,\quad g_{mn} = \frac14r^2\psi^2 d_{mn}.
\end{equation}
Then, Eq.~\eqref{eq:DoubRiemMetric} becomes
\begin{equation}
  \begin{split}
    ds^2_\text{Riem} &= \frac{r^2\psi^2}{4} d_{mn} p^m p^n +
    \frac{4}{r^2\psi^2} d^{mn} \Bigl(\frac14 n\psi^2 d_{mp}\qtilde^p\Bigr)
    \Bigl(\frac14 n\psi^2 d_{nq}\qtilde_q\Bigr)\\
    & = \frac{\hat n}{2} d_{mn}\Bigl(\frac{r^2}{n}p^m p^n +
    \frac{n}{r^2} q^m q^n\Bigr),\quad
    \hat n = \frac12 \psi^2 n.
  \end{split}
\end{equation}
As we will see below, the modulus $r^2$ is stabilized to $n$.
Therefore,
\begin{equation}
    ds^2_\text{Riem} 
    = \frac{\hat n}{2} d_{mn}\Bigl(p^m p^n + q^m q^n\Bigr)
    = -\frac{n}{4}\tr'\bigl(pp + qq\bigr)
    \qquad\text{($r^2=n$),}
\end{equation}
where $p = p^mt_m$ and $q = q^mt_m$ in terms of the generators $t_m$
of $G^\WZW$.  For the choice of moduli~\eqref{eq:ModuliChoice}, the
$O(d,d)$ metric becomes
\begin{equation}\label{eq:Oddofpq}
  ds^2_{O(d,d)} = \hat n d_{mn} p^m q^n
  = -\frac{\hat n}2 \tr'\bigl(pq\bigr).
\end{equation}

Observing that $\CP = (g_1^{-1}dg_1,g_2^{-1}g_2) = (\l_1,\l_2)$, and
comparing Eqs.~\eqref{eq:CPDoubWZW} and~\eqref{eq:PhiDoubWZW}, we see
that
\begin{equation}
  p = -h^{-1}\l_1h + h\l_2 h^{-1}
  \quad\text{and}\quad
  q = h^{-1}\l_1h + h\l_2 h^{-1},
\end{equation}
which, by the cyclic property of the trace, gives
\begin{align*}
  ds^2_\text{Riem} &= -\frac{n}2 \tr'\bigl(\l_1\l_1 + \l_2\l_2\bigr)
  \qquad\text{($r^2 = n$),}\\
  ds^2_{O(d,d)} &= \frac{n}2 \tr'\bigl(\l_1\l_1 - \l_2\l_2\bigr)
  \qquad\text{($r$ arbitrary),}
\end{align*}
as claimed in Eq.~\eqref{eq:DoubMetricsWZW}.

Following Hull and Reid-Edwards, we would like to interpret $p^m$ and
$\qtilde_m$ as local horizontal and vertical 1-forms on $\CG_{2d}$,
viewed as a fibration of $G_{\diag}$ over $G_\phys$.  Let us examine
these forms.  In the notation of Sec.~\ref{sec:DoubGenRecovery}, we
have
\begin{equation}
  r = (\r^m + \l^m)Z_m + (\r^m-\l^m)X_m
  \quad\text{and}\quad
  \tilde\ell = 2\tilde\l^mX_m,
\end{equation}
so indeed
\begin{align}
  p^m &= r_Z^m = \r^m + \l^m,\\
  \qtilde^m &= \tilde\ell_X^m + r_X^m = (2\ltilde + \r-\l)^m,
\end{align}
where, in terms of the structure constants $c_{mn}{}^p$ of $G_\WZW$,
\begin{equation}
  d\l^p + \tfrac12 c_{mn}{}^p\l^m\w\l^n = 0,\quad
  d\r^p - \tfrac12 c_{mn}{}^p\r^m\w\r^n = 0.
\end{equation}


\subsubsection*{Interpretation of $p$}

The projection $\pi$ takes $g = (g_1,g_2)$ to $\gphys = g_1^{-1}g_2 =
h^2$, so the left invariant 1-form on the physical base $G_\phys\cong
G_\WZW$ is
\begin{equation}
  \begin{split}
    \l_\phys &= \gphys^{-1}d\gphys = h^{-2} d(h^2)\\
    &= h^{-2} (dh\,h + h dh) = h^{-1}(h^{-1} dh + dh\,h^{-1})h\\    
    &= h^{-1}p h.
  \end{split}
\end{equation}
Thus, $p$ is just the left invariant global 1-form $\l_\phys$ on the
physical base $G_\phys$, up to the adjoint action of $h$.\footnote{In
  the end, only $g_\phys=h^2$ and not $h$ should appear in the
  physical metric.  The factor of $\Ad_h$ in $p^m =
  (\Ad_h)^m{}_n\l^n_\phys$ combines with similar factors in the
  vielbein $\CV = \Ad_{\iota h}$ to leave a result for $\CH$ that only
  depends on $h^2$ for all values of the moduli $g_{mn}$ and $b_{mn}$.
  This can also be seen from $ds^2_\CH = \CM_{MN}\CP^m\CP^N$ with $\CP
  = \frac12\bigl(g_\phys(\o - \lphys)g_\phys^{-1},\o +
  \l_\phys\bigl)$.}


\subsubsection*{Interpretation of $q$}

Likewise the global vertical 1-form
\begin{equation}
  \o = h^{-1}q h = 2\l_2 - \l_\phys
\end{equation}
defines the curvature 2-form $\O = d\o + \o\w\o$ of the fibration,
which can be shown to be
\begin{equation}
  \O = \frac12 (\o-\l_\phys)\w(\o-\l_\phys).
\end{equation}
The forms $\o$ and $\O$ on $\CG$ are related to the local potential
$\CA$ and field strength $\CF$ on the base $G_\phys$, via
\begin{equation}
  \CA = \s^*\o,\quad\text{and}\quad \CF = \s^*\O,
\end{equation}
where $\s$ is the corresponding choice of section.  For the local
analysis of Sec.~\ref{sec:DoubGenRecovery}, the choice of local
section is $\s\colon G_\phys\to \CG,\ h^2\mapsto \iota(h) =
(h^{-1},h)$.\bigskip

Note that the $O(d,d)$ metric expressed in terms of $p,q$ in
Eq.~\eqref{eq:Oddofpq} takes the same form when expressed in terms of
$\l_\phys,\o$,
\begin{equation}\label{eq:OddofLambdaOmega}
  ds^2_{O(d,d)} = \hat n d_{mn} \l_\phys^m\,\o^n
  = -\frac{\hat n}2 \tr'\bigl(\l_\phys\,\o\bigr).
\end{equation}
The change of variables $(\l_\phys,\omega) = h^{-1}(p,q)h$ is a local
transformation in $G^\WZW/C^\WZW\subset O(d)\subset O(d,d)$, where
$C^\WZW$ is the center of $G^\WZW$.  In terms of the chiral basis,
\begin{equation}\label{eq:12toHorVert}
  \lambda_\phys = \l_2 - g_\phys^{-1}\l_1 g_\phys
  \quad\text{and}\quad
  \o = \l_2 + g_\phys^{-1}\l_1 g_\phys,
  \quad g_\phys = g_1^{-1}g_2.
\end{equation}
%


\subsubsection*{Global recovery of group metric and $H$-flux}

We now verify that the local analysis of
Sec.~\ref{sec:DoubGenRecovery}, when expressed in terms of the global
horizontal and vertical 1-forms $\l_\phys$ and $\o$, gives the correct
group metric and $H$-flux on the physical space $G^\WZW$.  From
Eqs.~\eqref{eq:PhysMetricHull} and~\eqref{eq:HfluxHull} with
appropriately modified index placement, the physical metric and
$H$-flux are given by
\begin{gather}
  ds^2_\phys = G_{mn} p^m p^n,\\
  H = d\BH -\tfrac12 d(L_{mn}p^m\w\qtilde^n) + \tfrac12\CK.
  \label{eq:HmodifiedIndex}
\end{gather}
For the choice of moduli~\eqref{eq:ModuliChoice} above, this metric
indeed reproduces Eq.~\eqref{eq:PhysMetricG},
\begin{equation}\label{eq:CorrectPhysMetric}
  ds^2_\phys = \frac14 r^2\psi^2 d_{mn} \l^m_\phys \l^n_\phys
  = \frac14 r^2\tr'(\l_\phys\l_\phys).
\end{equation}
(Conjugation of $\l_\phys$ by $h$ leaves the trace invariant and
replaces $\l_\phys$ by $p$ in Eq.~\eqref{eq:CorrectPhysMetric}.)  It
is shown in App.~\ref{app:HfluxDerivation} that
Eq.~\eqref{eq:HmodifiedIndex} also reproduces the
$H$-flux~\eqref{eq:WZWHflux},
\begin{equation}
  H = \frac{\hat n}{12} c_{mnp}\l_\phys^m\w\l_\phys^n\w\l_\phys^p
    = -\frac{n}{12}\Tr'\bigl(\l_\phys\w\l_\phys\w\l_\phys\bigr).
\end{equation}  
Therefore, for the simple choice of moduli above, the doubled
description of Hull and Reid-Edwards correctly reproduces the physical
sigma model background for any WZW model.


\subsubsection{Stabilization of radial modulus}
\label{sec:DoubGModStab}


\subsubsection*{Worldsheet sigma model description at large radius}

For definiteness, this section treats the bosonic string, however, an
analogous discussion can be given for the common NSNS section in the
supersymmetric case.  The worldsheet description of the bosonic
$SU(2)$ WZW model is given in App.~\ref{app:Ws}.  The theory is
conformal provided the beta functions of the sigma model vanish.  At
large radius, it suffices to work to first order in $\a'$.  Allowing
only the overall volume modulus of the $S^3$ to vary, the first order
beta functions are (cf.~Eq.~(15.4.28) of
Ref.~\cite{Polchinski:1998rr}, with $q=2n\a'$)
\begin{subequations}\label{eq:WZWBetaFns}
  \begin{align}
    \b^G_{mn} &= 2 G_{mn}\left(\frac1{r^2} - \frac{n^2}{r^6}\right),\\
    \b^\Phi &= \frac12 -\frac{n^2}{r^6},
  \end{align}
\end{subequations}
where $H = 2n\o_{S^3}$.
Here, $n\in\IZ$ from flux quantization,
\begin{equation}
  \frac1{2\pi}\int_{S^3} H = 2\pi n \in 2\pi\IZ\qquad (2\pi^2 
  = \text{Volume of $S^3$}),
\end{equation}
so solving $\b^G_{mn} = 0$ determines the radius $r^2=n$.  The deficit
in central charge from $c = 6\b^\Phi <3$ is compensated by a surplus
in the noncompact dimensions from a linear dilaton in the radial
direction.  (Here, we have in mind the near horizon ``throat''
geometry $\IR^{6,1}\times S^3$ of a stack of $n$ NS 5-branes.  See
App.~\ref{app:Fivebrane}.)

What enters into the above beta functions on $S^3$ is the metric and
$H$-flux only, since there is no dilaton profile on $S^3$.  For the
full bosonic string theory, we have, in general, to first order in
$\a'$ (cf.~Eq.~(3.7.14) of Ref.~\cite{Polchinski:1998rq}),
\begin{subequations}
  \begin{align}
    \b^G_{MN} &= R_{MN} + 2\nabla_M\pd_N\Phi - \frac14 H_{MPQ}H_N{}^{PQ},\\
    \b^B_{MN} &= -\frac12 e^{2\Phi}\nabla^P\bigl(e^{-2\Phi}H_{PMN}\bigr),\\
    \b^\Phi &= \frac{(D-26)}{6\a'} -\frac12 \nabla^2\Phi + (\pd\Phi)^2 
               -\frac14\frac1{3!}H_{MNP}H^{MNP}.
  \end{align}
\end{subequations}
Setting $R_{mn} = (2/r^2)G_{mn}$, and $\frac12 H_{mpq}H_n{}^{pq} =
4n^2 G_{mn}/r^6$ on $S^3$ gives the quoted $\b^G_{mn}$ in
Eq.~\eqref{eq:WZWBetaFns}.  Setting $\frac1{3!}H_{mnp}H^{mnp} =
4n^2/r^6$ and including only the $S^3$ contribution of $3$ to
$(D-26)$, we obtain the quoted $\b^\Phi$.

A similar analysis can be performed when $SU(2)$ is replaced by an
arbitrary group $G^\WZW$, since the assumption of a homogeneous space
allows a corresponding simplification of the $\b$-function equations.
We will not provide that analysis here, however, the result is that
the first equation of~\eqref{eq:WZWBetaFns} is unchanged, and the
second is multiplied by $\dim(G^\WZW)/3$.


\subsubsection*{Exact CFT description}

As shown in App.~\ref{app:Ws}, when $r^2=n$ the WZW model possesses
chirally conserved currents on the worldsheet.  Upon quantization,
these currents generate a level $n$ affine $SU(2)$ algebra, with
respect to which the WZW model is a Sugawara model, whose Virasoro
algebra is constructed entirely from the currents.  This allows for
many exact statements, including the all $\a'$ order central
charge~\eqref{eq:cSugawara}, which agrees with the sigma model
analysis above to first order in $\a'$, or equivalently, first order
in the $1/n$ expansion.


\subsubsection*{Effective field theory description}

The scalar potential in Hull's doubled formalism was given in
Eq.~\eqref{eq:scalarpot} and
gives
\begin{equation}\label{eq:ExtremizeV}
  \begin{split}
    0 = 4\frac{\pd V(\CM)}{\pd \CM^{MQ}}
      &= \left(\CM^{NR}\CM^{PS}-L^{NR}L^{PS}\right)t_{MNP}t_{QRS}\\
      &= \CM^{NR}\CM^{PS}t_{MNP}t_{QRS}-\tilde D_{MQ},
  \end{split}
\end{equation}
Here, $\CM^{MN}$ is the inverse moduli matrix
\begin{equation}
 \CM^{MN}=
   \begin{pmatrix}
      g^{-1} & g^{-1}bL^{-1T}\\
      L^{-1}b^{T}g^{-1} & L^{-1}(g+b^Tg^{-1}b)L^{-1T}\\
    \end{pmatrix},
\end{equation}
with $L_{mn}$ given by Eq.~\eqref{eq:LforWZW}; $\tilde\CD_{MQ}$ is the
nonnormalized Killing form, given
by~\eqref{eq:CartanKillingTildeSumZX} in the $(Z_m,X_m)$ basis.

We focus on the simplified case~\eqref{eq:ModuliChoice} that only the
radial modulus is allowed to vary: $g_{mn} = \frac14 r^2\psi^2 d_{mn}$
and $b_{mn}=0$.  This gives
\begin{equation}
  \CM^{MN} = \frac2{\hat n}
  \diag\bigl(m^{mn},\tilde m^{mn}\bigr),
  \quad\text{where}\quad
  m^{mn} = (n/r^2)d^{mn},\quad \tilde m^{mn} = (r^2/n) d^{mn}.
\end{equation}
From the structure constants $t^M{}_{NP}$ of
Eq.~\eqref{eq:DoubWZWKfQR}, we find the following nonzero $t_{MNP}$:
\begin{equation}
  K_{mnp} = Q_{mnp} = \frac{\hat n}4 c_{mnp},
  \quad\text{where}\quad c_{mnp} = d_{mq}c^q{}_{np}.
\end{equation}
Then, using $c_{mp}{}^qc_{mq}{}^p = -\tilde d_{mn}$, the nontrivial
components of Eq.~\eqref{eq:ExtremizeV} are the upper left block
\begin{equation}\label{eq:ExtremizeNontrivial}
    \begin{split}
     0 &= m^{nr}m^{ps}K_{mnp}K_{qrs} 
     +  \tilde m^{nr}\tilde m^{ps}Q_{mnp}Q_{qrs} 
     - \frac12\tilde d_{mq}\\
     &= \frac14\bigl((n/r^2)^2 +  (r^2/n)^2 -2\bigr)\tilde d_{mq}\\
     &= \frac14\bigl(n/r^2 - r^2/n\bigr)^2\tilde d_{mq},
    \end{split}
\end{equation}
and lower right block
\begin{equation}
  \begin{split}
    0 &= m^{nr}\tilde m^{ps}Q_{npm}Q_{rsq} 
      +  \tilde m^{nr}m^{ps}Q_{pmn}Q_{sqr}
         - \frac12\tilde d_{mq}\\
      &= \frac14(1 + 1 -2)\tilde d_{mq} = 0,
  \end{split}
\end{equation}
with off-diagonal blocks vanishing identically.  Equality holds
Eq.~\eqref{eq:ExtremizeNontrivial} when
\begin{equation}
  r^2 = n,
\end{equation}
as desired.


\subsubsection{Other global polarizations}
\label{sec:DoubGPol}

We noted in Sec.~\ref{sec:DoubGenRecovery} that any maximal isotropic
subgroup $\tilde G_d\subset\CG_{2d}$, i.e., any dimension $d$ subgroup
that is null with respect to the $O(d,d)$ metric, defines a global
polarization on $\CG_{2d}$.  In addition to the diagonal subgroup of
$\CG_{2d}$, any conjugate group
\begin{equation}\label{eq:Gtildeb}
  \tilde G_\bb = \bb G_{\diag}\bb^{-1}
\end{equation}
satisfies this criterion.  Note that the same group is obtained from
any $\bb = (b_1,b_2)$ that differ only by right multiplication by an
element of $G_{\diag}$.  Therefore $G_\bb$ is determined by
$\pi(\bb^{-1}) = b_1b_2^{-1}$.  Let us define a projection
\begin{equation}
  \pi_\bb\colon \CG_{2d}\to G^\WZW_\phys,
  \quad\text{mapping}\quad
  \bg = (g_1,g_2)\mapsto g_\phys = (b_1^{-1}g_1^{-1}b_1)(b_2^{-1}g_2b_2).
\end{equation}
In this case, we have the (on-shell) identifications
\begin{equation}\label{eq:OnShellIdent}
  g_L(z) = b_1^{-1}g_1^{-1}b_1,\quad g_R(\zbar) = b_2^{-1}g_2b_2.
\end{equation}

Proceeding as in Sec.~\ref{sec:DoubGRecovery}, we have a new group
isomorphism
\begin{equation}
  \tilde\iota_\bb = \bb\circ\tilde\iota\circ\bb^{-1}
  \colon\quad G^\WZW\to \tilde G_\bb,
  \quad a\mapsto (b_1ab_1^{-1},b_2ab_2^{-1}).\\
\end{equation}
We write an arbitrary element of $\CG_{2d}$ as
\begin{equation}\label{eq:ghtildebh}
  \bg(x,\tilde x) = \tilde\bh_\bb(\tilde x)\circ \bh(x)
                = (b_1\tilde h b_1^{-1}h^{-1},b_2\tilde h b_2^{-1} h).
\end{equation}
Here,
\begin{equation}
  \tilde\bh_\bb = \tilde\iota_\bb(\tilde h) 
  = (b_1\tilde h b_1^{-1},b_2\tilde h b_2^{-1}) \in \Gtilde_\bb
  \quad\text{and}\quad
  \bh = \iota(h) = (h^{-1},h) \in G,
\end{equation}
where $\tilde h_\bb,h\in G^\WZW$, parametrized by coordinates $\tilde x$
and $x$, respectively.

The projection $\pi_\bb$ gives $\CG$ the structure of a $G^\WZW$
fibration over $G^\WZW$.  For each choice of $\bb$, it defines the
corresponding map from doubled group $\CG_{2d}=G_1\times G_2$ to the
physical group $G^\WZW_\phys$ of the standard WZW model.  For $\bb =
\bI$, the identity on $\CG_{2d}$, we recover the previously defined
projection and polarization.

The map
\begin{equation}
  \phi(h,\tilde h) = \iota(h)\circ\tilde\iota_\bb(\tilde h)
  \colon\quad G^\WZW\times G^\WZW\to \CG,
\end{equation}
gives a local trivialization of $\CG$ as a $G^\WZW$ fibration, where
we view $\tilde h(\tilde x)$ as the fiber coordinate and $h(x) =
g_\phys^{-1/2}$ as the base coordinate.  (See the corresponding
discussion in Sec.~\ref{sec:DoubGRecovery}, and
Footnote~\ref{foot:LocalSections}.)

For trivial discrete group $\G$, we again recover the physical space
$G^\WZW_\phys$ with metric~\eqref{eq:PhysMetricG} and $n$ units of
$H$-flux, proceeding as before for \emph{any} choice of global
polarization $\tilde G_b$.  When $\G$ is not containted in the center
of $\CG_{2d}$, only special choices of polarization will give rise to
a geometric compactification.  This is the condition of
``$\G$-invariant $Q$'' described in Sec.~\ref{sec:DoubGenRecovery}.


\subsubsection{Abelian T-duality}
\label{sec:DoubGTduality}

For toroidal compactifications, the T-duality group is the subgroup of
the $T^d\times \tilde T^d$ lattice automorphisms preserving the
$O(d,d)$ metric.  The $\tilde G_\bb$ polarization choices of
Sec.~\ref{sec:DoubGPol}, with $\bb$ varying over $\CG_{2d}$, are
expected to be related to nonabelian or Poisson-Lie T-duality, which
is not in general a symmetry beyond tree level in the genus expansion,
but relates also inequivalent conformal field
theories~\cite{Hull:2009sg,ReidEdwards:2010vp}.  However, for $\bb$
lying in a subgroup, which we now describe, the $\tilde G_\bb$
polarization choices correspond to ordinary abelian T-duality.

A choice of Cartan subalgebra of $\gfrak_{2d}$ gives
$\G\backslash\CG_{2d}$ the structure of a $T^r\times\tilde T^r$
fibration, where $r$ is the rank of $G^\WZW$.  The lattice of
$T^r\times\tilde T^r$ is the coroot lattice of $\CG_{2d}$ mod $\G$.
The restriction of the $O(d,d)$ metric to the fiber gives an $O(r,r)$
metric.  One expects that the abelian T-duality group (i.e., T-duality
relative to the Cartan torus as opposed to the full nonabelian group)
will be the subgroup of inner automorphisms
$\CG_{2d}\to\bb^{-1}\CG_{2d}\bb$ that restrict to automorphisms of the
lattice, and that preserve the $O(r,r)$ metric.

The T-duality inversion of a single $U(1)$ then acts as follows.
Given a choice of $U(1)_t\subset G^\WZW$ with generator $t$, we have
two groups $U(1)_Z,U(1)_X\subset\CG_{2d}$, generated by $Z =
\half(-t,t)$ and $X = \half(t,t)$, and consisting of elements of the
form $(\o^{-1},\o)$ and $(\o,\o)$, respectively.  Here, $\o$ is
obtained by exponentiating $t$ to some power.  Now, suppose that
$b^{-1} t b = -t$.  Then, conjugation of $\CG_{2d}$ by $\bb^{-1} =
(b^{-1},1)$ at fixed polarization $\tilde G = G_{\diag}$ interchanges
$Z$ and $X$ and hence the two $U(1)$s.  This is the active point of
view.  The passive point of view fixes $\CG_{2d}$ while conjugating
the polarization as in Eq.~\eqref{eq:Gtildeb}.

More generally, when the lattice of $T^r\times\tilde T^r$ also has a
discrete symmetry in the diagonal subgroup $O(r)_{\diag}\subset
O(r,r)$ containing an element $(b_2,b_2)$, we have $b_2^{-1}tb_2 = t$,
and can follow the conjugation of the previous paragraph with a
conjugation of by $(b_2^{-1},b_2^{-1})$ to obtain a general abelian
T-duality transformation.  This gives total conjugation of $\CG_{2d}$
by $(b_1,b_2)^{-1} = (b_2^{-1},b_2^{-1}) \circ (b_2b_1^{-1},1)$, where
$b_1b_2^{-1} = \pi(\bb^{-1})$ plays the role of $b$ in the previous
paragraph.  Again, this is the active point of view, and the passive
point of view conjugates the polarization from $G_{\diag}$ to
$\Gtilde_\bb$ at fixed $\CG_{2d}$.


\subsubsection{D-branes}
\label{sec:DoubGDbranes}

As noted in Ref.~\cite{Hull:2004in} and studied in detail in
Ref.~\cite{Lawrence:2006ma}, classically, the submanifolds on which we
can wrap D-branes in the fully doubled description are $d$ dimensional
submanifolds that are null with respect to the $O(d,d)$ metric, also
known as maximal isotropic submanifolds.  D-branes wrapped on these
submanifolds project to different combinations of lower dimensional
D-branes in the physical space, depending on the choice of
polarization.

The submanifolds $\Gtilde_\bb$ satisfy the requisite condition.  Let
us make contact with the well known results for D-branes in WZW
models~\cite{Kato:1996nu,Felder:1999ka,Stanciu:1999id,Maldacena:2001ky}.
We refer the reader to App.~\ref{app:WsWZWAffine} for a discussion of
the WZW model current algebra.  Writing $g_\phys = g_L(z)g_R(\zbar)$
in Eq.~\eqref{eq:JandJbar} of that appendix, and identifying $g_L =
g_1^{-1}$ and $g_R = g_2$ on-shell in the standard polarization, as in
Eq.~\eqref{eq:OnShellIdent}, we have
\begin{equation}
  J(z) = -\hat n g_1^{-1}\pd g_1,\quad
  \bar J(\zbar) = -\hat n g_2^{-1}\bar\pd g_2.
\end{equation}
Define $b = \pi(\bb^{-1}) = b_1b_2^{-1}$ and $b' = \pi(\bb) =
b_1^{-1}b_2$, which may be chosen independently.  Since the
submanifold $\Gtilde_\bb$ of \eqref{eq:Gtildeb} is characterized by
the condition $g_1 = b g_2 b^{-1}$, we have $J(z) = b \bar J(\zbar)
b^{-1}$, i.e.,
\begin{equation}\label{eq:BoundaryCondition}
  J^m = \bigl(\Ad_b\bigr)^m{}_n \bar J^n
  \quad\text{on}\quad
  \Gtilde_\bb\subset\CG_{2d}.
\end{equation}
Here, we have set $z=\zbar$ at the boundary of the worldsheet.  This
is precisely Dirichlet boundary condition characterizing D-manifolds
in the doubled geometry, and will project to an identical condition on
the physical D-manifold.  In the standard polarization $\Gtilde
=G_{\diag}$, the physical D-manifold $\pi(\Gtilde_\bb)$ obtained in
this way is the conjugacy class
\begin{equation}
  [b'] = \{g b' g^{-1} \mid g\in G^\WZW_\phys\},
\end{equation}
right multiplied by $b$.  Both
Eq.~\eqref{eq:BoundaryCondition} and the identification of D-manifolds
with conjugacy classes agree with the standard symmetry preserving
D-branes of WZW
models~\cite{Kato:1996nu,Felder:1999ka,Stanciu:1999id}.  Also well
understood are symmetry breaking branes satisfying twisted boundary
conditions $J(z) = b_\phys^{-1} \bar J(\zbar) f(b_\phys)$, where $f$
is a Dynkin diagram automorphism~\cite{Kato:1996nu,Felder:1999ka}.
These are similarly described in the doubled description.


\subsubsection{The discrete group $\G$}
\label{sec:DoubGDiscrete}


\subsubsection*{Conservative observations}

We have already noted in Sec.~\ref{sec:DoubGSym} that
$\IZ_2^\text{diag}\subset\G$ if we require the group of right
isometries of $\CX_{2d}$ to act faithfully as the gauge group.  Let us
assume the standard polarization $\tilde G =G_{\diag}$.  As argued in
the global geometry discussion of Sec.~\ref{sec:DoubGenRecovery}, a
restriction on $\G$ is that conjugation by arbitrary $\bgamma\in\G$
must map $\Gtilde$ to itself.  Let us write $\bgamma = (\g_1,\g_2)$.
Under conjugation by $\bgamma$, an element $(a,a)\in\Gtilde_d$ is
mapped to $(\g_1 a\g_1^{-1},\g_2 a\g_2^{-1})$, which lies in $\Gtilde$
iff $\g_1 a\g_1^{-1} = \g_2 a\g_2^{-1}$.  That is, $\g_2^{-1}\g_1$
commutes with $a$ for arbitrary $a\in G^\WZW$, from which
$\g_2^{-1}\g_1$ lies in the center $C^\WZW$ of $G^\WZW$.  Therefore,
$\bgamma = (\g_1,\g_1 c) = (\g_2 c^{-1},\g_2)$, for some $c\in C$,
i.e.,
\begin{equation}\label{eq:ConservativeGamma}
  \G\subset\Gtilde\times C_2 = \Gtilde\times C_1,
\end{equation}
where $C_1 = (C^\WZW,1)$ and $C_2 = (1,C^\WZW)$.  This result is also
obtained in the $\Gtilde_\bb$ polarizations.

\begin{description}
\item[Case 1: $\G\subset\Gtilde$.]  In this case, the projection $\pi$
  maps $\G$ to the identity, and the physical target space is
  $X_d^\phys = G^\WZW$.

\item[Case 2: $\G\not\subset\Gtilde$.]  In this case, the projection
  $\pi$ maps $\G$ to a nontrivial subgroup $\G^\phys$ of the center
  $C^\WZW$ of $G^\WZW$.  The physical target space is a geometric
  orbifold
  \begin{equation}\label{eq:ConservativeXphys}
    X_d^\phys = \G^\phys\backslash G^\WZW,\quad \G^\phys\subset C^\WZW.
  \end{equation}
\end{description}

The WZW model at level $n$ involves a choice of modular invariant: a
specification of which combinations of representations $R,\bar R$ of
$G^\WZW$ appear in the spectrum for left and right moving states,
respectively, subject to constraints from $\t\to\t+1$ and
$\t\to-1/\t$.  The simplest choice is the diagonal invariant $R=\bar
R$, which should correspond to Case 1 above.  The next simplest
choices are those constructed from outer automorphisms, which are
precisely the quotients by subgroups of the center of Case
2.\footnote{Additional modular invariant follow from the methods of
  conformal embeddings (e.g., $\widehat{su}(2)_{16}\oplus
  \widehat{su}_3\subset (\widehat E_8)_1$, for the $E_7$ modular
  invariant of the $SU(2)$ WZW model at level 16) and Galois
  permutations, however a general construction is lacking.  For a
  overview of modular invarants in WZW models, see Ch.~17 of
  Ref.~\cite{DiFrancesco:1997nk}.}


\subsubsection*{Speculative observations}

We now offer more speculative observations regarding the choice of
discrete group $\G$, focusing on the case $G^\WZW = SU(2)$.  If the
arguments leading to Eq.~\eqref{eq:ConservativeGamma} are correct,
then the result would similarly hold in any other $\tilde G_\bb$
polarization that leads to a geometric compactification.  It is hard
to see how there could then exist more than one polarization
compatible with $\G$, in order to obtain the T-dual physical spaces
$SU(2)$ and $SU(2)/\IZ_n$ depending on polarization.  Moreover,
$\G^\phys\subset C$ in Eq.~\eqref{eq:ConservativeXphys}, so quotients
by $\IZ_{n>2}$ would be impossible for $SU(2)$ with $C=\IZ_2$.
Therefore, let us relax the restriction that conjugation by an
arbitrary $\bgamma\in\G$ map $\tilde G_\bb$ to itself, and proceed
more heuristically.

By analogy to the T-fold discussion in Sec.~\ref{sec:TfoldG}, it is
natural to suppose that at level~$n$, the discrete group is
\begin{equation}\label{eq:GammaAnsatz}
  \G = (\IZ_n)^r \ltimes C^{\diag}\subset G_{\diag},
\end{equation}
where $r$ is the rank of $G^\WZW$ and $C$ is the center of
$\CG_{2d}$.\footnote{Even more desirable would be a quotient of ``all
  directions of $\tilde G$ by a factor of $n$,'' since this would
  allow the quotient to be defined independent of a choice of Cartan
  subalgebra.  This does not appear to be possible.  For example,
  viewing the physical fiber coordinate as $\tilde h^n$ rather than
  $\tilde h$ fails to do the trick, since the Maurer-Cartan form $\CP$
  would not be single valued on this $1/n$-fold cover.}  Here, the
semidirect product notation means that $\G$ is such that
$C^{\diag}\backslash\G = (\IZ_n)^r$. Indeed, for the polarization
choice $\tilde G = G_{\diag}$, the discrete group $\G$ then acts only
on the fibers, so that we have physical space $X_d = G^\WZW_\phys$ and
fibers
\begin{equation}
  \begin{split}
    \G\backslash G^{\diag}
    &\cong \bigl((\IZ_n)^r\ltimes C^\WZW\bigr)\backslash G^\WZW\\
    &\cong (\IZ_n)^r\backslash \bigl(G^\WZW\bigr)^*
    \quad\text{for simply laced $G^\WZW$,}
  \end{split}
\end{equation}
where $*$ denotes the dual of a group,\footnote{The dual of a group is
  obtained by interchanging its root and weight lattices.  For simply
  laced groups, $G^*\cong C\backslash G$, with $C = \IZ_{r+1}$ for
  $SU(r+1)$, $\IZ_4$ ($r$ odd) and $\IZ_2\times\IZ_2$ ($r$ even) for
  $SO(2r)$, and $\IZ_{3,2,1}$ for $E_{6,7,8}$.  The non simply laced
  groups $F_4$ and $G_2$ are self dual and the dual of $Sp(2r)$ is
  $SO(2r+1)$.  See App.~13.A. of Ref.~\cite{DiFrancesco:1997nk}.} in
agreement with the naive intuition that ``the fibers represent the
T-dual space.''  However, this last piece of naive intuition is
incorrect---the T-dual spaces arise from projections associated to
other permissible polarization choices, not from the fibers
themselves---so it is necessary to be more careful.

Suppose that $\bgamma\in\G$.  The identifications
\begin{equation}
  \bg\sim\bgamma\bg
\end{equation}
on $\CG_{2d}$ induce identifications
\begin{equation}\label{eq:gphysIdents}
  g_\phys \sim \pi_\bb(\bgamma\bg) 
\end{equation}
on $\pi_\bb\bigl(\CG_{2d}\bigr) = \Gtilde_\bb\backslash\CG_{2d}$, in
the $\tilde G_\bb$ polarizations of the previous section.
%
%
This gives
\begin{equation}\label{eq:WeakerIdents}
  g_\phys \sim \g_\phys g_\phys,
\end{equation}
%
%
along with other identifications that cannot be written in terms of a
$G^\WZW_L\times G^\WZW_R$ action on $g_\phys$.  Let us tentatively
ignore the others, although it seems unjustified to do so.  Indeed,
these additional identifications are present precisely when the
restriction mentioned above is violated.  Here,
$g_\phys = \pi_\bb(\bg)$ and $\g_\phys = \pi_b(\bgamma)$,
%
and the restriction is equivalent to requiring
\begin{equation}
  \pi_\bb(\bgamma\bg) = \pi_\bb(\bgamma)\pi_\bb(\bg),
\end{equation}
i.e., that the $\G$-action on $\CG_{2d}$ induces a group action on
equivalence classes in $\Gtilde_\bb\backslash\CG_{2d}$.  

We will instead explore the weaker condition implied by
Eq.~\eqref{eq:WeakerIdents}, that the $\gamma_\phys$ form a group,
\begin{equation}
  \pi_\bb(\bgamma_1\bgamma_2) = \pi_\bb(\bgamma_1)\pi_\bb(\bgamma_2).
\end{equation}
For definiteness, let us focus on $SU(2)$.  Then, for the appropriate
choice of $\G$, and a subset of the possible polarizations on $\CG_6$,
we expect to obtain physical space $X_3=SU(2)$ or
$SU(2)/\IZ_n$.\footnote{Other polarizations that are not suitably
  compatible with $\G$, are expected to lead to nongeometric
  compactifications.}  Under what conditions is
$\pi_\bb\colon\G\to\G_\phys$ a group homomorphism with
$\G_\phys=\IZ_n$?

One solution is as follows.  First, suppose that $\G$ is a cyclic
group $\IZ_m\subset G_{\diag}$ generated by $\bomega =
(\o,\o)$. Then,
\begin{equation}
  \o_\phys = (b_1^{-1}\o^{-1}b_1)(b_2^{-1}\o b_2),
\end{equation}
where each factor on the right hand side is a rotation by angle
$4\pi/m$ about some axis.\footnote{In the $SU(2)$ conventions of
  Sec.~\ref{sec:PhysBgSU}, the generators $t_m=-\frac{i}2\s_m$
  multiplied by $4\pi$ exponentiate to unity.}  By suitable choice of
$b_1$ and $b_2$, any desired axis can be obtained for either factor.
Thus, there exist $b_1$ and $b_2$ such that the two factors are
equal,\footnote{Another way to state this condition is $(\o^{-1},\o)
  \in \Gtilde_\bb\cap G$, where $G=\iota(G^\WZW)$ is the submanifold
  of $\CG_6$ of elements of the form $(h^{-1},h)$.  This means that
  $\Gtilde_\bb\cap G$ contains the whole $U(1)\ni(\o^{-1},\o)$.}
\begin{equation}\label{eq:bconstraint}
  b_1^{-1}\o^{-1}b_1=b_2^{-1}\o b_2 \equiv \omega'.
\end{equation}
This equation for $\pi(\bb^{-1}) = b_1b_2^{-1}$ is the condition for
the polarizations $G_\bb$ and $G_{\diag}$ to be related by an abelian
T-duality tranformation, as described in Sec.~\ref{sec:DoubGTduality}.
For any solution,
\begin{equation}
  \o_\phys = \o'^2
  \quad\text{where}\quad \o'^m = 1.
\end{equation}
Choosing $m=2n$, we obtain the desired homomorphism
\begin{equation}
  \pi_\bb\colon \IZ_{2n}\to \IZ_n,
  \quad\text{mapping}\quad
  (\o,\o)\mapsto \o_\phys,
\end{equation}
with kernel $\IZ_2^{\diag}$.

In summary, the tentative conclusion for $G^\WZW=SU(2)$ is as follows.
If there is room for relaxing the
condition~\eqref{eq:ConservativeGamma} and imposing only the
identifications \eqref{eq:WeakerIdents} on the physical space, then
the desired results for $G^\WZW=SU(2)$ are obtained from the choice
\begin{equation}\label{eq:TentativeResultSU}
  \CX_6 = \G\backslash\CG_6,
  \quad\text{where}\quad
  \CG_6 = SU(2)_1\times SU(2)_2
  \quad\text{and}\quad
  \G=\IZ_{2n}\subset SU(2)_{\diag},
\end{equation}
where $(\o,\o)$ denotes a generator of $\IZ_{2n}$.  This is indeed of
the general form~\eqref{eq:TentativeResultSU}.  For polarizations
choices $\tilde G_\bb$ with $\bb$ satisfying
Eq.~\eqref{eq:bconstraint}, the physical space is then the Lens space
$SU(2)_\phys/\G_\phys$, where $\G_\phys = \pi_\bb(\G)\cong\IZ_n$.  For
polarization choices such that $b_\phys$ commutes with $\o$, the
physical space is $SU(2)_\phys$.


\subsubsection*{ADE modular invariants}

At each level $n$, the $SU(2)$ WZW model gives rise to one, two, or
three different CFTs distinguished by a choice of $A$, $D$, or $E$
modular invariant.  The $A_{n+1}$ series exists at all levels $n$.
The $D_{n/2+2}$ series exists at all even levels $n$, and is orbifold
of the $A_{n+1}$ model quotiented by its $\IZ_2$ center, with target
space $SO(3)$.  The $E_6$, $E_7$, and $E_8$ models exist at levels
$n=10$, $16$, and $28$, respectively.  Other free orbifolds exist;
however, these are either equivalent to the ADE models,\footnote{As
  noted in Sec.~\ref{sec:TfoldSU} and App.~\ref{app:WsWZWTduality}, at
  level $n$, the $SU(2)$ and $SU(2)/\IZ_n$ models are equivalent
  as~CFTs.}  or do not have the full
$\IZ_2^{\diag}\backslash\bigl(SU(2)\times SU(2)\bigr)$
symmetry.\footnote{At level $n=n_1n_2$, the free orbifolds
  $SU(2)/\IZ_{n_1}$ and $SU(2)/\IZ_{n_2}$ are equivalent as
  CFTs~\cite{Maldacena:2001ky}.  See Ref.~\cite{Bordalo:2003wy} for
  further generalizations and a discussion of discrete torsion in this
  context.}  It is natural to seek to relate the choice of discrete
group $\G$ to the choice of ADE modular invariant.  Let us focus on
the $A_{n+1}$ and $D_{n/2+2}$ series, which have a large~$n$
semiclassical interpretation:

\begin{description}
\item[$A_{n+1}$ series.]  For the $A_{n+1}$ series, the target space
  is $SU(2)$, therefore $\G\in\Gtilde = SU(2)^\text{diag}$.  If the
  tentative choice~\eqref{eq:TentativeResultSU} is correct, this
  corresponds to $\G=\IZ_{2n}$ and
  $\G/\IZ_2^{\diag}=\IZ_n$.\footnote{From the point of view of the
    doubled space, the $A_{n+1}$ model would then appear to be related
    to the $A_1$ model via a generalized $\IZ_n$ orbifold, however
    this is not quite correct, since the $O(3,3)$ metrics also differ
    by a factor of $n$.}
\item[$D_{n/2+1}$ series.]  For the $D_{n/2+1}$ series to be a
  physical $\IZ_2$ orbifold of the $A_{n+1}$ series for even~$n$, with
  target space $SU(2)/\IZ_2^\text{center}$,
  Eq.~\eqref{eq:ConservativeXphys} implies that $\G$ is the extension
  of that of $A_{n+1}$ by the independent $\IZ_2$ in the
  $\IZ_2\times\IZ_2$ center of $SU(2)\times SU(2)$.
\end{description}

More generally, in the case that the physical space is $SU(2)$ rather
than $SU(2)/\IZ_2^\text{center}$, Eq.~\eqref{eq:ConservativeGamma}
implies that the discrete group $\G$ is a subgroup of
$SU(2)_\text{diag}$ containing $\IZ_2^\text{diag}$, so that
$\G/\IZ_2^\text{diag}\in SO(3)$ is a cyclic, dihedral, or polyhedral
finite group.  This includes many possibilities beyond the $\IZ_n$
cyclic case of our $A_n$ description above, whose physicality will be
explored in future work.


\section{Conclusions}
\label{sec:Conclusions}

\subsubsection*{Summary of results}

The two primary results of this paper are as follows:

\begin{enumerate}
\item A construction of the T-fold and fully doubled descriptions of
  WZW models in the formalism of Ref.~\cite{Hull:2009sg}, using $SU(2)$ as a guide.
\item A demonstration that the procedure given by Hull and
  Reid-Edwards in Ref.~\cite{Hull:2009sg} for recovering physical from
  doubled geometry indeed reproduces the physical WZW
  metric~\eqref{eq:PhysMetricG} and
  $H$-flux~\eqref{eq:WZWHflux}.\footnote{The recovery is trivial in
    the T-fold case.}
\end{enumerate}
Along the way, we have provided several additional details and
consistency checks of the formalism of Hull and Reid-Edwards:

\begin{enumerate}
\item For the T-folds, we have given an interpretation of the total
  space as the group manifold $(U(1)^r)_L\times G^\WZW_R$, where
  $U(1)^r$ is the Cartan torus, and have interpreted the physical
  $T^r$ fibration and dual $\tilde T^r$ fibrations in terms of
  explicit lattices and connection 1-forms.
\item In the fully doubled description, the total space is
  \begin{equation*}
    \CX_{2d} = \G\backslash\bigl(G_1\times G_2\bigr),
  \end{equation*}
  where $G_1$ and $G_2$ are two copies of $G^\WZW$.  
\item Polarization choices are determined by maximal isotropic
  submanifolds\footnote{A maximal isotropic submanifold is a
    $d$-dimensional submanifolds that are null with respect to the
    $O(d,d)$ metric.} of $G_1\times G_2$. A natural family is
  $\Gtilde_\bb = \bb G_{\diag}\bb^{-1}$.
\item Given a polarization choice, the doubled space projects to
  physical target space $\Gtilde\backslash\CG_{2d}$ (up to discrete
  identifications), which is in one-to-one correspondence with
  $G^\WZW$.
\item The map giving the projection is $\pi\colon \bg\mapsto g_\phys =
  g_1^{-1}g_2$ for the diagonal polarization, with suitable
  modification for the other $\Gtilde_\bb$ polarizations.  Thus, we
  can interpret $g_1^{-1}(z,\zbar)$ and $g_2(z,\zbar)$ as off-shell
  analogs of $g_L(z)$ and $g_R(\zbar)$ of the physical WZW model.  The
  appropriate left (right) moving constraint coming from gauging
  $\Gtilde$ in the worldsheet description of Hull and Reid-Edwards.
\item When expressed in terms of global horizontal and vertical
  1-forms $\l_\phys$ and $\o$ defined in Sec.~\ref{sec:DoubGRecovery},
  the local prescription of Hull and Reid-Edwards indeed globally
  reproduces the metric and $H$-flux of the WZW model.
\item As additional consistency checks, we have reproduced the moduli
  constraint $r^2 = n\a'$, the abelian T-duality group, and the
  classical D-brane spectrum, working solely in the doubled
  description.
\end{enumerate}

\subsubsection*{The discrete group $\G$ and recovery of WZW orbifolds}

The main unresolved question within the scope of this paper is the
choice of discrete group~$\G$.
%
%
%
A natural expectation from Secs.~\ref{sec:TfoldG}
and~\ref{sec:DoubGDiscrete} is $\G = \IZ_{2n}$ or $\IZ_n\times\IZ_2$
for the $SU(2)$ model with $A$ modular invariant, and more generally,
that
\begin{equation*}
  \begin{split}
    \G\backslash G^{\diag}
    &\cong \bigl((\IZ_n)^r\ltimes C^\WZW\bigr)\backslash G^\WZW\\
    &\cong (\IZ_n)^r\backslash \bigl(G^\WZW\bigr)^*
    \quad\text{for simply laced $G^\WZW$,}
  \end{split}
\end{equation*}
where $*$ denotes the dual group.  T-duality is known to relate the
physical $G^\WZW$ model and $G^\WZW/(\IZ_n)^r$ free orbifold at level
$n$ (and many intermediate orbifolds in between).  The doubled
description should reproduce all T-dual descriptions, depending on the
choice of polarization.\footnote{It is also tempting to try to relate
  $\G$ to ADE subgroups of $SU(2)$ for the $SU(2)$ model with ADE
  modular invariant.}  Unfortunately, the T-duality analysis in
Sec.~\ref{sec:DoubGDiscrete} combined with the
restriction~\eqref{eq:ConservativeGamma} on $\G$ suggests that $\G$
containing $\IZ_{n>2}$ is incompatible with a global polarization in
the T-dual frame that is expected to give target space $SU(2)/\IZ_n$.

Another possibility is that the $\IZ_n$ quotient is determined
dynamically.  The sigma model of Hull and Reid-Edwards involves a
chiral gauging of a $G_1\times G_2$ model that starts out with
$(G_1\times G_2)_L\times (G_1\times G_2)_R$ global symmetry.  A
subgroup conjugate to the diagonal (vector) subgroup of $(G_1\times
G_2)_L$ is gauged.  By analogy to the story described in
Ref.~\cite{Maldacena:2001ky}, one might expect this gauging to
generate an anomaly in the global antidiagonal (axial) subgroup of
$(G_1\times G_2)_L$, breaking the commutant of the field strength from
a $U(1)$ to a $\IZ_n$ at level $n$.\footnote{Note that the indices 1
  and 2 give the chirality in this context, not L and R.}  A third
possibility is that both stories are correct, with the vestiges of the
global symmetry providing an interpretation and/or means of
eliminating the unwanted identifications beyond
\eqref{eq:WeakerIdents} in Sec.~\ref{sec:DoubGDiscrete}.  A final
possibility is that it is simply not possible to describe the
$G^\WZW/(\IZ_n)^r$ models in this formalism, in a way that makes their
equivalence to the $G^\WZW$ model manifest.

Resolving this issue should provide additional insight into framework
of Ref.~\cite{Hull:2009sg}, perhaps at the level of its quantum
dynamics.  The following is also worth highlighting.  Given a choice
of polarization, the structure constants of $\CG_{2d}$ define the
$K,f,Q,R$ flux of Ref.~\cite{Shelton:2005cf}.  In that polarization,
the $R$-flux is traditionally identified with the obstruction to a
local geometric description (in terms of a standard sigma model), and
the $Q$-flux with an obstruction to global but not local
geometry.\footnote{These statements only apply relative to the
  \emph{particular polarization} used to decompose the structure
  constants into $K,f,Q,R$.  The say nothing about the \emph{existence
    of another polarization} in which a subset of the $K,f,Q,R$ might
  vanish.}  As we have argued in Sec.~\ref{sec:DoubGenRecovery}, the
$Q$-flux is only an obstruction to global geometry, when $Q$ is not
$\G$-invariant, i.e., when conjugation by $\G$ does not preserve the
subgroup
\begin{equation}
  \tilde G\subset\CG_{2d}\colon\quad [X^m,X^n] = Q^{mn}{}_p X_p.
\end{equation}
For polarizations reproducing the physical $SU(2)$ WZW model, it is
fairly clear that $\G\in\Gtilde$, so that this condition is satisfied.
And indeed, the model is geometric.  Unambiguous recovery of the
T-dual $SU(2)/\IZ_n$ from the doubled description will provide a good
probe of the validity of this criterion.

\subsubsection*{Broader questions for the future}

The broader goals toward which this investigation aims are: 

\begin{itemize}
\item[] Effective field theory goal: To generalize the notion of a
  string theory and supergravity compactification to accommodate
  generic gaugings (gauge group and gauge/matter couplings) of the low
  energy effective field theory.
\item[]Microscopic goal: To understand the NSNS sector topological and
  Riemannian choice \emph{defining} a string theory compactification
  in this generalized context.
\end{itemize}

\noindent As discussed in the introduction, there are currently at
least three different approaches toward these goals: the T-fold
description, doubled geometry, and generalized geometry.  As geometric
compactifications, WZW models can be consistently described in all
three formalisms.  Their doubled and generalized geometries are
nontrivial since the $O(d,d)$ structure is twisted relative to the
fiber-base decomposition of 1-forms.  Therefore, they possess features
more typical of nongeometric models, such as $Q$-flux, and should
provide a fruitful context for illuminating all three approaches as
well as the relations among them.  In this paper, we have presented
the T-fold and doubled descriptions of WZW models.  In companion
papers, we hope to present the description of WZW models via
generalized geometry, and the relation between all three descriptions,
building on Ref.~\cite{ReidEdwards:2010vp} which restricted its scope
to $f,K$ or $f,Q$ nonzero.\footnote{Even for T-folds and doubled
  geometry, the general relation is not clear.  For a $T^3$ with
  $H$-flux, the doubled geometry is a $U(1)^3$ fibration over $T^3$,
  and the T-fold seems to be obtained by partial projection.  For the
  chiral WZW models, the T-fold seems to be embedded in the doubled
  geometry as a subgroup.}

Ultimately, what is most interesting is the analogous global
description suitable for arbitrary gaugings of $\CN=2$ or $\CN=1$
supergravities.  The doubled geometry of Hull and Reid-Edwards
describes the common NSNS sector of gauged analogs of toroidal
reductions.  The physical times dual torus $T^d\times \tilde T^d$ is
replaced by a (similarly parallelizable) group manifold $\CG_{2d}$
with simultaneous $O(d,d)$ and Riemannian structure.  For $\CN=2$,
generalized geometry provides an excellent \emph{local} description in
terms of a geometry that doubles the tangent bundle rather then the
space
itself~\cite{Hitchin:2004ut,Gualtieri:2003dx,Jeschek:2004wy,Grana:2005ny,Grana:2006hr}.
However it is not valid \emph{globally} except for geometric
compactifications.  Calabi-Yau $n$-folds themselves have a natural
\hbox{T-fold} description as the fiber product of the two mirror
Strominger-Yau-Zaslow fibrations~\cite{Strominger:1996it} over the
same base.  But, what is the general description of the fully gauged
analog of a Calabi-Yau compactification and can the metric plus
$B$-field topological data be similarly geometrized?
In the $\CN=2$ context, it expected to be a rich structure integrating
the many beautiful results of complex and symplectic geometry and
replacing Hitchin's doubled tangent bundle with a doubling of the
manifold itself.  While the simple parallelizable context described by
Ref.~\cite{Hull:2009sg} might seem special and highly dependent on the
group structure and degree of homogeneity, it does suggest
generalizations,\footnote{For example, $K,f,Q,R$ become
  torsion~\cite{Grana:2005ny,Grana:2006hr,Lawrence:2007jb}.  The
  underlying integral structure could be naturally related to
  Leray-Hirsch spectral sequence, as in Ref.~\cite{Tomasiello:2005bp}.
  Likewise, one expects that the doubled geometry parametrizing
  twisted analogs of K3 or Calabi-Yau compactifications geometrizes
  the space of diffeomorphism and $B$-field transformations to
  $O(d,d)$-compatible diffeomorphisms on the doubled space.} and it
must be remembered that the simplest K3 surface and Calabi-Yau
manifolds are obtained as resolved or deformed orbifolds of
tori.\footnote{Indeed, due to its tractability, intersecting brane
  models have focused almost exclusively on $T^6/(\IZ_2\times\IZ_2)$.}
Might the doubled description of Hull and Reid-Edwards be orbifolded
to yield similar generalizations of reduced symmetry?  And if so, what
is the doubled geometry of K3 surface?

A direction that may serve as a guide is the development of a doubled
effective field theory, furnishing the equations that these doubled
spaces need to satisfy, and from which their structure potentially can
be
deduced~\cite{Hull:2009mi,Hull:2009zb,Hohm:2010jy,Hohm:2010pp,Hohm:2010xe,Hohm:2011ex}.
This effective field theory resembles Hitchin's generalized geometry
in that the physical $B$-field explicitly enters (in constrast to the
formalism of Ref.~\cite{Hull:2009sg}, c.f. Eq.~\eqref{eq:HfluxHull}).
On the other hand, it resembles the formalism of Hull and
Reid-Edwards, in that it is is a theory on a doubled space, not simply
the doubled tangent bundle.  Much headway has been made in this
direction over the past two years, and applications to examples will
likely further illuminate the formalism.  \bigskip


\bigskip\centerline{\bf Acknowledgements}\nobreak\medskip\nobreak

I am grateful to Washington~Taylor and Oliver DeWolfe for first
emphasizing the apparent tension between the fluxes in the physical
and doubled geometry of the $SU(2)$ WZW model, to Gianguido Dall'Agata
and Nikolaos Prezas for correspondence, and to Volker Braun, Nick
Halmagyi, Olaf Hohm, Yaron Oz, Ari Pakman, and Brian Wecht for
enjoyable and illuminating conversations.  I would like to thank the
Center for Theoretical Physics at MIT for its hospitality during a
Junior Faculty Research Leave, and Bryn Mawr College for its support
in making this leave possible.  In addition, I am indebted to KITP
Scholars program and the University of Pennsylvania for their
continued hospitality. This material is based upon work supported by
the National Science Foundation under Grant No.~PHY09-12219.  This
research was supported in part by the National Science Foundation
under Grant No.~PHY05-51164.


\appendix


\section{Lie algebra conventions}
\label{app:LieAlg}

Here, we establish the notation and conventions used for Lie algebras.
The discussion closely follows Sec.~13.1 of
Ref.~\cite{DiFrancesco:1997nk}.

   
\subsection{Basic definitions}
\label{app:LieAlgDef}

Given a basis $\{t_m\}$ for the Lie algebra $\gfrak$ of $G$, we write
\begin{equation}
    \bigl[t_m,t_n\bigr] = c_{mn}{}^p t_p,
\end{equation}
in terms of structure constants $c_{mn}{}^p$.  For $G$ semisimple, the
Killing form
\begin{equation}
  d(X,Y) = -\frac1{h^\vee\psi^2}\tr\bigl(\Ad X\,\Ad Y)
  \quad\Leftrightarrow\quad
  d_{mn} = -\frac1{h^\vee\psi^2}c_{mp}{}^q c_{nq}{}^p
\end{equation}
gives a positive definite inner product on $\gfrak$.  Here, $h^\vee$
is the dual Coxeter number of $\gfrak$, and $\psi^2 =
d^{mn}\psi_m\psi_n$ is the length squared of any long
root.\footnote{Given the root lattice of $\gfrak$, this definition
  determines $d_{mn}$ (and the length-squared of roots) only up to an
  overall rescaling, which is then fixed by specifying $\psi^2$.  In
  the standard normalization convention, $\psi^2=2$.}  A tilde denotes
the Killing form without the prefactors: 
\begin{equation}
  \tilde d_{mn} = -c_{mp}{}^qc_{nq}{}^p.
\end{equation}

Traces in all representations of $\gfrak$ are proportional.  In a
representation $R$, we define the Dynkin index $x_R$ via
\begin{equation}
  \tr_R(t_mt_n) = -\psi^2 x_R d_{mn}.
\end{equation}
The Dynkin index in the adjoint representation is the dual Coxeter
number $h^\vee$, by the definition of $d_{mn}$.\footnote{For
  reference, $h^\vee = N$ for $SU(N)$, $N+1$~for $Sp(2N)$, $N-2$~for
  $SO(N)$, $12,18,248$~for $E_6,E_7,E_8$, $9$~for $F_4$, and $4$~for
$G_2$; furthermore, $x_f=1$ in all of these cases except for $SU(N)$
where $x_f=1/2$.} It is convenient to define a representation
independent trace
\begin{equation}
  \tr'(T_mT_n) = \frac1{x_R}\tr_R(t_mt_n) = -\psi^2 d_{mn}.
\end{equation}
This is the trace that appears in the WZW action as described in
App.~\ref{app:Ws}.

For $SU(2)$, we have $h^\vee=2$, $x_f=1/2$, and conventionally choose
a Lie algebra basis such that $c_{mn}{}^p = \e_{mnp}$.  In the
fundamental (spinor) representation, we represent $t_m$ by
$-\frac{i}2\s_m$, where $\s_1,\s_2,\s_3$ are the Pauli matrices.  Then,
\begin{equation}
  \tr_f(t_mt_n) = -\frac12\d_{mn},\quad
  \tr_{\Ad}(t_mt_n) = -2\d_{mn},\quad
  \tr'(t_mt_n) = -\d_{mn},
\end{equation}
from which $\psi^2 d_{mn} = \frac12\d_{mn}$, $\psi^2\tilde d_{mn} =
2\d_{mn}$, and $G_{mn} = \frac14 r^2\d_{mn}$ for $SU(2)$, as in
Eq.~\eqref{eq:UnitThreeSphere}.


\subsection{Cartan-Weyl basis, roots, and inner products}
\label{app:LieAlgCartan}

For any Lie algebra $\gfrak$, we can choose a maximally commuting set
of generators $H_i$,
\begin{equation}
  [H_i,H_j] = 0,\quad m=1,\ldots,r,
\end{equation}
where $r$ is the rank of $G$.  This subalgebra of $\gfrak$ is called a
Cartan subalgebra $\hfrak$, and exponentiates to generate a maximal
torus $T^r\subset G$, for $G$ simply connected.  In a Cartan-Weyl
basis $\gfrak$, the remaining generators $E_\a$ are chosen so that
they are eigenvectors of the $H_i$,
\begin{equation}
  [H_i,E_\a] = \a_i E_\a.
\end{equation}
For a unitary representation, the $H_i$ are taken to be Hermitian and
$E_{-\a} = E_\a^\dagger$.  Here, we have labeled the generators by
their \emph{roots} $\a = (\a_1,\ldots,\a_r) = \a_i\e^i$, where $\e^i$
is the standard Cartesian basis.  Let $\D$ denote the set of all
roots.  The roots Lie in the dual vector space $\hfrak^*$, since $\a$
gives a natural map from any $\b^iT_i\in\hfrak$ to $\a_i\b^i$.  Using
the Jacobi identity, it can be shown that $[H_i,[E_\a,E_\b]] =
(\a_i+\b_i)E_{\a+\b}$, from which $[E_\a,E_\b]$ is: (i) in the Cartan
subalgebra when $\a+\b = 0$, (ii) proportional to $E_{\a+\b}$ when
$\a+\b\in\D$, and (iii) equal to zero otherwise.  The $E_\a$ are
normalized so that
\begin{equation}
    [E_\a,E_\b] =
    \begin{cases}
      N_{\a\b}E_{\a+\b} & 0\ne \a + \b \in\D,\\ 
      \frac2{|\a|^2}\a\cdot H & \a = -\b,\\
      0 & \text{otherwise},
    \end{cases}
\end{equation}
where $N_{\a\b} =$ constant.  Here $\a\cdot\b = d^{ij}\a_i\b_j$ and
$|\a|^2 = \a\cdot\a$, where $d_{ij}$ is the restriction of the
normalized Killing form to the Cartan subalgebra, and $d^{ij}$ is its
inverse.\footnote{In the Cartan-Weyl basis, it can be shown that
  $d(X,Y)$ is nonzero only for $(X,Y)$ equal to two elements of the
  Cartan subalgebra or $(E_\a,E_{-\a})$.}  Note that $d_{ij}$ gives an
isomorphism
\begin{equation}
  \hfrak\to \hfrak^*,
  \quad\text{mapping}\quad \a^iH_i\mapsto \a = \a_i\e^i,
  \quad\text{where}\quad
  \a_i=d_{ij}\a^j.
\end{equation}
We will refer to this map as the Killing isomorphism.


\subsection{Chevalley basis, coroots, and Cartan matrix}
\label{app:LieAlgChevalley}

It is possible to define a notion of positive roots $\D_+$ and
negative roots $\D_-$ such that $\D_- = -\D_+$ and $\D = \D_+\cup
\D_-$.  Simple roots are positive roots that cannot be written as the
sum of two positive roots.  There are $r$ simple roots $\a^{(i)},
i=1,\dots,r$.  Their inner products define the Cartan matrix
\begin{equation}
  A^i{}_j = \a^{(i)}\cdot\a_{(j)}^\vee.
\end{equation}
whose elements are integers.  Here, $\a^\vee = 2\a/|\a|^2$ is the
coroot associated to a root $\a$.  A Chevalley basis is defined as
follows.  For each simple root, define generators
\begin{equation}
  e_i = E_{\a_{(i)}},\quad f_m = E_{-\a_{(i)}},
  \quad\text{and}\quad
  h_i = \a_{(i)}^\vee\cdot H.
\end{equation}
Then, the Killing isomorphism maps
\begin{equation}
  h_i\mapsto \a_{(i)}^\vee
\end{equation}
and the commutation relations become
\begin{equation}
    \begin{split}
    [h_i,h_j] &= 0,\\
    [h_i,e_j] &= +A^j{}_ie_j\text{ (no sum),}\\
    [h_i,f_j] &= -A^j{}_if_j\text{ (no sum),}\\
    [e_i,f_j] &= \d_{ij}h_j\text{ (no sum).}
  \end{split}
\end{equation}
The remaining generators with roots in $\D_+$ ($\D_-$) are obtained
from multiple commutators of the $e_i$ ($f_i$) among themselves.  This
process terminates, due to the Serre relations
\begin{equation}
  \begin{split}
    [\Ad(e_i)]^{1-A^j{}_i}e_j &= 0,\\
    [\Ad(f_i)]^{1-A^j{}_i}f_j &= 0,
  \end{split}
\end{equation}
where $\Ad(a)b = [a,b]$.

It is conventient to let $e_\a$ ($f_{-\a}$) denote the full set of
generators obtained from the $e_i$ ($f_i$) in this way, including all
nonvanishing multiple commutators, with signs chosen so that $f_{-\a}
= e_{\a}^\dagger$ for a unitary representation.  Then, the Chevalley
basis is $\{h_i,e_{\a},f_{-\a}\}$, with integer structure constants.

The Cartan generators in the Chevalley basis satisfy the useful
property that
\begin{equation}
  \exp(2\pi i h_j) = 1,
  \quad\text{for $j=1,\dots,r$.}
\end{equation}
Thus, after accounting for the periodic identifications, the Killing
isomorphism gives a map between the Cartan torus $T^r$ and
$\hfrak^*/(2\pi\L^\vee)$,
\begin{equation}
  \exp(ix^jh_j) \mapsto x^j\a_{(j)}.
\end{equation}

For $SU(2)$, a Chevalley basis is $h = \s_3$, $e = \half(\s_1+i\s_2)$,
and $f = \half(\s_1-i\s_2)$, satisfying
\begin{equation}
  [e,f] = h,\quad [h,e] = 2e,\quad [h,f] = -2f. 
\end{equation}


\subsection{Lattices}
\label{app:LieAlgLat}

Three lattices are conventionally defined in the vector space
$\hfrak^*$.  The root lattice $\L$, coroot lattice $\L^{\vee}$, and
weight lattice $(\L^{\vee})^*$ are obtained by taking integer linear
combinations of the simple roots $\a^{(i)}$, simple coroots
$\a_{(i)}^\vee$, and weights $w^{(i)}$, respectively.  Here, the
weights are defined by
\begin{equation}
  w^{(i)}\cdot\a_{(j)}^\vee = \d_i{}^j,
\end{equation}
so that the weight lattice is dual to the coroot lattice.  For
$d^{ij}$ computed in the Chevalley basis, we have
$\o^{(i)}=d_{ij}\a_{(j)}^\vee$.

The weight lattice is a sublattice of the root lattice, with quotient
$(\L^\vee)^*/L = C$, the center of the group.  For simply laced groups
(e.g., the ADE groups), we have $(\a^{(i)})^2 = \psi^2$ for all roots.
Then, $\L^\vee = (2/\psi^2)\L$, so that in the standard convention
$\psi^2=2$, roots and coroots agree and $\L = \L^\vee$.


\section{Worldsheet description of the WZW
  model}
\label{app:Ws}

This Appendix reviews the basic results for WZW models.  We use a Lie
algebra convention in which group elements are obtained by
exponentiation of generators without additional factors of $i$.
Otherwise, the discussion closely follows Ch.~15 of
Ref.~\cite{DiFrancesco:1997nk}.  (See also Ref.~\cite{Gepner:1986wi}
and Ch.~15 of Ref.~\cite{Polchinski:1998rr}.)  As a preliminary,
Sec.~\ref{app:WsGeo} first describes the purely
geometric model, with no $H$-flux.  While this model exhibits a global
symmetry, it is not conformal, and there is no chirally conserved
current.  In Sec.~\ref{app:WsWZW}, we introduce $H$-flux via a
Wess-Zumino term to obtain the WZW model.  Allowing the overall volume
modulus to vary, we show obtain a conformal model with chirally
conserved currents at level $n$ when $r^2=n\a'$.  Finally, we describe
the chiral primary states of the $SU(2)$ WZW model at level $n$, and
the T-duality map that between the $SU(2)/\IZ_{p}$ and $SU(2)/\IZ_q$
models at level $n=pq$.


\subsection{Geometric sigma model}
\label{app:WsGeo}


\subsubsection{Action and symmetries}
\label{app:WsGeoAct}

The nonlinear sigma model action describing a string propagating on a
semisimple compact group manifold $G$ of radius~$r$ (relative to the
``unit metric'' $\tfrac14\psi^2d_{mn}$, as defined below) is
\begin{equation}
  S_0 = - \frac{r^2}{16\pi\a'}\int d^2\s \tr'(\pd_a g^{-1}\pd^a g)
  = -\frac{r^2}{8\pi\a'}\int d^2z \tr'(\pd g^{-1}\bar\pd g).
\end{equation}
This action has a natural $G\times G$ global symmetry from left and
right multiplication,
\begin{equation}\label{eq:LRaction}
  g(z,\bar z)\mapsto g' = \O_L\,g(z,\bar z)\O_R^{-1}.
\end{equation}
Here, $g(z,\bar z)$, $\O_L$, and $\O_R$ take values in a unitary
representation of $G$, and $\tr'$ denotes the representation
independent trace, as defined in App.~\ref{app:LieAlg}.

In terms of the left-invariant Maurer Cartan form $\l = \l^mt_m =
g^{-1}dg$ and coordinate fields $X^i$ on the group manifold, the sigma
model action becomes
\begin{equation}
  S_0  = \frac1{2\pi\a'}\int d^2z\,G_{mn}D\l^m\bar D\l^n,
  \quad\text{where}\quad
  G_{mn} = \tfrac14 r^2\psi^2d_{mn}.
\end{equation}
Here, $D\th^m = \l^m{}_i\pd X^i$ and $\bar D\l^m = \l^m{}_i\bar\pd
X^i$.


\subsubsection{Equations of motion and conserved currents}
\label{app:WsGeoEOM}

Under $g\mapsto g+\d g$, the variation of the action is
\begin{equation}\label{eq:nonWZWvariation}
  \begin{split}
    \d S &= -\frac{r^2}{8\pi\a'}\int d^2\s 
    \tr'\bigl(\d g\,g^{-1}\,\pd^a(\pd_ag\,g^{-1})\bigr)\\
    &= -\frac{r^2}{8\pi\a'}\int d^2\s 
    \tr'\bigl(g^{-1}\d g\,\pd^a(g^{-1}\pd_ag)\bigr),
  \end{split}
\end{equation}
from which the equations of motion are
\begin{equation}\label{eq:EOMGeo}
  \pd^a\bigl(\pd_ag\,g^{-1}\bigr) = 0,
  \quad\text{or equivalently,}\quad
  \pd^a\bigl(g^{-1}\pd_a g\bigr) = 0.
\end{equation}
Integration by parts in the first or second line of
Eq.~\eqref{eq:nonWZWvariation} shows that $\d S = 0$ for $\d g\,g^{-1}
=$ constant or $g^{-1}\d g =$ constant, respectively.  This is the
infinitesimal form of the global $G\times G$ symmetry,
\begin{equation}
  \d g(z,\bar z) = \e_L g(z,\bar z) - g(z,\bar z)\e_R,
\end{equation}
which agrees with Eq.~\eqref{eq:LRaction} for $\O_{L,R} =
\exp(\e_{L,R})$ and infinitesimal $\e_{L,R}$.  The corresponding
Noether current
\begin{equation}\label{eq:NoetherGeo}
  J_a = \frac{r^2}{4\a'}\Tr'\bigl(-\e_L\pd_a g\,g^{-1}+\e_R g^{-1}\pd_a g\bigr),
\end{equation}
is conserved, as a consequence of the equations of motion.  The
current conservation law combines $J_z$ and $J_{\bar z}$,
\begin{equation}
  \pd J_{\bar z} + \bar\pd J_z = 0.
\end{equation}
which are not separately conserved, nor is there a local symmetry.


\subsection{Wess-Zumino-Witten model}
\label{app:WsWZW}


\subsubsection{Action and symmetries}
\label{app:WsWZWAct}

It is possible to promote the global $G\times G$ symmetry of the
previous model to a local chiral symmetry
\begin{align}
  &g(z,\bar z)\mapsto g' = \O_L(z)g(z,\bar z)\O_R^{-1}(\bar z),
  \quad\text{(finite),}\\
  &\d g(z,\bar z) = \e_L(z) g(z,\bar z) - g(z,\bar z)\e_R(\bar z),
  \quad\text{(infinitesimal),}
\end{align}
with separately conserved left and right (holomorphic and
antiholomorphic) currents, through the addition of a Wess-Zumino term
proportional to
\begin{equation}
  \G = \frac{1}{24\pi}\int_M \tr'(\l^3),\quad \l = g^{-1}dg,
\end{equation}
where $M$ is any 3-manifold bounded by the worldsheet.  The complete
action is
\begin{equation}\label{eq:WZWaction}
  S = S_0 + n\G 
  = -\frac{r^2}{8\pi\a'}\int d^2z \tr'(\pd g^{-1}\bar\pd g)
  - \frac{n}{24\pi}\int_M \tr'(\l^3).
\end{equation}
The Wess-Zumino term contributes a boundary $H$-flux term to the
action,
\begin{equation}
  \frac1{2\pi\a'}\int_M H,
  \quad\text{where}\quad
  H = -\frac{n\a'}{12}\tr'(\l^3)
\end{equation}
which is well defined in the path integral (independent of the choice
of $M$), provided $n$ is an integer.  From
\begin{equation}
  \tr'(\l\w\l\w\l) 
  = \frac12\l^m\w\l^n\w\l^p\tr'([T_m,T_n]T_p)
\end{equation}
we can also write
\begin{equation}
  H = \frac{\hat n\a'}{12}c_{mnp}\l^m\w\l^n\w\l^p,
  \quad\text{where}\quad
  \hat n = \psi^2 n,\quad c_{mnp} = c_{mnq}d^{qp}.
\end{equation}


\subsubsection{Equations of motion and conserved currents}
\label{app:WsWZWEOM}

The equations of motion~\eqref{eq:EOMGeo} become
\begin{equation}
  \biggl(1 + \frac{n\a'}{r^2}\biggr)\pd\bigl(g^{-1}\bar\pd g\bigr) +
  \biggl(1 - \frac{n\a'}{r^2}\biggr)\bar\pd\bigl(g^{-1}\pd g\bigr) = 0,
\end{equation}
or, equivalently
\begin{equation}
  \biggl(1 - \frac{n\a'}{r^2}\biggr)\pd\bigl(\bar\pd g\,g^{-1}\bigr) +
  \biggl(1 + \frac{n\a'}{r^2}\biggr)\bar\pd\bigl(\pd g\,g^{-1}\bigr) = 0.
\end{equation}
The conserved current~\eqref{eq:NoetherGeo} becomes
\begin{equation}
  \begin{split}
  J_z &= \frac{r^2}{4\a'}\Tr'\biggl[
  -\e_L\biggl(1+\frac{n\a'}{r^2}\biggr)\pd g\,g^{-1}
  +\e_R\biggl(1-\frac{n\a'}{r^2}\biggr)g^{-1}\pd g
  \biggr],\\
  J_{\zbar} &= \frac{r^2}{4\a'}\Tr'\biggl[
  -\e_L\biggl(1-\frac{n\a'}{r^2}\biggr)\bar\pd g\,g^{-1}
  +\e_R\biggl(1+\frac{n\a'}{r^2}\biggr) g^{-1}\bar\pd g
  \biggr],
  \end{split}
\end{equation}

For $r^2 = n\a' > 0$, we obtain a conformal field theory with the
desired chiral conservation laws,\footnote{For $n<0$, the roles of $z$
  and $\bar z$ are interchanged, and $n$ is replaced by $|n|$ in what
  follows.}
\begin{equation}
  \bar\pd J^L_z = 0
  \quad\text{and}\quad
  \pd J^R_{\bar z} = 0,
\end{equation}
where
\begin{equation}\label{eq:LRcurrents}
  J^L_z(z) = -\frac{n}2\tr'\bigl(\e_L \pd g\,g^{-1}\bigr)  
  \quad\text{and}\quad
  J^R_{\bar z}(\bar z) = \frac{n}2\tr'\bigl(\e_R g^{-1}\bar\pd g\bigr),  
\end{equation}
associated with a $G\times G$ current algebra of central charge $n$.
This is the level $n$ WZW model with group $G$.

For $r^2 = n\a' > 0$, the general solution to the classical equations
of motion is
\begin{equation}
  g(z,\bar z) = g_L(z) g_R(\bar z),
\end{equation}
for arbitrary $g_L(z)$, $g_R(\bar z)$, analogous to $X(z,\bar z) =
X_L(z) + X_R(z)$ for a free boson.


\subsubsection{Affine Lie algebra}
\label{app:WsWZWAffine}

Let us write
\begin{align}\label{eq:JandJbar}
  J(z) &= J^m(z)T_m = \hat n \pd g\,g^{-1},\\
  \bar J(\zbar) &= \bar J^m(\zbar) T_m = -\hat n g^{-1}\bar\pd g,
\end{align}
where $\hat n = \psi^2 n/2$.  Then, in terms of $\e_{L,R}=\e^m_{L,R}
T_m$, the currents~\eqref{eq:LRcurrents} become
\begin{align}
  J^L_z(z) &
  = d_{mn}\e_L^mJ^n(z),\\
  J^R_z(\zbar) &
  = d_{mn}\e_R^m \bar J^n(\zbar).
\end{align}
The Laurent coefficients $J^m_k$ of the $J^m(z)$ generate an affine
Lie algebra
\begin{equation}\label{eq:AffineAlg}
  [J^m_k,J^n_l] = f^{mn}{}_p J^p_k - \hat n k d^{mn}\d_{k+l,0},
\end{equation}
with the $\bar J^m_k$ satisfying the same algebra.  Here, $f^{mn}{}_p$
is obtained from $f_{mn}{}^p$ by raising and lowering with $d_{mn}$.
In terms of $\hat n$, the level is $n = 2\hat n/\psi^2$ and is a
nonnegative integer.\footnote{Thus, $n = \hat n$ in the standard
  normalization convention $\psi^2=2$.}

Locally near the identity of $G$, we can expand
\begin{equation}
  g(z,\bar z) = \exp(X^mt_m) = 1 + X^m(z,\bar z)\,t_m + O(X^2)
\end{equation}
to write
\begin{gather}
  S = \frac1{4\pi}\int d^2z\, \hat n d_{mn}\pd X^m\bar\pd X^n + O(X^3),\\
  J^m(z) = \hat n\pd X^m + O(X^2),\quad
  \bar J^m(\zbar) = -\hat n\bar\pd X^m + O(X^2).
\end{gather}
In the neighborhood of the origin, we can treat $X^m$ as a free boson
to confirm that the central term of the $J^mJ^n$ OPE is indeed $\hat n
-d^{mn}/z^2$, in agreement with Eq.~\eqref{eq:AffineAlg}.


\subsubsection{Sugawara description and states}
\label{app:WsWZWStates}

The WZW model is a Sugawara model, which means that it is a CFT whose
stress tensor is constructed entirely from the currents.  The stress
tensor is
\begin{equation}
  T_{zz}(z) = -\frac1{\bigl(n+h^\vee\bigr)\psi^2}:d_{mn}J^m(z)J^n(z):,
\end{equation}
so that the Virasoro generators are
\begin{align}
  L_0 &= -\frac1{\bigl(n+h^\vee\bigr)\psi^2}
  d_{mn}\Bigl(J^m_0 J^n_0 + 2 \sum_{k=1}^\infty J^m_{-k}J^n_k\Bigr),\\
  L_k &= -\frac1{\bigl(n+h^\vee\bigr)\psi^2}
  d_{mn}\sum_{l=-\infty}^\infty J^m_lJ^n_{k-l},\quad k\ne0,
\end{align}
and similarly for $T_{\zbar\zbar}(\zbar)$ and $\bar L_k$

The central charge of the Sugawara model for a Lie algebra $\gfrak$
at level $n$ is
\begin{equation}\label{eq:cSugawara}
  c^{\gfrak,n} = (\dim{\gfrak})\Bigl(1-\frac{h^\vee}{n+h^\vee}\Bigr).
\end{equation}
At large level (large radius), where the semiclassical sigma model
interpretation of the WZW model is a good approximation,
$c^{\gfrak,n}$ falls short of the classical dimension $\dim{\gfrak}$
of the group manifold, by a deficit $(\dim{\gfrak})h^\vee/n +
O(1/n^2)$.  Therefore, in any critical string theory background, the
CFT in the remaining spacetime dimensions will need to compensate for
the deficit.  The simplest possibility is a linear dilaton background
(see App.~\ref{app:Fivebrane}).

The states are as follows.  The primaries $|R i\rangle\otimes |\bar
R\bar\imath\rangle$ are labeled by vectors $i$ and $\bar\imath$ in
representations $R$ and $\bar R$ of $G$, on which $J^m_0$ and $\bar
J^m_0$ act as
\begin{align}
  J^m_0|R i\rangle\otimes |\bar R\bar i\rangle 
  &= (T^m)^i{}_j|Rj\rangle\otimes |\bar R\bar i\rangle,\\
  \bar J^m_0|Ri\rangle\otimes |\bar R\bar i\rangle 
  &= (T^m)^{\bar\imath}{}_{\bar\jmath}|Ri\rangle\otimes |\bar R\bar\jmath\rangle,
\end{align}
where $T^m = d^{mn}T_n$.  The descendents are obtained by acting with
the raising operators $J^m_k$ and $\bar J^n_l$ for $k,l>0$.

Let us restrict to group $G=SU(2)$ and set $\psi^2=2$.  Then, $h^\vee
= 2$ and $d_{mn} = \frac12 \d_{mn}$, so that
\begin{equation}
  c = 3\bigl(1-\frac2{n+2}\bigr),
\end{equation}
and
\begin{align}
  L_0 &= -\frac1{4(n+2)} J^m_0 J^m_0 + N_\text{osc},
  \quad N_\text{osc} = -\frac2{4(n+2)} \sum_{k=1}^\infty J^m_{-k} J^m_k,\\
  \bar L_0 &= -\frac1{4(n+2)} \bar J^m_0 \bar J^m_0 + \bar N_\text{osc},
  \quad \bar N_\text{osc} = -\frac2{4(n+2)} \sum_{k=1}^\infty \bar
  J^m_{-k} \bar J^m_k.
\end{align}
A basis of primary states is given by 
\begin{equation}\label{eq:SUprimaries}
  |jm\rangle\otimes|\bar\jmath\bar m\rangle
  \qquad (G=SU(2)),
\end{equation}
where $j,m$ ($\bar\jmath,\bar m$) are the standard angular momentum
quantum numbers in the left (right) moving sectors, with $2j$ and
$2\bar\jmath$ nonnegative integers.  On the primaries,
\begin{equation}
  L_0 = \frac1{n+2}j(j+1),\quad  
  \bar L_0 = \frac1{n+2}\bar\jmath(\bar\jmath+1)\qquad
  \text{(on primaries).}
\end{equation}
At level $n$, it can be shown that $0\le j,\bar\jmath\le n/2$, so the
number of primaries is finite.  The only question is which pairs $j$
and $\bar j$ appear.  For the standard diagonal modular invariant at
level $n$ (the $A_{n+1}$ modular invariant), all pairs
$j=\bar\jmath\le n/2$ arise.  Aside from the maximum value of $j$,
this is exactly as expected.  The primaries carry the quantum numbers
of the non-oscillator zero modes.  Since $SU(2)\cong S^3$ is simply
connected, there is no winding, and the zero modes are simply the
momenta of a point particle moving on $S^3$.  There are two commuting
momentum components $iJ^3_0=m$ and $i\bar J^3_0=\bar m$, which
genenerate left and right multiplication by $U(1)_{\s_3}$ (i.e.,
shifts of the coordinates $\phi^2$ and $\phi^3$ of
Sec.~\ref{sec:PhysBg}).  For $S^3$ embedded in $\IC^2$ as in
App.~\ref{app:5dTfold}, $m\pm\bar m$ are the angular momenta
corresponding to rotation in each $\IC^1$ (i.e., shifts of $\xi_3$ and
$\xi_2$).

For $n$ even, it is possible to orbifold the $A_{n+1}$ model by the
$\IZ_2$ symmetry $(-1)^{2j}$ to obtain the $D_{n/2+2}$ model.  This
gives the theory with target space $SU(2)/\IZ_2^\text{center} =
SO(3)$.  From the point of view of the $A_n$ primaries, this projects
out half-integer $j=\bar\jmath$, and introduces a twisted sector with
$\bar\jmath = n/2-j$.  The twisted sector consists of integer
$j,\bar\jmath$ for $n\equiv 0\mod 4$ and half-integer $j,\bar\jmath$
for $n\equiv 2\mod 4$.  Finally, at the special levels $n=10,16,28$
there are exceptional models with $E_6,E_7,E_8$ modular invariants
(c.f.~Ref.~\cite{DiFrancesco:1997nk}).


\subsubsection{Orbifolds and T-duality}
\label{app:WsWZWTduality}

For $G=SU(2)$, in contrast to toroidal compactifications, there are
two physical momentum quantum numbers and no winding quantum numbers:
the left and right momenta $m$ and $\bar m$ are eigenvalues of the
state under the generators of physical motions on the group manifold
generated by the $-i\s_3/2$ action by left or right multiplication.
Since $SU(2)$ is simply connected, there are no winding sectors.  A
particle moving on the $SU(2)$ group manifold would have the same
quantum numbers as the primaries~\eqref{eq:SUprimaries}, with the
diagonal constraint $j=\bar\jmath$.

At level $n$, T-duality maps the $SU(2)_n$ WZW model to the freely
acting $\IZ_n$ WZW orbifold of $SU(2)_n$ on the left or
right.\footnote{It is possible to perform either a left or right
  T-duality, corresponding to the choice of $O(2,2)$ elements with
  $\pm1$ on the off-diagonal.} For definiteness, let us assume that it
is the $\IZ_n$ generated by $\o = \exp\bigl(-(2\pi i/n)\s_3\bigr)$
acting on the right, as in Secs.~\ref{sec:PhysBgSU}
and~\ref{sec:TfoldSU}.  More generally, at level $n=pq$, T-duality
maps the orbifold $SU(2)_{pq}/\IZ_p$ to
$SU(2){pq}/\IZ_q$~\cite{Gaberdiel:1995mx,Maldacena:2001ky,Distler:2006}.

The untwisted sector primaries of the orbifold $SU(2)_{pq}/\IZ_p$
consists the subset of states~\eqref{eq:SUprimaries} such that $\bar
m$ is divisible by $p$.  For $p=pq$, only the $\bar m=0$ state
satisfies this condition.  For $p$ a proper divisor of $pq$ there will
be other states as well.  The manifold $SU(2)/\IZ_p$ has fundamental
group $\IZ_p$.  The $\s_3$ Hopf fiber, which is a boundary in
$SU(2)$\footnote{The Hopf fiber of $S^3$ is the boundary of the half
  ``large 2-sphere'' discussed in Footnote~\ref{foot:Hemisphere} of
  App.~\ref{app:5dTfold}.} is now a $\IZ_p$ torsion cycle.  The
twisted sectors are labeled by winding numbers $\bar w=1,\dots,p-1$,
and carry shifted momenta.  The right moving chiral primary states of
the $SU(2)_{pq}/\IZ_p$ model are thus labeled as
\begin{equation}
  |\bar\jmath,\bar m,\bar w\rangle
  \qquad\qquad\bigl(SU(2)_{pq}/\IZ_p\ \text{model}\bigr).
\end{equation}
The $SU(2)_{pq}$ and $SU(2)_{pq}/\IZ_p$ models are \emph{not}
equivalent as conformal field theories for $q\ne1$.  On the other
hand, the $SU(2)_{pq}/\IZ_p$ and $SU(2)_{pq}/\IZ_q$ models \emph{are}
equivalent.  The precise mapping of states can be found in
Refs.~\cite{Maldacena:2001ky,Distler:2006}.  The duality naturally
generalizes to an arbitrary group by choosing independent $\bar
m_i,\bar n_i, p_i, q_i$, $i=1,\dots, r$, for each Cartan generator,
where $r$ is the rank of the group~\cite{Gaberdiel:1995mx}.


\section{$S^3$ as a $T^2$ fibration}
\label{app:5dTfold}

A 3-sphere can also be viewed as a $T^2$ fibration over the interval
$I_1 =[0,\pi]$, with the first $S^1$ shrinking at one end of the
interval and the second $S^1$ shrinking at the other end.  Written in
terms of the coordinates $\phi^1$, $\xi^2 = \ha(\phi^2-\phi^3)$ and
$\xi^3 = \ha(\phi^2+\phi^3)$, the metric on the unit $S^3$ becomes
\begin{equation}\label{eq:Hopfcoords}
  ds^2_{S^3} = \qu (d\phi^1)^2
  + \sin^2\Bigl(\frac{\phi^1}2\Bigr)(d\xi^2)^2
  + \cos^2\Bigl(\frac{\phi^1}2\Bigr)(d\xi^3)^2.
\end{equation}
The coordinates $(\phi^1\!/2,\xi^2,\xi^3)$ are known as Hopf
coordinates.  They arise naturally from the point of view of the
embedding $S^3\subset \IC^2$.  The locus
\begin{equation}\label{eq:SinC}
  S^3 = \Bigl\{(z_1,z_2)\in\IC^2 
  \Bigm| |z_1|^2+|z_2|^2 = 1\Bigr\},
\end{equation}
can be parametrized by
\begin{equation}\label{eq:zxi}
  z_1 = e^{i\xi^2}\sin\Bigl(\frac{\phi^1}2\Bigr)
  \quad\text{and}\quad
  z_2 = e^{i\xi^3}\cos\Bigl(\frac{\phi^1}2\Bigr).
\end{equation}
The usual $\IC^2$ metric $ds^2 = |dz_1|^2 + |dz_2|^2$ restricted to
the $S^3$ gives the metric \eqref{eq:Hopfcoords}.\footnote{\label{foot:Hemisphere}In this
  description, the $\eta^2$ fibration at fixed $\eta^3$ (or vice
  versa) gives half of a great 2-sphere.  For example, consider the
  great 2-sphere $\im z_2 =0$: the hemisphere $\re z_2\ge 0$ is
  obtained from $\eta^3 = 0$ and the hemisphere $\re z_2\le 0$ is
  obtained from $\eta^3=\pi$.}

Since circles in the $T^2$ fiber vanish at the endpoints of the base
$I_1$, the T-dual circles blow-up at these points, and the 5d T-fold
description with $T^2\times \tilde T^2$ fiber over $I_1$ is singular.


\section{The near horizon geometry of an NS5-brane}
\label{app:Fivebrane}

The supergravity background describing an NS5-brane as a soliton in an
ambient 10D flat spacetime of parallel metric $\eta_{\m\n} =
\diag(-1,+1,\dots,+1)$ and transverse metric $G_{mn}$ is
\begin{align}
  ds^2 &= \eta_{\m\n}dx^\m dx^\n + g_{mn} dy^m dy^n,\quad g_{mn} = e^{2\Phi}G_{mn},\\
  e^{2\Phi} &= e^{2\Phi_\infty} + \frac{n}{y^2},\quad y^2 = G_{mn}y^m y^n,\\
  H_{mnp} &=-\text{Vol}_{(g)}{}^q{}_{mnp}\,\pd_q(2\Phi) 
  \quad\Leftrightarrow\quad
  H = 2n\,\o_{S^3}.
\end{align}
Here a subscript $(g)$ denotes the metric $g_{mn}$.  The last equation
can be written as $*_{(g)} H = e^{2\Phi} d\bigl(e^{-2\Phi}\bigr)$,
which corresponds to a generalized calibration of $1$, in the sense of
Ref.~\cite{Gauntlett:2003cy}\footnote{This is as expected, since the
  NS5-brane ``wraps'' a point, with volume form $1$, in the transverse
  $\IR^4$.}  Here our orientation convention is that the volume form
in the metric $G_{mn}$ is $y^3dy\w\o_{S^3}$, where $y$ is the
transverse radial coordinate.  The flux through the angular $S^3$, and
the Bianchi identity for $H$ are
\begin{equation}
  \frac1{2\pi}\int_{S^3}H = 2\pi n,\quad d*\bigl(e^{-2\Phi}H\bigr) = 0.
\end{equation}
Dirac quantization requires that $n\in\IZ$.  For generic constant
$G_{mn}$, the $S^3$ is not round, but has metric defined by
\begin{equation}
  G_{mn} dy^m dy^n = dy^2 + y^2 ds^2_{(S^3)},
  \quad\text{i.e.,}\quad
  ds^2_{(S^3)} = \frac1{y^2}\bigl(G_{mn}dy^mdy^n - dy^2\bigr).
\end{equation}

After compactifying from 10D to 7D, we can think of the NS5-brane as a
domain wall in 7D,
\begin{equation}
  ds^2 = ds^2_7 + e^{2\Phi} y^2 ds^2_{(S^3)},
  \quad\text{where}\quad
  ds^2_7 = \eta_{\m\n}dx^\m dx^\n + e^{2\Phi} dy^2.
\end{equation}
In the 7D interpretation, $y\ge0$ is the coordinate transverse to the
domain wall.

Near $y=0$, we have $e^{2\Phi} \simeq n/y^2$, so the near horizon
geometry is an infinite throat, with 10D metric neatly
factorizing as $\IR^{6,1}\times S^3$:
\begin{equation}
  ds^2\simeq ds^2_{\IR^{6,1}} + ds^2_{(S^3)},
  \quad\text{where}\quad
  ds^2_{\IR^{6,1}} = \eta_{\m\n}dx^\m dx^\n + dx^7 dx^7,
  \quad x^7 = \sqrt{n}\log(y/\sqrt{n}).
\end{equation}
In terms of $x^7$, the near horizon dilaton is linear,
\begin{equation}\label{eq:LinearDilaton}
  \Phi\simeq -x^7/\sqrt{q}.
\end{equation}
The additional central charge from the varying dilaton in $\IR^{6,1}$
exactly compensates for the deficit in central charge from $S^3$.  For
example, in the bosonic theory, the deficit in $c = 6\b^\Phi$ from
Eq.~\eqref{eq:WZWBetaFns} is $6n^2/r^6$ to leading order in $\a'$, and
the surplus from linear dilaton~\eqref{eq:LinearDilaton} is $6/n$, so
the two contributions indeed cancel for $r^2 = n$.

For $G_{mn}\propto \d_{mn}$, the worldsheet CFT describing the near
horizon background factorizes as the $SU(2)_N$ WZW model times a
linear dilaton theory.  Generic $G_{mn}$ are obtained by marginal
deformation of the WZW CFT away from the WZW point.


\section{Derivation of $H$-flux from doubled geometry}
\label{app:HfluxDerivation}

The physical $H$-flux determined by the doubled description of the WZW
model is given by Eq.~\eqref{eq:HmodifiedIndex},
\begin{equation*}
  H = dB^\CM -\half d(L_{mn}p^m\w\qtilde^n) + \half\CK.
\end{equation*}
For $L_{mn}$, $p^m$ and $q^n$ as defined in
Sec.~\ref{sec:DoubGRecovery}, we have
\begin{equation}
  \half L_{mn} p^m\w\qtilde^n = \tfrac{\hat n}{4} d_{mn} p^m\w\qtilde^n
  = -\tfrac{n}8\tr'(p\w\qtilde).
\end{equation}
It can be shown that
\begin{align}
  dp &= \half(p\w\qtilde + \qtilde\w p) - (p\w\ltilde + \ltilde\w p),\\
  dq &= \half(p\w p + \qtilde\w\qtilde) -(\qtilde\w\ltilde + \ltilde\w\qtilde).
\end{align}
Therefore,
\begin{align}
  -d(p\w\qtilde) &= -dp\w\qtilde +p\w\qtilde\\
  &= \half (p\w p\w p - \qtilde\w p\w\qtilde) 
  + \tilde\lambda\w p\w\qtilde -p\w q\w\ltilde,
\end{align}
and
\begin{equation}
  -\half d(L_{mn}p^m\w\qtilde^n)
  = -\tfrac{n}{16}\tr'(p\w p\w p -\qtilde\w p\w\qtilde), 
\end{equation}
using the cyclic property of the trace.  

Next, from Eq.~\eqref{eq:DoubWZWKfQR} for the $t_{MNP}$, we have
\begin{equation}
  \begin{split}
    \CK &= \tfrac16 t_{MNP}\CP^M\CP^N\CP^P = \tfrac16 t_{MNP}\Phi^M\Phi^N\Phi^P\\
    &= \tfrac16 K_{mnp}p^m\w p^n\w p^p 
    + \tfrac12 Q_{mnp}p^m\w\qtilde^n\w\qtilde^p\\
    &=\tfrac{\hat n}4 c_{mnp}
    \bigl(\tfrac16 p^m\w p^n\w p^p + \tfrac12 p^m\w\qtilde^n\w\qtilde^p\bigr)\\
    &= -\tfrac{n}8\tr
    \bigl(\tfrac13 p\w p\w p + p\w\qtilde\w\qtilde\bigr)
  \end{split}
\end{equation}
Combining the last two results, Eq.~\eqref{eq:HmodifiedIndex} becomes
\begin{equation}
  H = d\BH - \tfrac{n}{12}\tr'\bigl(p\w p\w p\bigr)
  = d\BH + \tfrac{\hat n}{12} c_{mnp} p^m\w p^n\w p^p.
\end{equation}
Since $p = h\l_\phys h^{-1}$, the cyclic property of the trace allows
us to eliminate the factors of $h$, leaving the desired result
\begin{equation}
  H    = d\BH - \tfrac{n}{12}\tr'\bigl(\l_\phys\w \l_\phys\w \l_\phys\bigr)
  = d\BH + \tfrac{\hat n}{12} c_{mnp} \l_\phys^m\w \l_\phys^n\w \l_\phys^p.
\end{equation}

For $G_\WZW = SU(2)$, with $\hat n = n$ and $d_{mn}=\half\d_{mn}$ as
in App.~\ref{app:LieAlgDef}, we have $c_{mnp} = 2\e_{mnp}$ in terms of
the totally antisymmetric tensor
\begin{equation*}
  \e_{mnp} \equiv 
  \begin{cases}
    +1 & mnp = \text{cyclic permutation of 123,}\\
    -1 & mnp = \text{anticyclic permutation of 123,}\\
    0  & \text{otherwise}.
  \end{cases}
\end{equation*}
The previous result becomes
\begin{equation}
  \begin{split}
    H &= dB^\CM + \tfrac{n}{24}\e_{rst} \lphys^r\w\lphys^s\w\lphys^t\\
      &= dB^\CM + \tfrac{n}{4}\,8\o_{S^3},
   \end{split}     
\end{equation}
in agreement with Eq.~\eqref{eq:PhysBg}, for a choice of moduli such
that $B^\CM=0$.


\end{document}